\PassOptionsToPackage{nameinlink}{cleveref}
\documentclass[a4paper,UKenglish,cleveref, autoref, thm-restate,numberwithinsect]{lipics-v2021}

\nolinenumbers
\hideLIPIcs

\usepackage{amsfonts,amsmath,amssymb,amsthm}
\usepackage{xspace}
\usepackage{paireddelimiters}
\usepackage{comment}
\usepackage{setspace}
\usepackage{namedtheoremref}

\usepackage{refcommand}
\usepackage{xspace}

\newcommand{\FPT}{\textsf{\textup{FPT}}\xspace}

\newcommand{\NP}{\textsf{\textup{NP}}}

\newcommand{\bigO}{\mathcal{O}}
\newcommand{\polynomial}{\textup{poly}}
\newcommand{\polylog}{\textup{poly}\log}

\newcommand{\nat}{\mathbb{N}}

\usepackage{tikz}
\usetikzlibrary{arrows.meta, positioning}

\tikzset{
  mynode/.style={rectangle, rounded corners, draw, align=center, minimum width=2.8cm},
  myarrow/.style={-{Latex}, thick},
}

\newtheorem*{lemma:main:statement}{Lemma \ref{fact:main-lemma}}
\newcommand{\mainlemma}{Let $\F$ be a finite set of (unlabeled) connected graphs, let $X$ be a set of labels, let~$\Q$ be a~$(\min_{H \in \F}|V(H)|)$-saturated set of connected~$X$-labeled graphs of at most~$\max_{H \in \F}|E(H)|+1$ vertices each, and let~$C$ be an $X$-labeled graph. If all optimal solutions to \Fdeletion on~$C$ leave a $\Q$-minor, then there is a subset~$\Q^*\subseteq\Q$ whose size depends only on~$\F$ and $\EDF(C)$,
such that all optimal solutions leave a~$\Q^*$-minor.}

\newrefcommand{\ED}[1]{\mathsf{ed}_{#1}}
\newrefcommand{\tw}{\mathsf{tw}}
\newrefcommand{\td}{\mathsf{td}}
\newcommand{\F}{\mathcal{F}}
\newcommand{\EDF}{\ED{\F}}
\newcommand{\Q}{\mathcal{Q}}
\newcommand{\R}{\mathcal{R}}
\renewcommand{\H}{\mathcal{H}}
\newcommand{\C}{\mathcal{C}}
\newrefcommand{\opt}{\ensuremath{\textsc{opt}_{\F}}}
\newrefcommand{\optin}[1]{\ensuremath{\textsc{opt}_{#1}\textsc{in}}}
\newrefcommand{\optsol}[1][\F]{\ensuremath{\textsc{optsol}_{#1}}}
\newrefcommand{\optsolin}[1]{\ensuremath{\textsc{optsol}_{#1}\textsc{in}}}
\newrefcommand{\optsolst}{\ensuremath{\textsc{optsol}_{\F}\textsc{st}_{\Q}}}
\newcommand{\defaultoptsolst}{\optsolst(G_A, G_B, G_C, \Pi_A, \Pi_B, \Pi_C, R_B)}
\newrefcommand{\forget}{\textsc{forget}}
\newrefcommand{\folio}{\textsc{folio}}
\newrefcommand{\folioqt}{\textsc{folio}^*_{\Q, t}}
\newrefcommand{\pcs}{\textsc{pcs}}
\newcommand{\pieces}{\pcs}
\newrefcommand{\mpcs}{\textsc{mpcs}}
\newcommand{\multipieces}{\mpcs}
\newcommand{\mpcsplus}[1]{\mpcs_{+#1}}
\newrefcommand{\ext}[1]{\textsc{ext}_{+#1}}
\newcommand{\Fdeletion}{\texorpdfstring{$\F$}{F}-{\sc Minor Deletion}\xspace}
\newcommand{\EliminationDistanceToF}{{\sc Elimination Distance to} $\F$-{\sc Minor-Free}\xspace}
\newrefcommand{\minorleq}{\preceq_m}
\newrefcommand{\cc}{\textsc{cc}}
\newrefcommand{\lab}{\mathsf{Labels}}
\newrefcommand{\bound}{b}
\newrefcommand{\iscon}{\textsc{isCon}}
\newcommand*{\yes}{\textsf{yes}\xspace}
\newcommand*{\fixedspacingcite}[2][]{\!\!\ifthenelse{\equal{#1}{}}{\cite{#2}}{\cite[#1]{#2}}}

\newcommand{\ig}[1]{\textcolor{red}{[Ig: #1]}}
\newcommand{\eric}[1]{{\color{brown}[Eric: #1]}}

\usepackage{scalefnt,cite}
\usepackage{xspace}
\usepackage{amsmath}
\usepackage{graphicx,url}
\usepackage{booktabs}
\usepackage{float}
\usepackage{algpseudocode}
\usepackage{soul}
\usepackage{etoolbox}
\usepackage{cancel}

\usepackage[english,algoruled,longend,ruled,vlined,linesnumbered]{algorithm2e}

\floatname{algorithm}{Listing}

\usepackage{amssymb}
\usepackage{bm}
\usepackage{dsfont}
\usepackage{marvosym}

\usepackage{todonotes}

\usepackage{tikz}
\usetikzlibrary{calc}
\usepackage{boxedminipage}
\usepackage{framed}
\usepackage{xspace}
\usepackage{xifthen}
\usepackage{tabularx}
\usepackage{color}
\usepackage{enumitem}

\definecolor{RED}{RGB}{255,0,0}

\newlength{\RoundedBoxWidth}
\newsavebox{\GrayRoundedBox}
\newenvironment{GrayBox}[1]%
   {\setlength{\RoundedBoxWidth}{.93\textwidth}
    \def\boxheading{#1}
    \begin{lrbox}{\GrayRoundedBox}
       \begin{minipage}{\RoundedBoxWidth}}%
   {   \end{minipage}
    \end{lrbox}
    \begin{center}
    \begin{tikzpicture}%
       \node(Text)[draw=black!20,fill=white,rounded corners,%
             inner sep=2ex,text width=\RoundedBoxWidth]%
             {\usebox{\GrayRoundedBox}};
        \coordinate(x) at (current bounding box.north west);
        \node [draw=white,rectangle,inner sep=3pt,anchor=north west,fill=white]
        at ($(x)+(6pt,.75em)$) {\boxheading};
    \end{tikzpicture}
    \end{center}}

\newenvironment{defproblemx}[2][]{\noindent\ignorespaces%
                                \FrameSep=6pt%
                                \parindent=0pt%
                \vspace*{-1.5em}
                \ifthenelse{\isempty{#1}}{%
                  \begin{GrayBox}{\textsc{#2}}%
                }{%
                  \begin{GrayBox}{\textsc{#2} parameterized by~{#1}}%
                }
                \begin{tabular*}{\textwidth}{@{\hspace{.1em}} >{\itshape} p{1.8cm} p{0.8\textwidth} @{}}%
            }{
                \end{tabular*}%
                \end{GrayBox}%
                \ignorespacesafterend
            }

\newcommand{\defproblema}[3]{%
  \begin{defproblemx}{#1}
    {\bf Instance:}  & #2 \\
    {\bf Question:} & #3
  \end{defproblemx}
}%

\newcommand{\defproblemaparam}[4]{%
  \begin{defproblemx}{#1}
    {\bf Instance:}  & #2 \\
    {\bf Parameter:}  & #3 \\
    {\bf Question:} & #4
  \end{defproblemx}
}%

\newtheorem*{customprop*}{Proposition}
\newtheorem*{customlmm*}{Lemma}
\newtheorem*{customdef*}{Definition}
\newtheorem*{customthm*}{Theorem}
\newtheorem*{customclaim*}{Claim}
\newtheorem*{customcor*}{Corollary}
\newtheorem*{customobs*}{Observation}

 \hypersetup{
     colorlinks, linkcolor={dark-blue},
    citecolor={mid-green}, urlcolor={dark-blue}}

  \definecolor{mid-green}{rgb}{0.15,0.65,0.15}
 \definecolor{dark-green}{rgb}{0.15,0.25,0.15}
 \definecolor{dark-red}{rgb}{0.7,0.15,0.15}
 \definecolor{dark-blue}{rgb}{0.15,0.15,0.9}
 \definecolor{medium-blue}{rgb}{0,0,0.5}
 \definecolor{gray}{rgb}{0.5,0.5,0.5}
 \definecolor{color-Ig}{rgb}{0.15,0.7,0.15}
 \definecolor{darkmagenta}{rgb}{0.30, 0.0, 0.30}
 \definecolor{blue}{rgb}{0.15,0.15,0.9}

\renewcommand{\NP}{{\sf NP}\xspace}
\newcommand{\coNP}{{\sf coNP}\xspace}
\newcommand{\poly}{{\sf poly}\xspace}

\AtBeginEnvironment{claimproof}{} % Fix \qedhere in claimproofs

\newcommand{\FdeletionHittingLabeledQ}{\Fdeletion \textsc{Hitting Labeled $\Q$}\xspace}
\newcommand{\FdeletionHittingQ}{\Fdeletion \textsc{Hitting $\Q$}\xspace}

\title{Kernelization dichotomies for hitting minors under structural parameterizations}

\author{Marin Bougeret}{LIRMM, Université de Montpellier, CNRS, Montpellier, France \and \url{https://www.lirmm.fr/~bougeret/} }{marin.bougeret@lirmm.fr}{https://orcid.org/0000-0002-9910-4656}{French project \textsc{ELiT} (ANR-20-CE48-0008-01).}

\author{Eric Brandwein}{Departamento de Computación, FCEyN, Universidad de Buenos Aires, Argentina \and \url{https://eric.com.ar/}}{ebrandwein@dc.uba.ar}{https://orcid.org/0009-0003-2559-7173}{French project \textsc{ELiT} (ANR-20-CE48-0008-01).}

\author{Ignasi Sau}{LIRMM, Université de Montpellier, CNRS, Montpellier, France \and \url{https://www.lirmm.fr/~sau/} }{ignasi.sau@lirmm.fr}{https://orcid.org/0000-0002-8981-9287}{French project \textsc{ELiT} (ANR-20-CE48-0008-01).}

\authorrunning{M. Bougeret, E. Brandwein, and I. Sau}

\Copyright{Bougeret, Brandwein, and Sau}



\relatedversion{An extended abstract presenting the kernelization results of this paper appeared in the \emph{Proceedings of STACS 2026, volume 364 of LIPIcs, pages 17:1--17:19}. This extended abstract did not contain the lower bounds presented in \autoref{sec:lower-bounds} and \autoref{sec:lower-bounds-treelike} of the current submission, nor the analysis of the size of the kernels presented in \autoref{sec:explicit-upper-bounds}.}

\begin{document}

\maketitle

%\ig{WATCH OUT: I left some left remarks such as this one in the text. Take a look}

\begin{abstract}
For a finite collection of connected graphs $\F$, the \Fdeletion problem consists in, given a graph $G$ and an integer $\ell$, deciding whether $G$ contains a vertex set of size at most $\ell$ whose removal results in an $\mathcal{F}$-minor-free graph. We lift the existence of (approximate) polynomial kernels for \Fdeletion by the solution size to (approximate) polynomial kernels parameterized by the vertex-deletion distance to graphs of bounded elimination distance to $\mathcal{F}$-minor-free graphs. This results in exact polynomial kernels for every family $\mathcal{F}$ that contains a planar graph, and an approximate polynomial kernel for \textsc{Planar Vertex Deletion}. Moreover, combining our result with a previous lower bound, we obtain the following infinite set of dichotomies, assuming $\NP \not\subseteq \coNP/\poly$: for any finite set $\F$  of biconnected graphs on at least three vertices containing a planar graph, and any minor-closed class of graphs ${\mathcal C}$, \Fdeletion admits a polynomial kernel parameterized by the vertex-deletion distance to ${\mathcal C}$ if and only if ${\mathcal C}$ has bounded elimination distance to $\F$-minor-free graphs. For instance, this yields dichotomies for \textsc{Cactus Vertex Deletion}, \textsc{Outerplanar Vertex Deletion}, and \textsc{Treewidth-$t$ Vertex Deletion} for every integer $t \geq 0$.  We also prove that the same dichotomies hold for two different infinite classes of families $\F$ that contain non-biconnected graphs, by establishing appropriate lower bounds. Prior to our work, such dichotomies were only known for the particular cases of \textsc{Vertex Cover} and \textsc{Feedback Vertex Set}.

Additionally, via a slight modification of known lower bounds, we note that, when $\C$ is not only minor-closed but also closed under taking disjoint union, there exist no polynomial kernels of size $\bigO(n^{d - \varepsilon})$ for any $\varepsilon > 0$, where $d$ grows with the elimination distance to $\F$-minor-free graphs of $\C$, assuming $\NP \not\subseteq \coNP/\poly$.

Our approach for the polynomial kernels builds on the techniques developed by Jansen and Pieterse [Theor. Comput. Sci. 2020] and also uses adaptations of some of the results by Jansen, de Kroon, and W{\l}odarczyk [STOC 2021]. For the lower bounds, we adapt the techniques used by Dekker and Jansen [Discrete Applied Mathematics 2024].

%\ig{IMPORTANT: speak about lower bounds here as well. Take some ``blabla'' from \autoref{sec:lower-bounds} once it is stable, and from what I wrote in the intro}\eric{Done, please take a look to see if we should add something else.}
\end{abstract}

\newpage

\section{Introduction}
\label{sec:new-intro}
The field of \emph{parameterized complexity} studies the computational complexity of problems when a parameter $k \in \nat$ is given in addition to the input. One of the main objectives of the field is to find efficient preprocessing algorithms called \emph{kernelization algorithms} (or \emph{kernels}), which are polynomial-time algorithms that transform an instance of a parameterized problem into an equivalent instance whose size is bounded by a function of the parameter $k$. Of particular interest are \emph{polynomial kernels}, which are kernels that produce instances of size bounded by a polynomial in $k$. See~\cite{FominLSZ19,CyganFKLMPPS15,DowneyF13,Niedermeier06,FlumG06} for monographs on the area.

A very active direction within kernelization deals with so-called \emph{structural parameters}. The idea is, for a given problem $\Pi$, to unveil the ``smallest'' parameter (usually related to the structure of the input graph) for which $\Pi$ admits a polynomial kernel. Ideally, the holy grail is to find a dichotomy describing which parameterizations allow for a polynomial kernel and which do not, subject to reasonable complexity assumptions. Not surprisingly, finding such dichotomies turns out to be quite hard, as we proceed to discuss.

The \textsc{Vertex Cover} problem, which consists in deciding whether a graph $G$ contains a vertex set of size at most $\ell$ that intersects all edges, has usually served as a testbed for new techniques in parameterized complexity, and in particular in kernelization with structural parameters. Given that \textsc{Vertex Cover} is well-known to admit a polynomial kernel parameterized by the size of the desired solution~\cite{FominLSZ19}, the challenge is to find parameters, potentially smaller than the size of a minimum vertex cover (which is called the \emph{vertex cover number}), that still permit to obtain polynomial kernels. A very convenient and robust way of describing such structural parameters is by considering the \emph{vertex-deletion distance} of the input graph $G$ to a fixed graph class ${\mathcal C}$, defined as the minimum size of a vertex set $X$ such that $G \setminus X \in {\mathcal C}$; such a set $X$ is called a \emph{modulator}  to ${\mathcal C}$. Note that the vertex cover number corresponds to the vertex-deletion distance to the class of empty graphs.

Bodlaender and Jansen~\cite{JansenB13} proved a very influential result in this direction, namely a polynomial kernel for \textsc{Vertex Cover} parameterized by the {\em feedback vertex number} of the input graph, that is, the vertex-deletion distance to the class of forests. This result triggered a number of polynomial kernels for \textsc{Vertex Cover} parameterized by the vertex-deletion distance to other graph classes, such as graphs of
maximum degree two~\cite{MajumdarRR18}, graphs of constant treedepth~\cite{vertex-cover-treedepth-poly-kernel-Algorithmica}, pseudo-forests~\cite{FominS16}, or $d$-pseudo-forests~\cite{HolsK17}. It is worth noting that all the classes ${\mathcal C}$ mentioned so far are {\em minor-closed}, that is, if a graph is in ${\mathcal C}$, then any graph obtained from it by removing vertices or edges, or by contracting edges, is also in ${\mathcal C}$. Bougeret, Jansen, and Sau~\cite{vertex-cover-bridge-depth} culminated this line of research by proving the following dichotomy: assuming $\NP \not\subseteq \coNP/\poly$ (which is the standard hypothesis in this area), \textsc{Vertex Cover} parameterized by the vertex-deletion distance to a minor-closed graph class ${\mathcal C}$ admits a polynomial kernel if and only if ${\mathcal C}$ has bounded {\em bridge-depth}. Here, bridge-depth is a newly introduced graph parameter that can be seen as a common generalization of feedback vertex number and tree-depth, in the sense that it is (functionally) smaller than both of them; see~\cite{vertex-cover-bridge-depth} for the precise definition.

It is worth mentioning that, using randomized
algorithms with a small error probability, polynomial kernels for \textsc{Vertex Cover} are also known for several
parameterizations by the vertex-deletion distance to graph classes that are not minor-closed,
such as K{\"o}nig graphs~\cite{KratschW12}, bipartite graphs~\cite{KratschW12}, and parameterizations based on the linear programming relaxation of \textsc{Vertex Cover}~\cite{HolsKP19,Kratsch18}. However, for non-minor-closed graph classes, we are still far from a dichotomy. Thus, if one aims at obtaining similar dichotomies for generalizations of \textsc{Vertex Cover}, it is reasonable to stick to parameterizations defined as the vertex-deletion distance to a {\sl minor-closed} graph class, and this is what we do in this article.

A very natural way of generalizing the \textsc{Vertex Cover} problem is by fixing a finite family of graphs ${\mathcal F}$ and considering the \Fdeletion problem, defined as follows: given a graph $G$ and an integer $\ell$, the goal is to decide whether at most $\ell$ vertices can be removed from $G$ so that the resulting graph does not contain any of the graphs in ${\mathcal F}$ as a minor. Note that \textsc{Vertex Cover} corresponds to the case ${\mathcal F}=\{K_2\}$. The \Fdeletion problem has attracted great interest in the last years within the parameterized complexity community~\cite{planar-F-deletion-kernel,GiannopoulouJLS17,vertex-planarization-approximate-kernel,deterministic-approximation-for-planar-F-deletion,minor-representative-families,SauST23,SauST22,LPRRSS16}, in particular in kernelization. Namely, when parameterizing by the solution size, Fomin et al.~\cite{planar-F-deletion-kernel} showed that \Fdeletion admits a {\sl randomized} polynomial kernel whenever $\F$ contains at least one planar graph. It can be checked that the only randomized step in their kernel is a constant-factor approximation for the problem. In a subsequent work,
Gupta et al.~\cite{deterministic-approximation-for-planar-F-deletion} provided a
{\sl deterministic} constant-factor approximation for \Fdeletion, which together with the proof of Fomin et al.~\cite{planar-F-deletion-kernel} yield a deterministic polynomial kernel for \Fdeletion parameterized by the solution size when ${\mathcal F}$ contains a planar graph. For collections ${\mathcal F}$ containing only non-planar graphs, the existence of polynomial kernels for \Fdeletion is one of the most notorious open problems in the field of kernelization~\cite{planar-F-deletion-kernel,GiannopoulouJLS17,vertex-planarization-approximate-kernel,FominLSZ19,quo-vadis25,open-problems-worker}.

Probably, the most relevant open case is the case ${\mathcal F}=\{K_5,K_{3,3}\}$, commonly known as \textsc{Planar Vertex Deletion}, which is conjectured to admit a polynomial kernel~\cite{quo-vadis25}. On the positive side, recently Jansen and W{\l}odarczyk~\cite{vertex-planarization-approximate-kernel} presented an {\sl approximate} kernel for \textsc{Planar Vertex Deletion} by using intricate topological arguments.

Given the apparent hardness of finding polynomial kernels for the general \Fdeletion problem parameterized by the solution size, it is natural to consider structural parameters that are not necessarily smaller than the solution size. In an article that is crucial to our work, Jansen and Pieterse~\cite{minor-hitting} provided a polynomial kernel for \Fdeletion, when ${\mathcal F}$ contains only connected graphs, parameterized by the vertex-deletion distance to a graph of constant treedepth. Note that this parameter may indeed be larger than the solution size.

So far, there is only one more particular case of \Fdeletion, other than \textsc{Vertex Cover}, for which a kernelization dichotomy is known. This ``outlier'' is \textsc{Feedback Vertex Set}, corresponding to the case ${\mathcal F}=\{K_3\}$, and for which polynomial kernels by the solution size are well-known~\cite{FominLSZ19}. One may expect that, similarly to the dichotomy for \textsc{Vertex Cover} discussed above~\cite{vertex-cover-bridge-depth}, the dichotomy for \textsc{Feedback Vertex Set} is also determined by (some variation of) bridge-depth. But somehow surprisingly, Dekker and Jansen~\cite{FVS-via-EDF-DAM} recently showed that, assuming $\NP \not\subseteq \coNP/\poly$, \textsc{Feedback Vertex Set} parameterized by the vertex-deletion distance to a minor-closed graph class ${\mathcal C}$ admits a polynomial kernel if and only if ${\mathcal C}$ has bounded  elimination distance to a forest.
%\marin{In the following I would change notation ${\mathcal C}$ by something else, to avoid confusion with previous ${\mathcal C}$. Maybe ${\mathcal G}$ or ${\mathcal C'}$ ?}\ig{you are right! I changed it to $\mathcal{H}$}
The \emph{elimination distance} to a graph class ${\mathcal H}$ is a parameter introduced by Bulian and Dawar~\cite{isomorphism-elimination-distance, elimination-distance} and defined as the minimum number of rounds needed to recursively delete one vertex from each connected component of the current graph until obtaining a graph that belongs to ${\mathcal H}$ (see~\autoref{sec:preliminaries} for the formal definition). Note that treedepth corresponds to the particular case where ${\mathcal H}$ is the class of empty graphs. Note also that, for any collection of graphs ${\mathcal F}$ and any graph $G$, a solution to \Fdeletion in $G$ is a modulator to the class of graphs of elimination distance zero to ${\mathcal F}$-minor-free graphs. Thus, a polynomial kernel for \Fdeletion parameterized by the size of a solution is a weaker result than a polynomial kernel parameterized by the vertex-deletion distance to graphs of bounded elimination distance to  ${\mathcal F}$-minor-free graphs.

\subparagraph{Our results.} In a nutshell, our contribution is to lift polynomial
kernels for \Fdeletion parameterized by the solution size (if they exist) to polynomial kernels parameterized by the vertex-deletion distance to graphs of bounded elimination distance to ${\mathcal F}$-minor-free graphs. This result also holds for approximate kernels by preserving the same approximation factor. We first provide a formal statement of our result and then discuss some of its consequences. We use the notation $\EDF(G)$ to denote the elimination distance of a graph $G$ to the class of $\F$-minor-free graphs.

\begin{restatable}{theorem}{polykernel}
  \label{fact:poly-kernel}
    For every fixed finite set $\F$ of connected graphs, every integer $\eta \geq 0$, and every positive constant $\alpha$, if \Fdeletion parameterized by the size of a given solution admits a polynomial ($\alpha$-approximate) kernel, then \Fdeletion parameterized by the size of a given modulator to graphs with $\EDF \leq \eta$ admits a polynomial ($\alpha$-approximate) kernel.
\end{restatable}

Recall that if $\F$ contains a planar graph, then \Fdeletion is known to admit a polynomial kernel parameterized by the solution size~\cite{planar-F-deletion-kernel,deterministic-approximation-for-planar-F-deletion}. Thus, \autoref{fact:poly-kernel}
  implies polynomial kernels for \Fdeletion parameterized by the size of a given modulator to graphs with bounded $\EDF$ whenever $\F$ contains at least one planar graph (in \autoref{sec:consequences} we discuss that, in fact, the hypothesis that the modulator is given is not necessary). Prior to our work, this was only known for \textsc{Vertex Cover} ($\F = \{K_2\}$)~\cite{vertex-cover-treedepth-poly-kernel-Algorithmica} and \textsc{Feedback Vertex Set} ($\F = \{K_3\}$)~\cite{FVS-via-EDF-DAM}. Some relevant problems covered by our result are \textsc{Cactus Vertex Deletion}~\cite{FioriniJP10,Tsur23,AoikeGHKKKKO22}, \textsc{Outerplanar Vertex Deletion} ($\F = \{K_4, K_{2,3}\}$)~\cite{DonkersJW22}, \textsc{Pumpkin Hitting Set}~\cite{JoretPSST14}, \textsc{$d$-Pseudoforest Deletion}~\cite{r-pseudoforest-deletion}, or \textsc{Treewidth-$t$ Vertex Deletion}~\cite{treewidth-modulator-parameterized-by-treewidth-modulator,treewidth-2-deletion} for every integer $t \geq 0$, that is, the problem of finding a smallest modulator to graphs of treewidth at most $t$ (note that the cases $t=0$ and $t=1$ correspond, respectively, to \textsc{Vertex Cover} and \textsc{Feedback Vertex Set}). For this latter problem, it is easy to verify that, for every $t \geq 0$, all minor obstructions to graphs of treewidth at most $t$ are biconnected, and that at least one of them is planar. Other examples of problems encompassed by \autoref{fact:poly-kernel} are \textsc{Pathwidth-$t$ Vertex Deletion}, \textsc{Treedepth-$t$ Vertex Deletion}, and \textsc{Branchwidth-$t$ Vertex Deletion} for every integer $t \geq 0$.

  It turns out that \autoref{fact:poly-kernel} yields infinitely many kernelization dichotomies for \Fdeletion. Indeed, Dekker and Jansen~\cite{FVS-via-EDF-DAM} proved that, assuming that $\NP \not\subseteq \coNP/\poly$, for any finite collection $\F$ of biconnected graphs on at least three vertices containing at least one planar graph\footnote{As we discuss in \autoref{sec:consequences}, the statement of \cite[Theorem 2]{FVS-via-EDF-DAM} requires {\sl all} the graphs in $\F$ to be planar, but the same proof goes through if only one of them is planar, as acknowledged by one of the authors~\cite{Bart-personal25}.}, the \Fdeletion problem does not admit a polynomial kernel parameterized by the size of a given modulator to a graph of unbounded $\EDF$. Thus, this lower bound combined with \autoref{fact:poly-kernel} yields the following result.

\newrefcommand{\G}{\mathcal{G}}
\begin{restatable}{theorem}{dichotomy}
    \label{fact:dichotomy}
    Let ${\mathcal C}$ be a minor-closed class of graphs and let $\F$ be a finite set of biconnected graphs on at least three vertices containing at least one planar graph. Assuming that $\NP \not\subseteq \coNP/\poly$, \Fdeletion admits a polynomial kernel in the size of a ${\mathcal C}$-modulator if and only if ${\mathcal C}$ has bounded elimination distance to the class of $\F$-minor-free graphs.
\end{restatable}

\autoref{fact:dichotomy} can be seen as a vast generalization of the dichotomy for \textsc{Feedback Vertex Set} by Dekker and Jansen~\cite{FVS-via-EDF-DAM}, which was the only one known so far other than \textsc{Vertex Cover}~\cite{vertex-cover-bridge-depth}. Concrete examples of other problems covered by \autoref{fact:dichotomy} are \textsc{Cactus Vertex Deletion}, \textsc{Pumpkin Hitting Set}, \textsc{Outerplanar Vertex Deletion}, \textsc{Treewidth-$t$ Vertex Deletion} and \textsc{Branchwidth-$t$ Vertex Deletion} for every $t \geq 0$, or \textsc{$C_p$ Hitting Set}~\cite{GokeMM20} for every  $p \geq 3$ (that is, the problem of hitting all cycles of length at least $p$).

On the other hand, plugging the approximate kernel for \textsc{Planar Vertex Deletion} by Jansen and W{\l}odarczyk~\cite{vertex-planarization-approximate-kernel} in \autoref{fact:poly-kernel} we get the following result, which is a significant strengthening of their kernel~\cite{vertex-planarization-approximate-kernel}, corresponding to the case $\eta = 0$.

\begin{restatable}{theorem}{approxkernelplanar}
  \label{fact:planar-poly-approx-kernel}
    For every integer $\eta \geq 0$, the {\sc Planar Vertex Deletion} problem parameterized by the size of a given modulator to a graph of elimination distance to planar graphs at most $\eta$ admits a polynomial $\alpha$-approximate kernel, for some constant $\alpha > 1$.
\end{restatable}

Donkers and Jansen~\cite{DonkersJ21} asked whether, for every collection $\F$, the \Fdeletion problem admits a polynomial kernel when parameterized by the vertex-deletion distance to a linear forest, that is, a disjoint collection of paths. \autoref{fact:poly-kernel} provides a positive answer to their question for every collection $\F$ of connected graphs that are not paths containing a planar graph (indeed, in that case, no graph in $\F$ is a minor of a path, so a linear forest has elimination distance zero to the class of $\F$-minor-free graphs).

  Finally, let us mention another interpretation of our results. Agrawal et al.~\cite{AgrawalKLPRSZ22delet} proved,
among other results, that for every hereditary target graph class $\mathcal{C}$ satisfying some mild
assumptions, parameterizing by the vertex-deletion distance to $\mathcal{C}$ and by the elimination
distance to $\mathcal{C}$ are equivalent from the point of view of the existence of fixed-parameter tractable
algorithms. \autoref{fact:poly-kernel} implies, in particular, that the same kind of equivalence holds with
respect to the existence of polynomial (approximate) kernels in this ``distance from triviality'' setting, namely
for problems defined by the exclusion of connected minors.

\medskip

In \autoref{sec:explicit-upper-bounds} we give an explicit upper bound on the size of the kernels obtained in \autoref{fact:poly-kernel}. On the negative side, we present two different types of lower bounds, both relevant for different reasons that we proceed to discuss.

On the one hand, in \autoref{sec:lower-bounds-not-biconnected} we show that the same dichotomy of \autoref{fact:dichotomy} still holds for infinitely many families $\F$ that contain graphs that are {\sl not} biconnected. We show this by modifying appropriately the reduction of Dekker and Jansen~\cite{FVS-via-EDF-DAM}, which was in turn inspired by other similar reductions~\cite{vertex-cover-bridge-depth,HolsKP22}). We give another class of non-biconnected families $\F$ for which the dichotomy holds in \autoref{sec:lower-bounds-treelike}. We point out that the relevance of these lower bounds is not given by the actual families $\F$ to which it applies, but as a proof of concept that such families $\F$ containing non-biconnected graphs do exist.

On the other hand, in \autoref{sec:lower-bounds-degree} we show how essentially the same reduction can be used to show a lower bound on the degree of our kernels when the elimination distance is bounded but large. We also discuss how the elimination distance and the collection $\F$ appear in the degree of our kernels, by relating our positive and negative results to the ones by Giannopoulou et al.~\cite{GiannopoulouJLS17}
(parameterized by the solution size, which corresponds to elimination distance zero).

\subparagraph{Organization.} In \autoref{sec:new-summary} we present a summary of our techniques for our positive result (i.e., the kernelization algorithm), which are based on proving two main ingredients, and a road map of the whole proof. In \autoref{sec:preliminaries} we introduce the necessary definitions and notation, including the concepts of labeled graphs, the extension of the minor relation to these graphs, and the particular kind of reductions needed later for our lower bounds. %\ig{The sentences that follow are slightly out of context here, since we have not mentioned any technical detail of our approach at all. It is not dramatic, but quite strange}\eric{changed them}
In \autoref{sec:poly-kernel}, assuming that the two main ingredients are proved, we provide the proof of \autoref{fact:poly-kernel}, and its consequences mentioned above. The proof of the first ingredient is presented in \autoref{sec:compute-solution-breaking-Q}, while the proof of the second one is presented in \autoref{sec:bounding-Q}. In \autoref{sec:explicit-upper-bounds} we thoroughly analyze the proof of the second ingredient to give an explicit upper bound on the size of our kernels. We present our lower bounds in \autoref{sec:lower-bounds} and \autoref{sec:lower-bounds-treelike}. Finally, we conclude the article in \autoref{sec:conclusions} with some directions for further research.

\section{Summary of our techniques for the kernelization algorithm}
\label{sec:new-summary}
Our techniques for obtaining the kernel of \autoref{fact:poly-kernel} are strongly based
on those used by Jansen and Pieterse in~\cite{minor-hitting}, and in this section we explain the main ideas of this approach and which are our main technical contributions that allow us to obtain \autoref{fact:poly-kernel}. We start in \autoref{sec:overview-classical-approach} by surveying which is the most common strategy used in the literature for kernelization with structural parameters, and we abstract it in terms of two main ingredients, which we explain in \autoref{sec:overview-ingredient-1}
 and \autoref{sec:overview-ingredient-2}, respectively, for our particular setting.

 Given the required amount of technical definitions and lemmas, the entire proof of these two ingredients spans over a number of pages in \autoref{sec:compute-solution-breaking-Q} and \autoref{sec:bounding-Q}, and the goal of this section is to provide some insight on the most important notions involved in the proofs, as well as highlighting our main technical novelties with respect to the proof in~\cite{minor-hitting}. As it will become clear in this section, many of the ingredients that we need are either borrowed directly from~\cite{minor-hitting}, or follow from the corresponding results in~\cite{minor-hitting} with very minor modifications. In the latter case, for the sake of completeness we provide (in \autoref{sec:compute-solution-breaking-Q} or \autoref{sec:bounding-Q}) both a full proof and a sketch of proof in which we just list which parts of the proof in~\cite{minor-hitting} need to be changed.

We conclude this section with a road map of the whole proof (\autoref{fig:road-map}), where one can see how our main technical contributions fit within the structure of the proof.

\subsection{Typical approach for kernelization with structural parameters}
\label{sec:overview-classical-approach}

Let us start by explaining how kernels
usually work for hitting problems parameterized by the size of modulator to trivial classes $\C$. More precisely, we consider here \Fdeletion problems where the input is $(G,X,k)$, and we have to decide whether at most $k$ vertices can be removed from $G$ so that the resulting graph does not contain any of the graphs in $\F$ as a minor. The modulator $X \subseteq V(G)$ given in the input is such that $m(G\setminus X) \le \eta$, for a fixed integer $\eta \geq 0$ and graph measure $m$, and the parameter is $|X|$. We assume that $m$ is such that for disjoint graphs $G_1,G_2$, $m(G_1 \cup G_2) \le \max(m(G_1),m(G_2))$, and for any connected graph $G$, there exists $v \in G$ such that $m(G\setminus\set{v}) < m(G)$.
Recall that this setup encapsulates several previous works, for example \cite{BougeretS17} where $\F=\{K_2\}$ and $m$ is the treedepth, \cite{FVS-via-EDF-WG} where $\F=\{K_3\}$ and $m$ is the elimination distance to a forest,
and \cite{minor-hitting} where $\F$ is arbitrary and $m$ is the treedepth.

Let $\mathcal{D}$ be the set of connected components of $G\setminus X$ and $n_D = |\mathcal{D}|$.
All kernels in the mentioned articles follow the same two steps:
\begin{enumerate}
    \item\label{enum:introstep1} Remove some connected components of $\mathcal{D}$ until $n_D$ becomes polynomial in $|X|$. This leads to an equivalent instance $(G',X,k')$. Let $\mathcal{D'}$ denote the set of connected components of $G'\setminus X$.
    \item For each $C \in \mathcal{D'}$, find a vertex $v_C$ such that $m(C\setminus\set{v_C}) < m(C)$. Define $X' = X \cup \bigcup_{C \in \mathcal{D'}}V(C)$ and recurse on  $(G',X',k')$.
\end{enumerate}
Informally, we moved vertices from $G' \setminus X$ to the modulator to get a slightly larger modulator $X'$, but such that
$m(G' \setminus X') < \eta$, implying that we can apply a recursive argument on the measure.
Notice that when following this approach, the only challenge is to achieve \autoref{enum:introstep1}.
As our setting for the \Fdeletion problem and $m=\EDF$ fit into this framework, we also follow these two steps, and our only goal is to prove the following lemma, which formalizes \autoref{enum:introstep1}. This lemma corresponds exactly to \cite[Lemma 6]{minor-hitting} (which provides a polynomial kernel for \Fdeletion where $\td(G\setminus X) \le \eta$, instead of $\EDF(G\setminus X) \le \eta$ in our case), where $\td$ is replaced by $\EDF$.

\newtheorem*{fact:reduce-components}{Lemma \ref{fact:reduce-components}}
\begin{namedlemma}{Reduce Components Lemma}[-- Generalized version of {\cite[Lemma 6]{minor-hitting}}]
\label{fact:reduce-components}
    Let $\F$ be a finite set of connected graphs and let $\eta \geq 0$ be a constant. There is a polynomial-time algorithm that, given a graph $G$ along with a modulator $X \subseteq V(G)$ such that $\EDF(G \setminus X) \leq \eta$, outputs an induced subgraph $G'$ of $G$ together with an integer $\Delta$ such that $\opt(G) = \opt(G') + \Delta$ and $G' \setminus X$ has at most $\abs{X}^{\bigO_{\F,\eta}(1)}$ connected components. Moreover, a set $Y'$ that hits all $\F$-minors in $G'$ can be extended in polynomial time to a set $Y$ of size $\abs{Y'} + \Delta$ that hits all $\F$-minors in $G$.
\end{namedlemma}

%Simplest ex of Step 1 could be VC/k (distance to td=0) and high degree rule and isolated vertices

Assuming the above lemma, \autoref{fact:poly-kernel} follows immediately by induction on $\eta$ (see \autoref{sec:poly-kernel} for the details), and thus in this overview we only focus on this lemma.
\autoref{fact:reduce-components} requires two ingredients: \autoref{fact:compute-solution-breaking-Q}
%(that we could call the "existence checking of a great optimal solution", as we check if a given connected components $C$ of $G\setminus X$, there exists a local optimal of $C$ that in additions hits other small structures),
and \autoref{fact:main-lemma}.
%the blocking set lemma 43 and the checking freeness lemma 15),
These lemmas also correspond to the two ingredients required by \cite[Lemma 6]{minor-hitting}, where $\td$ is replaced by $\EDF$ here. Generalizing these two ingredients to $\EDF$ is the contribution of this paper, and we now aim at explaining the challenges and new ideas behind this generalization.
To do so, and following the formalism of \cite{minor-hitting}, we first need to introduce the notion of labeled minor (see \autoref{sec:preliminaries} for formal definitions). For a set $X$, an {\em $X$-labeled graph} $G$ is a graph where each vertex $v$ is equipped with a set $\lab_G(v) \subseteq X$ of labels. Given two $X$-labeled graphs $G$ and $H$, we say that $H$ is a {\em labeled minor} of $G$ if $H$ is a ``classical minor'' -- ignoring the labels -- of $G$,
certified by a minor model $\phi$, that additionally satisfies that for any $v \in V(H)$, $\lab_H(v) \subseteq \bigcup_{u \in \phi(v)} \lab_G(u)$.

\begin{figure}[t]
    \centering
    \vspace{-.3cm}
    \includegraphics[width=0.47\textwidth]{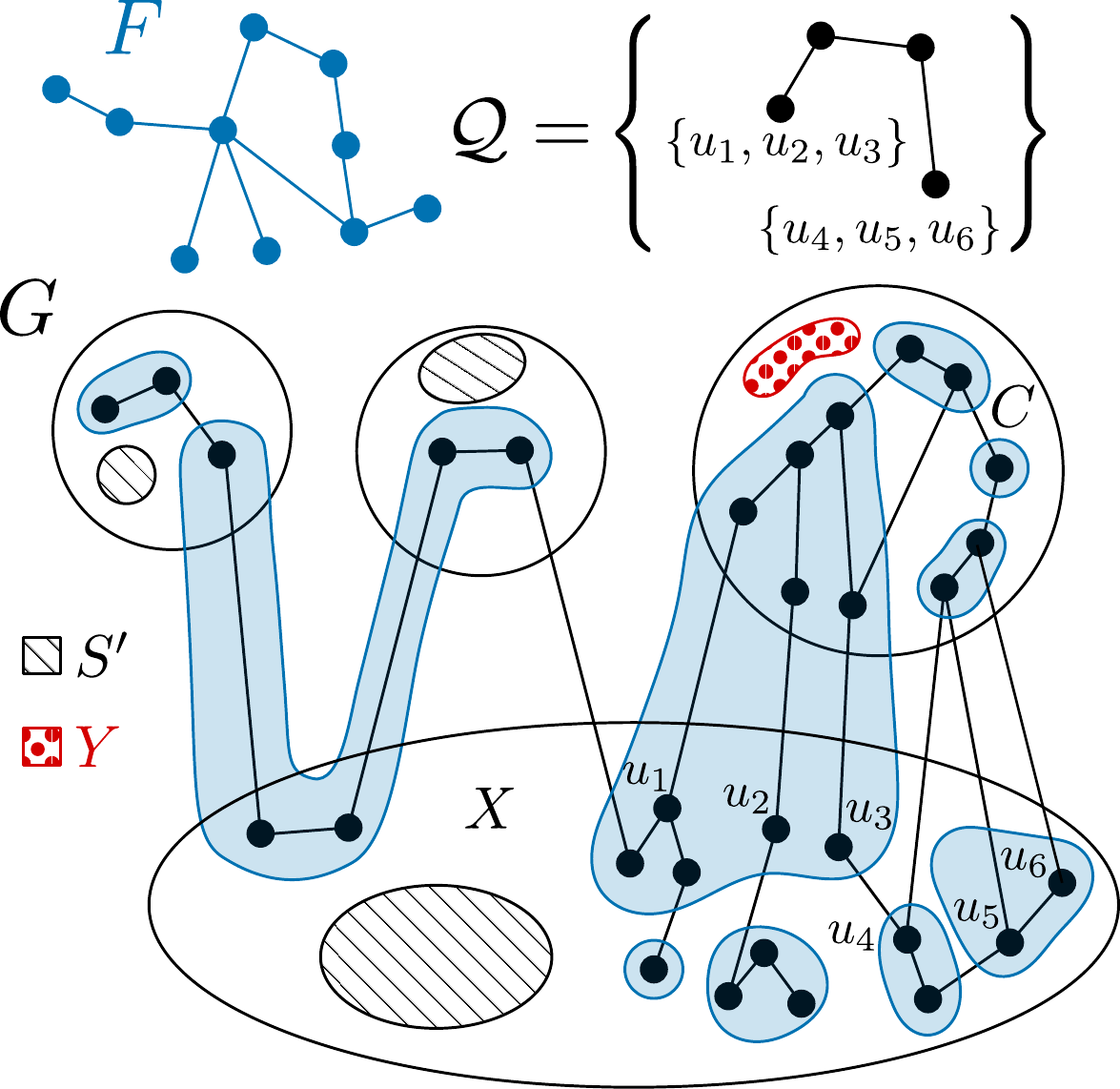}
    \caption{Example where adding $Y$, a local optimal solution to \Fdeletion in $G[C]$, to $S'$, an optimal solution in $G\setminus C$, misses an $F$-model. This particular model would have been hit if $Y$ was also required to hit the graph in $\Q$ (as a labeled minor).}
    \label{fig:ourTechniques}
\end{figure}

Let us now discuss how labeled minors appear in the kernelization algorithm for \Fdeletion (we also refer the reader to introduction of \cite{minor-hitting} for additional intuition on the role of labeled minors).
Suppose that, given an instance $(G,X,k)$ of \Fdeletion,  we want to remove a connected component $C$ of $G\setminus X$ by defining $G'=G\setminus C$ and $k'=k-\opt(G[C])$ (where $\opt$ denotes the smallest solution size for the \Fdeletion problem).
To prove that $(G',X,k')$ being a {\sf yes}-instance implies that $(G,X,k)$ is as well, a common approach is to consider a solution $S'$ of $(G',X,k')$, and find $Y \subseteq C$ of size $\opt(G[C])$ such that $S = S' \cup Y$ is a solution of $G$.
However, using an arbitrary local optimal solution $Y$ of $C$ (that ``only'' hits all $\F$-minors in $G[C]$) may not be enough, as there could be $\F$-minor models in $G$ whose fragments intersect both $C$ and $V(G)\setminus C$ (see \autoref{fig:ourTechniques}).

Typically, if we consider the $F$-minor model and the labeled graph $Q$ of \autoref{fig:ourTechniques}, to prevent this particular model of $F$, we need that there exists a local optimal solution  $Y \subseteq C$ of size $\opt(G[C])$ which also hits in $C$ the labeled minor $Q$, called {\em fragment}. In the real setting, a local optimal solution of $G[C]$ may be even asked to hit {\sl a set} $\Q$ of labeled minors, corresponding to all possible fragments of an $\F$-minor model in $G$.
Checking the existence of such special optimal solution in a connected component of $G\setminus X$ is precisely what we achieve in the first following ingredient.

\subsection{Ingredient 1: checking for the existence of special optimal \Fdeletion solutions}
\label{sec:overview-ingredient-1}

In \autoref{sec:compute-solution-breaking-Q} we prove the following lemma.

\begin{lemma}[Generalized version of {\cite[Lemma 5]{minor-hitting}}]
\label{fact:compute-solution-breaking-Q}
    Let $\F$ be a fixed set of connected (unlabeled) graphs, let $\eta \geq 0$ be a constant, and let $X$ be a set. For any set $\Q$ of connected $X$-labeled graphs and $X$-labeled graph $C$ with $\EDF(C) \leq \eta$, one can:
    \begin{enumerate}
        \item\label{item:compute-opt} compute $\opt(C)$ in $\bigO_{\F, \eta}(\abs{V(C)})$ time;
        \item\label{item:compute-opt-that-breaks-Q} determine whether there is a solution $Y \in \optsol(C)$ such that $C \setminus Y$ has no labeled $\Q$-minors, in time $f(\F,L,\sum_{H \in \Q}\abs{V(H)}, \eta) \cdot \abs{V(C)}^{\bigO(1)}$ for some function $f$.
    \end{enumerate}
    Here, $L$ is defined as the number of elements of $X$ that appear in the labelset of at least one vertex in at least one graph of $\Q$.
\end{lemma}

The proof of Lemma 5 in \cite{minor-hitting} quickly follows from the fact that $\td(C) \le \eta$ implies $\tw(C) \le \eta$ (where $\td$ denotes the treedepth and $\tw$ the treewidth), and that the problem of finding labeled minors can be expressed as an MSOL formula.
However, in our setting, a graph $C$ with $\EDF(C) \le \eta$ may have unbounded treewidth (because of the subgraphs induced by the leaves in the elimination distance decomposition), and we rather
rely on the following reduction.

\begin{figure}[ht]
    \centering
    \begin{minipage}{0.49\textwidth}
    \includegraphics[width=\textwidth]{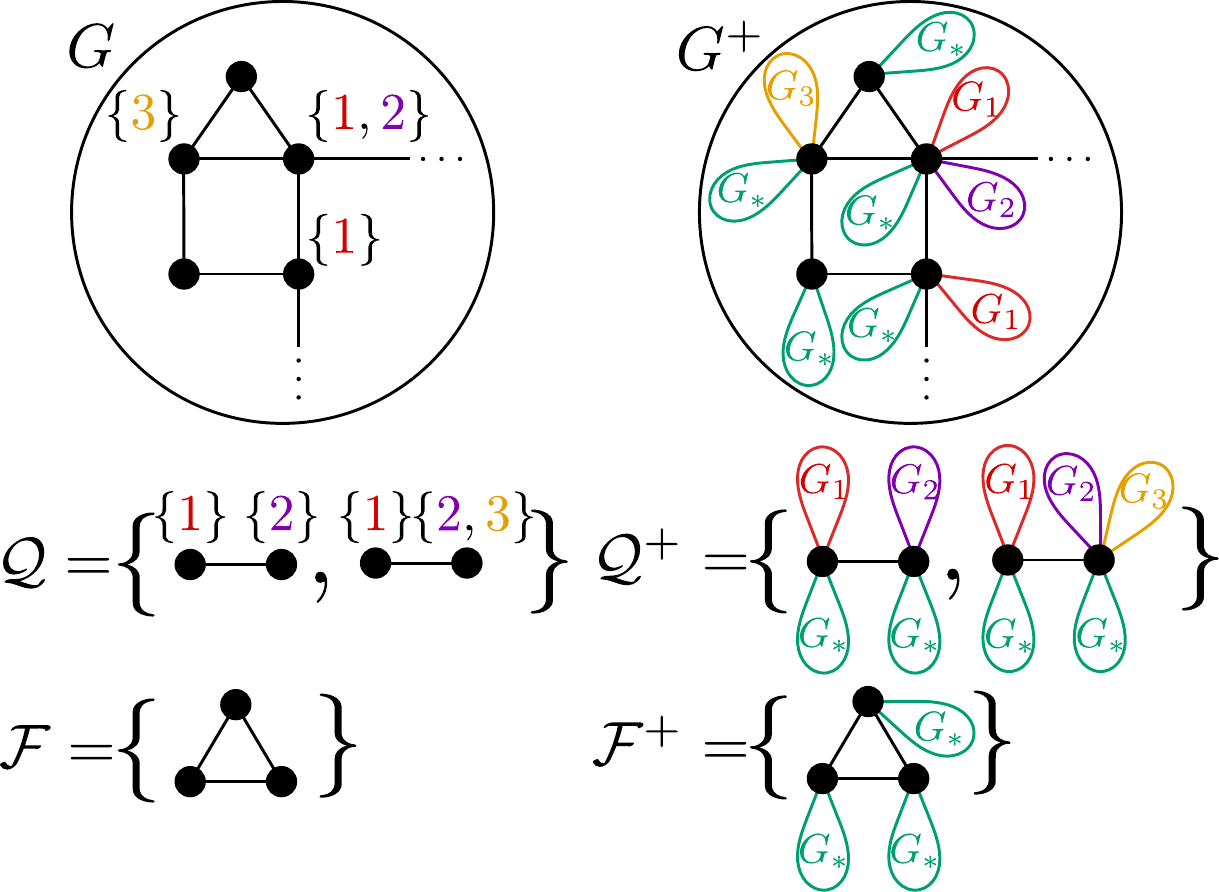}
    \end{minipage}
    \begin{minipage}{0.49\textwidth}
    \includegraphics[width=\textwidth]{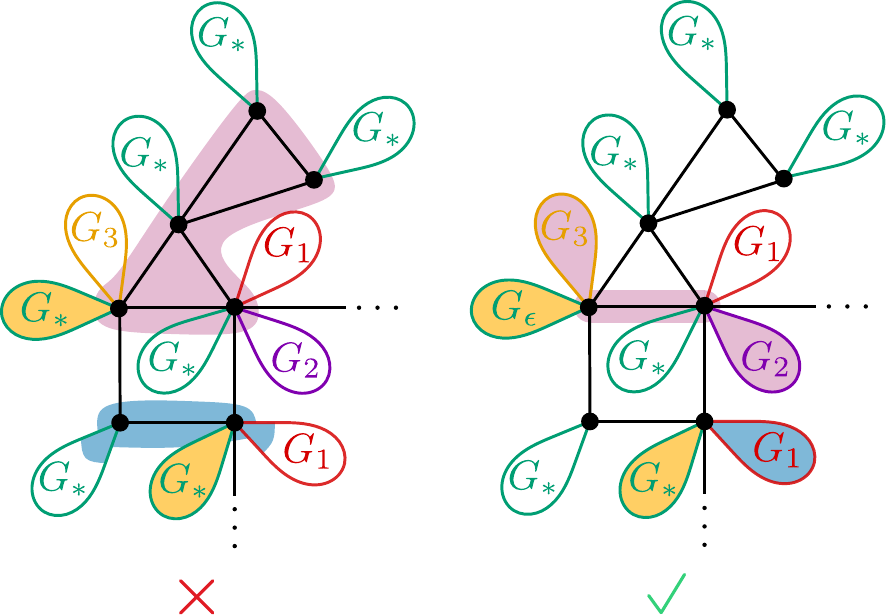}
    \end{minipage}
    \caption{Left: How we compute $(G^+,\F^+,\Q^+)$ from $(G,\F,\Q)$ to reduce to the unlabeled version. Right: Example of a ``bad'' minor model of the second graph in the set $\Q^+$ for two reasons. The first reason is that the pink branch set, which is a minor of $G_2$ glued with $G_3$, uses vertices of $G$ instead of the attached $G_{\ell}$. The second reason is that the blue branch set, which is a minor of $G_1$, uses partially vertices of some $G_\ell$. The orange branch sets of this bad model behave as expected. A ``good'' minor model is depicted on the right.}
    \label{fig:gluing-gadgets}
\end{figure}

The problem we need to solve is what we later call the \FdeletionHittingLabeledQ problem,  where given a labeled graph $G$ with $\EDF(G) \le \eta$, and a set of labeled graphs $\Q$, one has to decide if there exists an optimal \Fdeletion solution for $G$ that also hits all $Q \in \Q$. To solve this problem (in \FPT time parameterized by the total size of $\F$ and $\eta$), we first reduce (in \autoref{sec:transform-labeled-to-unlabeled}) to the unlabeled version called \FdeletionHittingQ.
The idea of the reduction is the following.
We define a gadget graph $G_\ell$ for each label $\ell$ in $X$, and glue to each vertex $v \in G$ all gadgets corresponding to labels of $v$ (see the left part of \autoref{fig:gluing-gadgets}). We do the same for each graph in $\Q$. Moreover, we also add a last gadget $G_\epsilon$ that we glue to every vertex of $G$, every vertex of a graph in $\Q$, and every vertex of a graph in $\F$.
 Let $G^+$, $\Q^+$, and $\F^+$ denote, respectively, the obtained graphs.
To guarantee that $(G,\F,\Q,k)$ is equivalent to $(G^+,\F^+,\Q^+,k)$, we need to control how models of
$\Q^+$ and $\F^+$ live in $G^+$. For example, we want to avoid models of $\F^+$ or $\Q^+$ that invade partially  a gadget $G_\ell$, or models of $\Q^+$ where the part of the model corresponding to a gadget lives in $V(G)$
 (see the right part of \autoref{fig:gluing-gadgets}). This is achieved through the notion of {\em nice gadgets}, which are informally (see \autoref{def:nice-gadgets}) biconnected graphs that are pairwise incomparable with respect to the minor relation, and that are not minors of the host graph $G$. Notice that it is also crucial to guarantee that $\ED{\F^+}(G^+) \le \EDF(G)$, as this implies by the assumption on $G$ that $\ED{\F^+}(G^+) \le \eta$, as we wish.

It remains now to solve \FdeletionHittingQ parameterized by the size of $\F$ and $\eta$, which we handle in \autoref{sec:solve-unlabeled-version} by adapting the algorithm of Jansen, de Kroon, and W{\l}odarczyk\cite[Theorem 1.2]{vertex-deletion-parameterized-by-elimination-distance} for \Fdeletion parameterized by ${\mathcal H_\F}$-treewidth, where ${\mathcal H_\F}$ is the class of $\F$-minor-free graphs. This parameter generalizes both treewidth and $\EDF$. However, the simple strategy consisting in solving directly $(\F \cup \Q)$-{\sc Minor Deletion} using \cite{vertex-deletion-parameterized-by-elimination-distance} as a black box is not possible,
as the input graph $G$ may have unbounded ${\mathcal H_{\F \cup \Q}}$-treewidth, even if its ${\mathcal H_\F}$-treewidth is at most $\eta$.
Thus, we adapt their result to our case by proving that all key lemmas involved in the dynamic programming algorithm of \cite{vertex-deletion-parameterized-by-elimination-distance} can be generalized to allow for the presence of $\Q$.

\subsection{Ingredient 2: bounding the size of minimal blocking sets}
\label{sec:overview-ingredient-2}

Let us now turn to the second ingredient, which has a purely combinatorial flavor and is our main technical contribution.
Recall that our goal is to prove \autoref{fact:reduce-components}, which removes some connected components of $G\setminus X$.
To achieve this, we rely on the marking algorithm  of \cite{minor-hitting}. We chose not to provide the details of this algorithm here (as we do not add any modification to it or to its proof), but only explain  why this algorithm requires the second ingredient.
To prevent scenarios like the one depicted in \autoref{fig:ourTechniques}, the algorithm has to check, before removing a connected component $C$, that for any set $\Q$ of labeled fragments, $C$ is a {\sf yes}-instance of \FdeletionHittingLabeledQ. Notice that for each $\Q$, Ingredient~1 allows us to perform this check in polynomial time, but the problem is that the list of all possible sets $\Q$ may have size exponential in $|X|$.
Ingredient 2 (given by \autoref{fact:main-lemma} below, which we prove in \autoref{sec:bounding-Q}) exactly fulfills this needs, by proving that if there exists a large set $\Q$ such that $C$ is a {\sf no}-instance of \FdeletionHittingLabeledQ, then there exists also a constant-sized subset $\Q^\star \subseteq \Q$ such that $C$ remains a {\sf no}-instance of \Fdeletion \textsc{Hitting $\Q^\star$}.
In that way, the marking algorithm will only enumerate (in polynomial time) such sets $\Q^\star$ of constant size.

\begin{namedlemma}{Main Lemma}[-- Generalized version of  {\cite[Lemma 3]{minor-hitting}}]\label{fact:main-lemma}
    \mainlemma
\end{namedlemma}

We point out that there is an additional (very helpful) hypothesis required on $\Q$ in the above lemma that we did not discuss so far: the \emph{saturated} property. Indeed, even if this hypothesis is unavoidable (see \cite[Figure 11]{minor-hitting}), and may seem counter-intuitive (for example, the set $\Q$ in \autoref{fig:ourTechniques} does not satisfy it), the fact that we are allowed to assume it comes from the details of the proof of the marking algorithm of \cite{minor-hitting}, which we prefer to keep as a black box in this high-level summary.

We would also like to mention that such a set $\Q$ as in \autoref{fact:main-lemma} that ``affects'' the behavior of all optimal solutions is often referred as a {\em blocking set} \cite{vertex-cover-treedepth-poly-kernel-Algorithmica, vertex-cover-bridge-depth, HolsKP22}, and \autoref{fact:main-lemma} can be rephrased as bounding the size of an inclusion-wise minimal blocking set (as invoking the lemma with an inclusion-wise minimal $\Q$ leads to $\Q^\star = \Q$, thus bounding $|\Q|$). One can also observe that the marking algorithm of \cite{minor-hitting} for \Fdeletion corresponds to a generalized version of the marking algorithms used in
\cite{JansenB13, vertex-cover-treedepth-poly-kernel-Algorithmica, vertex-cover-bridge-depth}, that is, for \textsc{Vertex Cover} parameterized by the feedback vertex number, the distance to constant treedepth, and the distance to constant bridge-depth, respectively.

Let us now discuss what differs between our proof of \autoref{fact:main-lemma} and the proof of~\cite[Lemma 3]{minor-hitting}.
The proof of Lemma 3 in~\cite{minor-hitting} is inductive on the depth of the treedepth decomposition of $C$, and we also follow this approach. However, in our setting, $C$ does not necessarily have bounded treedepth, but it does have bounded $\EDF$. Given the similarities between the two parameters, the only challenge is  to add a base case for the leaves of the $\F$-elimination forest of $C$, which are $\F$-minor-free graphs instead of empty graphs. We consider this new base case, \namedtheoremref{fact:extra-base-case}, to be our main technical contribution. The statement of this lemma is very technical and falls beyond the scope of this overview but, intuitively, it bounds the size of two objects that are relevant to the induction. The first one (denoted by $\R_N$ in the lemma) deals with the number of possible ``remainders'' of solutions that do {\sl not} leave a $\Q$-minor, while the second one (denoted by $\R_\Q$ in the lemma) corresponds to the $\Q^{\star}$ discussed above. Let us now say a few words about the proof of the second item, since the first one is more technical and would require additional preliminaries.

To simplify the presentation, let us start by stating a simplified version of the setup of \autoref{fact:extra-base-case}.
Consider a graph $G$ and subgraphs $G_A$, $G_C$ of $G$ (we use these notation to match the notation of \autoref{fact:extra-base-case}, where $G_B$ is assumed to be empty here) such that:
\begin{itemize}
    \item $S := V(G_A) \cap V(G_C)$ is a separator in $G$, with $|S| \le \eta$.
    \item $G'_A  :=G_A \setminus S$ is $\F$-minor-free.
\end{itemize}
Given a set $\Q$ of connected $X$-labeled graphs, the simplified goal of this second item is to define a set $\Q^\star \subseteq \Q$ whose size only depends on $\F$ and $\eta$, and such that for any optimal solution $Y$ of the \Fdeletion problem in $G$, if $G'_A \setminus Y$ leaves a $\Q$-minor, then it also leaves a $\Q^\star$-minor.

In order to bound the size of $\Q^{\star}$, our strategy is to identify a subset of labels $X' \subseteq X$ of size depending only on $\F$ and $\eta$, such that any such solution leaving a $\Q$-minor will also leave a $\Q$-minor that only uses labels from $X'$. Then, $\Q^*$ will be defined as all graphs of $\Q$ using only labels from $X'$, implying immediately that $|Q^\star|$ only depends on $\F$ and $\eta$.

To identify such a restricted set $X'$ of labels,
we consider a set called $\mathsf{Breaker}$ corresponding to a minimum-size set in $G'_A$ that hits all $\Q$-minors in $G'_A$.
This setup can be seen as a generalization of the proof of \cite[Lemma 4]{FVS-via-EDF-DAM} by Dekker and Jansen, where in their setting (where there are no labels) $\Q$ corresponds to  a set $T=\{(u_i,v_i)\}$ (encoding that solutions must hit all $(u_i,v_i)$-paths) and they consider a minimum-size set $Z$ hitting all these paths.

Then, we use the crucial fact that any optimal solution $Y$ of the \Fdeletion problem in $G$ is such that $|Y \cap V(G'_A)| \le |S|$. Indeed, if $Y$ used strictly more vertices in $G'_A$, then restructuring the solution by removing all vertices from $G'_A$ and adding the whole of $S$ instead would result in a smaller solution, as all graphs in $\F$ are connected; see \autoref{fact:Y-has-at-most-neighbors-vertices-in-S}. We point out that this is the only place where the $\F$-minor-freeness of leaves is used, meaning in particular that we do not need to invoke any complex property on the structure of $\F$-minor-free graphs.

With this set $\mathsf{Breaker}$ and the property that $|Y \cap G'_A| \le |S|$ at hand, we perform a marking scheme (that generalizes the marking scheme of \cite[Lemma 4]{FVS-via-EDF-DAM}) aiming at keeping only marked labels. This marking scheme marks some labels for each $v \in \mathsf{Breaker}$ (implying that we must ensure that $|\mathsf{Breaker}|$ is small), and uses the fact that, as $|Y \cap V(G'_A)| \le |S|$, we only need to mark a small number of labels to ensure that one of the $\Q$-minors in $G'_A \setminus Y$ only uses marked labels.

Finally, let us just mention that our proof of the first item additionally introduces the definition of a \emph{mandatory} vertex, which is a vertex that appears in every \Fdeletion solution meeting some conditions. The set $M$ of all such vertices is helpful in imposing a nice structure in the set of solutions that we need to consider (cf. \autoref{fig:components-in-GAB-minus-M}).

In \autoref{fig:road-map} we provide a road map of the whole proof, by highlighting our main technical contributions.

\begin{figure}[h]
\centering
\resizebox{0.9\textwidth}{!}{
\begin{tikzpicture}[node distance=1cm and 0.3cm]

% Top nodes
\node[mynode] (thm) {\autoref{fact:poly-kernel}\\ (polynomial kernel)};
\node[mynode, right=1cm of thm] (lemma) {\autoref{fact:reduce-components} \\(remove connected components)};

% Ingredient 1
\node[mynode, very thick, fill=blue!15, below=of thm] (ing1) {
\autoref{fact:compute-solution-breaking-Q}
(Ingredient 1):\\ Solve \FdeletionHittingLabeledQ};

% Fragment 2
\node[mynode, right=of ing1] (frag2) {
\autoref{fact:main-lemma}
(Ingredient 2):\\ Main Lemma (bounding blocking sets)};

% Inductive proof node (above transmission/base)
\node[mynode, below=of frag2] (ind) {
\autoref{fact:inductive-main-lemma-ed}\\
Inductive version of Main Lemma};

% Transmission and Base case below inductive proof
\node[mynode, below left=of ind] (trans) {Inductive step};
\node[mynode, very thick, fill=blue!15, below =of ind] (base) {
\autoref{fact:extra-base-case}\\
$\F$-minor-free base case};

% Section 4
%\node[mynode, left=of ing1] (sec4) {Section 4:\\ reductions + algo};

% Reversed arrows (bottom -> top)
\draw[myarrow] (lemma) --  (thm);
\draw[myarrow] (ing1) -- (lemma);
\draw[myarrow] (frag2) -- (lemma);
\draw[myarrow] (ind) -- (frag2);
\draw[myarrow] (trans) -- (ind);
\draw[myarrow] (base) -- (ind);
%\draw[myarrow] (ing1) -- (sec4);

\end{tikzpicture}}
\caption{Road map of the proof, which has the same structure as the proof in \cite{minor-hitting}. Our two contributions correspond to the thicker blues boxes. Results in all other boxes can be obtained directly from the proofs of \cite{minor-hitting} by minor modifications, essentially by replacing $\td$ with $\EDF$.\label{fig:road-map}}
\end{figure}

\section{Preliminaries}
\label{sec:preliminaries}
In this section we present
 some standard definitions used in \autoref{sec:poly-kernel} and \autoref{sec:lower-bounds}. The additional (sometimes, rather technical) definitions that are needed in \autoref{sec:compute-solution-breaking-Q} and \autoref{sec:bounding-Q} are deferred to the corresponding section as well.

 %% Change for the full version

\subparagraph{Parameterized complexity.}
A \emph{parameterized problem} is a decision problem where each instance is a pair $(I, k)$, with $I$ the input and $k \in \mathbb{N}$ the parameter. A problem is \emph{fixed-parameter tractable} (\FPT) if it can be solved in time $f(k)\cdot |I|^{\bigO(1)}$ for some computable function $f$.

A \emph{kernelization algorithm} (or \emph{kernel}) for a parameterized problem is a polynomial-time algorithm that transforms any instance $(I, k)$ into an equivalent instance $(I', k')$ (i.e., $(I, k)$ is a \yes-instance if and only if $(I', k')$ is), such that $\abs{I'} + k' \leq g(k)$ for some function $g$. It is known that a parameterized problem is \FPT if and only if it admits a kernel~\cite{FPT-equals-kernel}. If $g$ is a polynomial, the kernel is called a \emph{polynomial kernel}.

A related notion introduced by Lokshtanov et al.~\cite{lossy-kernelization} is that of an \emph{$\alpha$-approximate kernel} for a parameterized optimization problem, for some constant $\alpha \geq 1$. The goal of a parameterized optimization problem is to find a solution of minimum \emph{cost}. In the case of \Fdeletion, the cost of a solution is its size. The formal definition of approximate kernelization can be found in \cite{lossy-kernelization}; below we give a simpler formulation that is enough for our purposes.

An $\alpha$-approximate kernel is a polynomial-time algorithm that transforms any instance $(I, k)$ of the parameterized optimization problem into an instance $(I', k')$ such that:
\begin{itemize}
    \item $(I', k')$ has size at most $g(k)$ for some function $g$ (as in standard kernelization), and
    \item any solution $s'$ to $(I', k')$ of cost $\beta$ times the optimum for $(I', k')$ can be transformed in time polynomial in $\abs{I}$, $k$, $\abs{I'}$, $k'$, and $s'$ into a solution $s$ to $(I, k)$ whose cost is at most $\alpha \cdot \beta$ times the optimum for $(I, k)$.
\end{itemize}
If $g$ is a polynomial, the kernel is called a \emph{polynomial $\alpha$-approximate kernel}.

\newrefcommand{\subsets}[1]{2^{#1}}
\newrefcommand{\subsetsWithSize}[2]{\binom{#1}{#2}}

%\ig{IMPORTANT: we need to define polynomial parameter transformation (PPT) here, and all the other definitions that we need for the lower bounds of \autoref{sec:lower-bounds}. I added only the following paragraph, but we also need to define the types of reductions that give lower bounds on the DEGREES of the kernels: }\eric{Done!}

To transfer the non-existence of polynomial kernels (under reasonable complexity assumptions), we use the notion of \emph{polynomial parameter transformations} (PPT for short), introduced by Bodlaender, Thomass{\'{e}}, and Yeo~\cite{BodlaenderTY11}.  A polynomial parameter transformation from a parameterized problem $P$ to a parameterized problem $Q$ is an algorithm that, given an instance $(x,k)$ of $P$, computes in polynomial time an equivalent instance $(x',k')$ of $Q$ such that $k'$ is bounded by a polynomial depending only on $k$. It follows easily from the definition that if $P$ does not admit a polynomial kernel under some complexity assumption, then the same holds for $Q$.

Similarly, to transfer the non-existence of polynomial kernels of degree $d - \varepsilon$ for any $\varepsilon > 0$ (i.e., kernels of size $\bigO(k^{d - \varepsilon})$) under reasonable complexity assumptions, we use the notion of \emph{linear parameter transformations} (LPT for short), introduced by Hermelin and Wu~\cite{linear-parameter-transformation}. A linear parameter transformation from a parameterized problem $P$ to a parameterized problem $Q$ is a PPT where the parameter $k'$ of the output instance is instead bounded by a linear function of $k$. It follows from the definition that if $P$ does not admit a polynomial kernel of size $\bigO(k^{d - \varepsilon})$ for any $\varepsilon > 0$ under some complexity assumption, then the same holds for $Q$.

\subparagraph{Graphs.} For graph notions not defined here, we refer the reader to standard textbooks such as~\cite{Diestel16}.
For a set $S$, we use $\subsets{S}$ to denote the set of all subsets of $S$, and $\subsetsWithSize{S}{k}$ to denote the set of all subsets of $S$ of size $k$. All graphs we consider are finite, undirected, and simple. A graph~$G$ consists of a vertex set~$V(G)$ and edge set~$E(G) \subseteq \subsetsWithSize{V(G)}{2}$. The open neighborhood of a vertex~$v$ is denoted~$N_G(v)$. For a vertex set~$S \subseteq V(G)$, its open neighborhood is~$N_G(S) \coloneqq \bigcup_{v \in S} N_G(v) \setminus S$. For an edge~$\{u,v\}$ in a graph~$G$, \emph{contracting~$\{u,v\}$} results in the graph~$G'$ obtained from~$G$ by removing~$u$ and~$v$, and replacing them by a new vertex~$w$ with~$N_{G'}(w) = N_G(\{u,v\})$. For a vertex set~$S \subseteq V(G)$, we use~$G \setminus S$ to denote the graph obtained from~$G$ by deleting all vertices in~$S$ and their incident edges. We use $G - e$ to denote the graph obtained from~$G$ by deleting the edge~$e \in E(G)$. The subgraph of~$G$ induced by vertex set~$S$ is denoted~$G[S]$. We use $\cc(G)$ to denote the set of connected components of~$G$. A graph is \emph{biconnected} if it is connected and does not contain a \emph{cut vertex}, i.e., a vertex whose removal increases the number of connected components of the graph. A \emph{biconnected component} of a graph~$G$ is a maximal biconnected subgraph of~$G$. A \emph{grid graph of width $w$ and height $h$} for two constants $w,h \geq 1$ is the graph $G$ where $V(G)$ consists of the tuples $(a, b)$ where $1 \leq a \leq w$ and $1 \leq b \leq h$, and two vertices $(a_1, b_1)$ and $(a_2, b_2)$ are adjacent if and only if $\abs{a_1 - a_2} + \abs{b_1 - b_2} \leq 1$.

\let\oldnorm\norm
\renewrefcommand{\norm}[1]{\oldnorm{#1}}
For a set of graphs $\F$, we use $\norm\F$ as a shorthand for $\max_{H\in \F} \abs{V(H)}$.
We say that a set $Y \subseteq V(G)$ \emph{hits} all $\F$-minors in $G$ if $G \setminus Y$ is $\F$-minor-free.

\definedhere{\opt,\optsol}
We denote the size of an optimal \Fdeletion solution on~$G$ by~$\opt(G)$, and the set of optimal solutions by~$\optsol(G)$. In our bounds, we use the notation~$\bigO_{z}(1)$ for a list of values~$z$ to denote a constant that only depends on~$z$. For instance, $\bigO_{\F,\Q}(1)$ denotes a constant that only depends on the sets of graphs $\F$ and $\Q$.

\begin{definition}[treedepth] \label{def:treedepth}
    \definedhere{\td}
    A \emph{treedepth decomposition} of a connected graph $G$ is a rooted tree $T$ such that $V(T) = V(G)$, and for every edge $\{u, v\} \in E(G)$, vertex $u$ is an ancestor of vertex $v$ in $T$, or vice versa. A treedepth decomposition of a disconnected graph is just a disjoint union of treedepth decompositions for its connected components.

    The \emph{treedepth} of $G$, denoted by $\td(G)$, is the minimum depth (in number of vertices) of a treedepth decomposition of $G$.
\end{definition}

The following definition was introduced by Bulian and Dawar~\cite{isomorphism-elimination-distance, elimination-distance}. In fact, the general definition is for a general target graph class $\H$, but since we only deal with $\F$-minor-free classes, we provide only the following more restricted definition.

\begin{nameddefinition}{elimination distance}
    \label{def:elimination-distance}
    \newrefcommand{\bags}{\chi}
    \definedhere{\ED}
    Let $\F$ be a finite collection of graphs. An \emph{$\F$-elimination forest} of a connected graph $G$ is a pair $(T, \bags)$ of a rooted tree $T$ and a function $\bags \colon V(T) \to \subsets{V(G)}$, called the \emph{bags of $T$}, such that:
    \begin{enumerate}
        \item $\bigcup_{t \in V(T)} \bags(t) = V(G)$.
        \item For every two nodes $\set{t, t'} \subseteq V(T)$ we have that $\bags(t) \cap \bags(t') = \emptyset$.
        \item Every internal node $t \in V(T)$ is such that $\abs{\bags(t)} = 1$.
        \item For every edge $\{u,v\} \in E(G)$ either:
        \begin{itemize}
            \item there exists a leaf $t \in V(T)$ such that $\set{u,v} \subseteq \bags(t)$; or
            \item there exist two nodes $\set{t, t'} \subseteq V(T)$ such that $u \in \bags(t)$, $v \in \bags(t')$, and $t$ and $t'$ are in an ancestor-descendant relationship in $T$.
        \end{itemize}
        \item For every leaf $t$ of $T$, the induced subgraph $G[\bags(t)]$ is connected and $\F$-minor-free.
    \end{enumerate}
    An $\F$-elimination forest of a disconnected graph $G$ is just a disjoint union of $\F$-elimination forests for its connected components.

    The \emph{elimination distance} of $G$ to the class of $\F$-minor-free graphs is the minimum depth (in number of edges) of an $\F$-elimination forest of $G$. We denote this distance by $\ED{\F}(G)$.
\end{nameddefinition}

Note that the treedepth is exactly the elimination distance to the class of empty graphs, which is why the treedepth is measured in number of vertices, while the elimination distance is measured in number of edges.

\begin{definition}[minor model] \label{def:minor-model}
    A \emph{minor model} of a graph $H$ in a graph $G$ is a mapping $\varphi \colon V(H) \rightarrow \subsets{V(G)}$ assigning a \emph{branch set} $\varphi(v) \subseteq V(G)$ to each vertex $v \in V(H)$, such that:
    \begin{itemize}
    \item $G[\varphi(v)]$ is non-empty and connected for all $v \in V(H)$,
    \item $\varphi(v) \cap \varphi(u) = \emptyset$ for all $u \neq v \in V(H)$, and
    \item if $\{u,v\} \in E(H)$, then there exist $u' \in \varphi(u)$ and $v' \in \varphi(v)$ such that $\{u',v'\}\in E(G)$.
    \end{itemize}
    For a vertex set $S \subseteq V(H)$, we define $\varphi(S) \coloneqq \bigcup_{v \in S} \varphi(v)$ to be the branch set of $S$.
    A minor model is \emph{minimal} if there is no minor model that results from removing a single vertex from a branch set $\varphi(v)$ for some $v \in V(H)$.
\end{definition}

\subparagraph{Labeled graphs.}
We will be annotating the vertices of graphs with labels from a set $X$. This set $X$ will be the modulator $X$ that will be the parameter in our kernelization of \Fdeletion. The labels in each vertex will encode the adjacency of that vertex to the modulator $X$.

\begin{nameddefinition}{labeled graph}[{\cite[Definition 2]{minor-hitting}}]
    \definedhere{\lab}
    Let $X$ be a set. An \emph{$X$-labeled graph} $G$ is a graph $G$ together with label function $\lab_G \colon V(G) \rightarrow \subsets{X}$, assigning a (potentially empty) subset of labels to each vertex in $G$.
\end{nameddefinition}

We will in fact be looking for minors in these labeled graphs which are connected to the modulator $X$ in certain ways. This is captured by the following definition.

\begin{nameddefinition}{labeled minor model}[{\cite[Definition 4]{minor-hitting}}] \label{def:labeled-minor-model}
    A \emph{labeled minor model} of an $X$-labeled graph $H$ in an $X$-labeled graph $G$ is a mapping~$\varphi$ as in \defref{def:minor-model}, that additionally satisfies that for all $v \in V(H)$ and~$\ell \in \lab_H(v)$ there exists $v' \in \varphi(v)$ such that $\ell \in \lab_G(v')$.
\end{nameddefinition}

\definedhere{\minorleq}
If~$G$ contains a (labeled) minor model of~$H$, then we say that~$G$ contains~$H$ as a (labeled) minor and denote this as~$H \minorleq G$. Observe that~$G$ contains~$H$ as a (labeled) minor if and only if~$H$ can be obtained from~$G$ by deleting edges and vertices (and potentially labels), and contracting edges (merging the labelsets of the corresponding vertices).

\section{Proof assuming the two ingredients and consequences}
\label{sec:poly-kernel}

In \autoref{sec-proof-assuming-ingredients} we provide the proof of
\autoref{fact:poly-kernel} assuming that we have at hand the two  ingredients described in \autoref{sec:new-summary}, and in \autoref{sec:consequences} we discuss some of its consequences.

\subsection{Proof of the polynomial kernel assuming the two ingredients}
\label{sec-proof-assuming-ingredients}

In this section we assume that \autoref{fact:compute-solution-breaking-Q} and \autoref{fact:main-lemma} hold, and we prove
the following lemma and \autoref{fact:poly-kernel}. Unlike in all other results that we need to slightly modify from~\cite{minor-hitting}, for which we provide both a full proof and a sketch, we only provide a sketch of the proof of the following lemma, since it is almost exactly the same as the (very long) proof in~\cite{minor-hitting}.

\begin{fact:reduce-components}[Reduce Components Lemma -- Adaptation of {\cite[Lemma 6]{minor-hitting}}]
    % \reduceComponentsLemmaStatement
    Let $\F$ be a finite set of connected graphs and let $\eta \geq 0$ be a constant. There is a polynomial-time algorithm that, given a graph $G$ along with a modulator $X \subseteq V(G)$ such that $\EDF(G \setminus X) \leq \eta$, outputs an induced subgraph $G'$ of $G$ together with an integer $\Delta$ such that $\opt(G) = \opt(G') + \Delta$ and $G' \setminus X$ has at most $\abs{X}^{\bigO_{\F,\eta}(1)}$ connected components. Moreover, a set $Y'$ that hits all $\F$-minors in $G'$ can be extended in polynomial time to a set $Y$ of size $\abs{Y'} + \Delta$ that hits all $\F$-minors in $G$.
\end{fact:reduce-components}
\begin{proof}[Sketch of proof]

    Modify the proof of Lemma 6 in \cite{minor-hitting} by replacing:
    \begin{itemize}
        \item treedepth with $\EDF$;
        \item usage of their Lemma 3 with \namedtheoremref{fact:main-lemma}; and
        \item usage of their Lemma 5 by our \autoref{fact:compute-solution-breaking-Q}. Note that $\mathcal{H}$ in their proof consists of connected graphs, so the added requirement in \autoref{fact:compute-solution-breaking-Q} that $\Q$ is connected is not a problem.
    \end{itemize}
    \vspace{-.7cm}
\end{proof}

We are now ready to prove \autoref{fact:poly-kernel}, but before that, two comments are in place. The first one is that the proof is almost the same as the proof of~\cite[Theorem 1]{minor-hitting}, but we provide it here because we need to use a polynomial ($\alpha$-approximate) kernel by the solution size in the case $\eta=0$, and an extra argument to bound the approximation factor in the case of approximate kernels. The second one is that, for the sake of generality, we state the theorem for the case where the required modulator is {\sl given} along with the input. In the first application discussed in \autoref{sec:consequences} we show how to get rid of this hypothesis.

%\polykernel*
\begin{proof}[Proof of \autoref{fact:poly-kernel}]
    Consider an input $(G, X, k)$ to \Fdeletion parameterized by the size of a given modulator $X$ to graphs with $\EDF \leq \eta$, where $k$ is the size of the sought solution. The proof is by induction on $\eta$.

    ($\eta = 0$) If $\EDF(G \setminus X) = 0$, set $X$ is a modulator to an $\F$-minor-free graph. Thus, $(G, X, k)$ is an instance of \Fdeletion parameterized by the size of a given solution. We apply the polynomial ($\alpha$-approximate) kernelization from the hypothesis to the input to obtain an instance $(G', X', k')$ of \Fdeletion parameterized by the size of a given solution, which serves as our kernel.

    ($\eta \geq 1$) We apply \namedtheoremref{fact:reduce-components} on the input to obtain $G'$ and $\Delta$. We will augment the modulator $X$ into a superset $X'$ to ensure that $\EDF(G' \setminus X') < \eta$. To this end, we consider each connected component $C$ of $G' \setminus X$. Since the \EliminationDistanceToF problem parameterized by the target width is fixed-parameter tractable \cite{elimination-distance-fpt}, and $\eta$ is a constant, we can decide if $\EDF(C)$ is smaller than $\eta$ in $f(\eta)\cdot n^{\bigO(1)}$ time.
    If it is, we do not need to add any vertex from $C$ to $X'$. Otherwise, by the definition of the elimination distance there is a vertex $x_C$ such that $\EDF(C \setminus \{x_C\}) < \EDF(C)$. We find such a vertex $x_C$ by trying all options for $x_C$ and computing the elimination distance to an $\F$-minor-free graph of the resulting graph, again in $f(\eta)\cdot n^{\bigO(1)}$ time. We initialize $X'$ as $X$. For each component $C$ of $G' \setminus X$ with $\EDF(C) = \eta$, we add the corresponding vertex $x_C$ to $X'$.

    Since \namedtheoremref{fact:reduce-components} guarantees that the number of connected components of $G' \setminus X$ is polynomial in $\abs{X}$ for fixed $\F$ and $\eta$, the resulting modulator $X'$ has size polynomial in $\abs{X}$. Moreover, it guarantees that $\EDF(G' \setminus X') < \eta$. Hence we now have an instance $(G', X', k - \Delta)$ of \Fdeletion parameterized by a modulator to $\EDF \leq \eta - 1$, with the same answer as $(G, X, k)$. Using the inductive hypothesis, we apply the ($\alpha$-approximate) kernel for the parameterization by a modulator to $\EDF \leq \eta - 1$, which outputs an instance $(G^\star, X^\star, k^\star)$. By induction, the size of $G^\star$ is bounded by some polynomial in $\abs{X'}$, which is in turn bounded by a polynomial in $\abs{X}$. Hence $G^\star$ has size $\abs{X}^{\bigO_{\F,\eta}(1)}$ which (since $\F$ and $\eta$ are constants) is bounded by $\bigO(\abs{X}^c)$ for some suitably chosen constant $c$, and we output $(G^\star, X^\star, k^\star)$ as the result of the kernelization.

    If the kernelization algorithm from the hypothesis is exact, then $(G^\star, X^\star, k^\star)$ has the same answer as $(G', X', k - \Delta)$ and therefore as $(G, X, k)$, which gives an exact kernelization.

    \newcommand{\solutionLiftingAlgorithm}{\mathcal{A}}
    Otherwise, following the definition of approximate kernelization due to Lokshtanov et al. in the case of a structural parameterization~\cite[Section 2.1]{lossy-kernelization}, and since we parameterize by the size of a solution that we assume to be given in the input (and not by the size of sought solution), there exists a solution lifting algorithm $\solutionLiftingAlgorithm$ that transforms a solution $Y^\star$ for $(G^\star, X^\star, k^\star)$ into a solution $Y'$ for $(G', X', k - \Delta)$ such that \[
    \frac{\abs{Y'}}{\opt(G')} \leq \alpha\cdot\frac{\abs{Y^\star}}{\opt(G^\star)}.
    \]
    Here, we follow the guidelines for problems with structural parameterizations given in \cite{lossy-kernelization}, defining the value of a valid solution $Y$ for an instance $(G^\star, X^\star, k^\star)$, where $X^\star$ is a valid modulator to graphs with bounded $\EDF$, as $\abs{Y}$.

    We construct our solution lifting algorithm by combining $\solutionLiftingAlgorithm$ with the algorithm described in \namedtheoremref{fact:reduce-components}, which outputs a solution $Y$ of size $\abs{Y'} + \Delta$ for $G$ in polynomial time. As $\opt(G') + \Delta = \opt(G)$, we get that \[
    \frac{\abs{Y}}{\opt(G)} = \frac{\abs{Y'} + \Delta}{\opt(G') + \Delta} \leq \frac{\abs{Y'}}{\opt(G')} \leq \alpha\cdot\frac{\abs{Y^\star}}{\opt(G^\star)}.
    \] It follows that $(G^\star, X^\star, k^\star)$ is an $\alpha$-approximate polynomial kernel for $(G, X, k)$ as well.
\end{proof}

Lokshtanov et al. \cite{lossy-kernelization} additionally introduce the notion of \emph{strict} approximate kernelization and \emph{$\alpha$-safe rules} to facilitate the usage of \emph{reduction rules} commonly used in the literature. We could have used these notions for our proof as well, but as our kernelization does not use reduction rules that are applied exhaustively, we opted for the more direct approach above.

\subsection{Consequences of \autoref{fact:poly-kernel}}
\label{sec:consequences}

As said earlier, Fomin et al.~\cite{planar-F-deletion-kernel} showed that \Fdeletion parameterized by the size of a solution admits a randomized polynomial kernel whenever $\F$ contains at least one planar graph. The reliance on randomization of their algorithm lies in the use of a randomized constant-factor approximation algorithm for \Fdeletion~\cite[Theorem 1]{planar-F-deletion-kernel}. Gupta et al.~\cite[Corollary 1.1]{deterministic-approximation-for-planar-F-deletion} gave later a deterministic constant-factor approximation algorithm for the problem, which gives us the following corollary of \autoref{fact:poly-kernel}.

\begin{corollary}
    \label{fact:poly-kernel-planar-F-deletion}
    For every fixed finite set $\F$ of connected graphs containing at least one planar graph and every constant $\eta$, \Fdeletion parameterized by the size of a modulator to graphs with $\EDF \leq \eta$ admits a polynomial kernel.
\end{corollary}

Notice that we do not require that the modulator is given as part of the input in \autoref{fact:poly-kernel-planar-F-deletion}. This is because we can actually compute such a modulator (of size slightly larger) in polynomial time using the constant-factor approximation algorithm by Gupta et al.~\cite[Corollary 1.1]{deterministic-approximation-for-planar-F-deletion} mentioned earlier. Indeed, when $\F$ contains at least one planar graph, the class of graphs with $\EDF \leq \eta$ is characterized by a finite set of forbidden minors that also contains a planar graph. This can be seen, for instance, by observing that if $\EDF(G) \leq \eta$, then $\tw(G) \leq t_\ell + \eta$, where $t_\ell$ is the maximum treewidth of a graph induced by the bag of a leaf in an $\F$-elimination forest of $G$. As the minor obstructions of graphs of bounded treewidth contain at least one planar graph~\cite{excluded-minor-theorem}, the claim follows. We point out that a similar argument is used in~\cite[Section 3]{vertex-cover-bridge-depth} to compute a modulator to bounded bridge-depth.

On the other hand, Dekker and Jansen~\cite[Theorem 2]{FVS-via-EDF-DAM} showed that, assuming $\NP \not\subseteq \coNP/\poly$, \Fdeletion does not admit a polynomial kernel when parameterized by the size of a modulator to graphs with unbounded $\EDF$ if $\F$ is a finite set of biconnected planar graphs on at least three vertices. The requirement that all graphs in $\F$ are planar exclusively comes from their Lemma 10. They define a structure called an \emph{$\F$-necklace}, and state that every graph $G$ that does not contain large $\F$-necklaces as minors has bounded treewidth. They prove this by assuming that $G$ has large treewidth, and then using the Excluded Grid Theorem~\cite{excluded-minor-theorem,tighter-excluded-grid-theorem} to show that $G$ contains a large grid minor, which in turn contains every large enough planar graph as a minor. As $\F$-necklaces are planar if all graphs in $\F$ are planar, this shows that $G$ contains a large $\F$-necklace as a minor, which is a contradiction. This argument still holds if not every graph in $\F$ is planar, as having just one planar graph guarantees that there exist planar $\F$-necklaces of arbitrarily large size~\cite{Bart-personal25}. Thus, $G$ still has an $\F$-necklace as a minor in this case. Hence, their proof can be adapted to show the following theorem, which together with \autoref{fact:poly-kernel-planar-F-deletion} yield \autoref{fact:dichotomy}.

\begin{theorem}[cf. {\cite[Theorem 2]{FVS-via-EDF-DAM}}]
    \label{fact:lower-bound}
    Let ${\mathcal C}$ be a minor-closed family of graphs and let $\F$ be a finite set of biconnected graphs on at least three vertices containing at least one planar graph. If ${\mathcal C}$ has unbounded elimination distance to an $\F$-minor-free graph, then \Fdeletion does not admit a polynomial kernel in the size of a  ${\mathcal C}$-modulator, unless $\NP \subseteq \coNP/\poly$.
\end{theorem}

Finally, combining \autoref{fact:poly-kernel} with the polynomial $\alpha$-approximate kernel for \textsc{Planar Vertex Deletion} of Jansen and W{\l}odarczyk~\cite{vertex-planarization-approximate-kernel} gives us \autoref{fact:planar-poly-approx-kernel} as stated in the introduction.

\section{Ingredient 1: Computing an \texorpdfstring{\Fdeletion}{F-Minor Deletion} solution hitting all labeled \texorpdfstring{$\Q$}{Q}-minors}
\label{sec:compute-solution-breaking-Q}

In this section we show how to obtain our first ingredient. Note that, contrary to Lemma 5 in \cite{minor-hitting}, our next lemma assumes $\F$ and $\Q$ to be sets of {\sl connected} graphs. Fortunately, this does not generate any problem when using it to prove \namedtheoremref{fact:reduce-components}, as the fragments of graphs in $\F$ that we will consider will be connected.

\newtheorem*{fact:compute-intro}{Lemma \ref{fact:compute-solution-breaking-Q}}
\begin{fact:compute-intro}[Computation of optimal solution hitting fragments -- Generalized version of {\cite[Lemma 5]{minor-hitting}}]
    Let $\F$ be a fixed set of connected (unlabeled) graphs, let $\eta \geq 0$ be a constant, and let $X$ be a set. For any set $\Q$ of connected $X$-labeled graphs and host $X$-labeled graph $C$ with $\EDF(C) \leq \eta$, one can:
    \begin{enumerate}
        \item Compute $\opt(C)$ in $\bigO_{\F, \eta}(\abs{V(C)})$ time;
        \item Determine whether there is a solution $Y \in \optsol(C)$ such that $C \setminus Y$ has no labeled $\Q$-minors, in time $f(\F,L,\sum_{H \in \Q}\abs{V(H)}, \eta) \cdot \abs{V(C)}^{\bigO(1)}$ for some function $f$.
    \end{enumerate}
    Here, $L$ is defined as the number of elements of $X$ that appear in the labelset of at least one vertex in at least one graph of $\Q$.
\end{fact:compute-intro}

\newcommand{\hh}{\ensuremath{\mathcal{H}}\xspace}

Item \ref{item:compute-opt} was already shown to be true by Jansen, de Kroon, and W{\l}odarczyk\cite[Theorem 1.2]{vertex-deletion-parameterized-by-elimination-distance}. In fact, they prove it for a stronger parameter than $\EDF$, namely the $\H$-treewidth, where $\H$ is the class of $\F$-minor-free graphs. We will be adapting their proof of this fact to prove \autoref{item:compute-opt-that-breaks-Q}.
In other words, we will prove that the following problem admits an algorithm with the claimed running time. Here, $\F$ and $\Q$ are fixed, and $\Q$ is a set of labeled graphs.

\defproblema{\FdeletionHittingLabeledQ}{
  A labeled graph $G$.
}{
    Is there an optimal \Fdeletion solution for $G$ that also hits all labeled $\Q$-minors?
}

As \cite[Theorem 1.2]{vertex-deletion-parameterized-by-elimination-distance} deals with \emph{unlabeled} graphs, we will first show how to reduce \FdeletionHittingLabeledQ to a similar problem in which $\Q$ is instead a set of unlabeled graphs. This is done in \autoref{sec:transform-labeled-to-unlabeled}. Then, we will show how to solve the unlabeled version of the problem in \autoref{sec:solve-unlabeled-version}, by adapting \cite[Theorem 1.2]{vertex-deletion-parameterized-by-elimination-distance}. Finally, we tie everything together in \autoref{sec:tying-everything-together}. Before all this, we provide some required preliminaries about boundaried graphs in \autoref{sec-prelim-boundaried}.

\subsection{Boundaried graphs}
\label{sec-prelim-boundaried}

We will  be using the concept of a \emph{boundaried graph}, which is similar to a labeled graph, but where the vertices have each at most one \emph{boundary index}, and the boundary indices are not repeated between vertices.

\newrefcommand{\boundary}{\mathsf{Boundary}}
\newrefcommand{\boundaryIndex}{\mathsf{BoundaryIndex}}
\begin{nameddefinition}{boundaried graph}[{\cite[Definition 5]{minor-hitting}}]
    \definedhere{\boundary,\boundaryIndex}
    For a positive integer $t$, a \emph{$t$-boundaried graph} $G$ is a graph with a boundary set $\boundary(G) \subseteq V(G)$ together with an injective boundary index function $\boundaryIndex_G \colon \boundary(G) \rightarrow \{1,\dots,t\}$. For $S \subseteq V(G)$, let \[\boundaryIndex_G(S) \coloneqq \{\boundaryIndex_G(u) \mid u \in (S \cap \boundary(G))\}\] be the (possibly empty) set of boundary indices that are present in $S$.
\end{nameddefinition}

We will also be interested in finding minors in boundaried graphs.

\begin{nameddefinition}{boundaried minor model}[{\cite[Definition 9]{minor-hitting}}] \label{def:boundaried-minor-model}
A \emph{boundaried minor model} of a $t$-boundaried graph $H$ in a $t$-boundaried graph $G$ is a mapping~$\varphi$ as in \defref{def:minor-model}, that additionally satisfies the following for all $v \in V(H)$:
\begin{equation*}
    \boundaryIndex_G(\varphi(v)) =
    \begin{cases}
        \emptyset, & \text{if } v \notin \boundary(H) \\
        \{\boundaryIndex_H(v)\}, & \text{otherwise.}
    \end{cases}
    \end{equation*}
\end{nameddefinition}
Having at most one boundary index in each branch set forces the branch sets to have at most one boundary vertex, in contrast with labeled minors, which allow to have more than one labeled vertex in a branch set. This can be seen as forbidding to contract edges between two boundary vertices when trying to find boundaried minors.

A \emph{boundaried labeled minor model} simultaneously satisfies the conditions of \autoref{def:labeled-minor-model,def:boundaried-minor-model}.

The boundary, instead of encoding the adjacency of a vertex to another graph, will encode which vertex corresponds to which other vertex in another graph. This is captured by the following definition.

\begin{nameddefinition}{$\oplus$}[{\cite[Definition 14]{minor-hitting}}] \label{def:oplus}
    Let~$G_1, G_2$ be two $X$-labeled $t$-boundaried graphs. Then~$G_1 \oplus G_2$ is defined as the $X$-labeled $t$-boundaried graph obtained from the disjoint union of~$G_1$ and~$G_2$ by identifying vertices $u \in V(G_1)$ and $v \in V(G_2)$ whenever $\boundaryIndex_{G_1}(u) = \boundaryIndex_{G_2}(v)$. The labelset of the new vertex is $\lab_{G_1}(u) \cup \lab_{G_2}(v)$. We stress that no parallel edges are introduced in this step.

    For a set~$\mathcal{G} = \{G_1, \ldots, G_k\}$ of $t$-boundaried graphs, define~$\bigoplus_{G \in \mathcal{G}} G$ as~$G_1 \oplus G_2 \oplus \ldots \oplus G_k$.
\end{nameddefinition}

\begin{nameddefinition}{isomorphism}[{\cite[Definition 6]{minor-hitting}}]\label{def:isomorphism} We extend the definition of \emph{graph isomorphism} to boundaried labeled graphs as follows. We say two $t$-boundaried $X$-labeled graphs $G$ and $G'$ are \emph{isomorphic} if there is an isomorphism $f \colon V(G) \rightarrow V(G')$ for which the following additional conditions hold.
\begin{itemize}
\item A vertex $v \in V(G)$ belongs to $\boundary(G)$ if and only if $f(v) \in \boundary(G')$.
\item For all $v \in \boundary(G)$, we have $\boundaryIndex_G(v) = \boundaryIndex_{G'}(f(v))$.
\item For all $v \in V(G)$, $\lab_G(v) = \lab_{G'}(f(v))$.
\end{itemize}
\end{nameddefinition}

\subsection{Transforming labeled minors into unlabeled minors}
\label{sec:transform-labeled-to-unlabeled}

We will be reducing the \FdeletionHittingLabeledQ problem to the problem where $G$ and $\Q$ instead do not have labels.

\defproblema{\FdeletionHittingQ}{
  A graph $G$.
}{
    Is there an optimal \Fdeletion solution for $G$ that also hits all unlabeled $\Q$-minors?
}

To reduce the \textsc{$\F$-Minor Deletion Hitting Labeled $\Q$} problem to \textsc{$\F$-Minor Deletion Hitting $\Q$}, we will be transforming the graph $G$, and the graphs in $\F$ and $\Q$, to unlabeled graphs. The transformation is such that there exists an $\F$- (or $\Q$-) minor in $G$ if and only if the transformed $\F$ (or $\Q$) contains a minor of the transformed graph $G$. This will be done by performing an operation we will call \emph{graph extension}.

\begin{definition}[gluing a graph to a vertex]
    Let $G$ be a (possibly labeled) graph and $H$ be a $1$-boundaried graph. The \emph{gluing} of $H$ to a vertex $v \in V(G)$ is the operation of adding a copy $H'$ of $H$ to $G$ and identifying the boundary vertex of $H'$ with $v$, removing it from the boundary.
\end{definition}

\newrefcommand{\extend}[2]{#1^{+#2}}
\newrefcommand{\gadgets}{\mathsf{Gadgets}}
\newrefcommand{\defaultGadgetSymbol}{*}
\begin{definition}[graph extension]
    \label{def:graph-extension}
    \definedhere{\extend}
    Let $X$ be a set, let $G$ be an $X$-labeled graph, and let $\defaultGadgetSymbol$ be a label not appearing in $X$. Let $\gadgets$ be a function that maps each label $\ell \in X \cup \{\defaultGadgetSymbol\}$ to a different $1$-boundaried graph. The \emph{extension of $G$ with $\gadgets$} is the unlabeled graph $\extend{G}{\gadgets}$ obtained by performing the following steps on every vertex $v \in V(G)$:
    \begin{enumerate}
        \item Glue $\gadgets(\defaultGadgetSymbol)$ to $v$.
        \item For every label $\ell \in \lab(v)$, glue $\gadgets(\ell)$ to $v$.
        \item Remove the labelset of $v$.
    \end{enumerate}
\end{definition}

Note that if $X = \emptyset$, the above operation just glues a copy of $\gadgets(\defaultGadgetSymbol)$ to every vertex in $G$.
We will be extending $G$ and all the graphs in $\F$ and $\Q$ with a special kind of $\gadgets$ function, which we will call a \emph{nice gadgets function}.
\renewrefcommand{\gadgets}{\mathsf{Gadgets}}
\renewrefcommand{\defaultGadgetSymbol}{*}
\begin{definition}[nice gadgets function]
    \label{def:nice-gadgets}
    Let $X$ be a set, let $G$ be an $X$-labeled graph, and let $\defaultGadgetSymbol$ be a label not appearing in $X$. Let $\gadgets$ be a function that maps each label $\ell \in X \cup \{\defaultGadgetSymbol\}$ to a different $1$-boundaried graph. We say $\gadgets$ is a \emph{nice gadgets function for $G$ and $X$} if for every label $\ell \in X \cup \{\defaultGadgetSymbol\}$:
    \begin{itemize}
        \item $\gadgets(\ell)$ is biconnected;
        \item $\gadgets(\ell)$ is not a minor of $G$; and
        \item for every other label $\ell' \in X \cup \{\defaultGadgetSymbol\}$, the graphs $\gadgets(\ell)$ and $\gadgets(\ell')$ are not minors of each other.
    \end{itemize}
\end{definition}

The following lemma shows that there exist nice gadget functions for a graph $G$ with gadgets of size not depending on the size of $G$.

\begin{lemma}
    \label{fact:existence-of-nice-gadgets}
    Let $\F$ be a set of graphs, and $\eta \in \nat$. There exists an algorithm that, given a set $X$ and an $X$-labeled graph $G$ such that $\EDF(G) \leq \eta$, outputs a nice gadgets function for $G$ and $X$ in time $\bigO(g(\abs{X}, \norm{\F}, \eta))$ for some function $g$.
\end{lemma}
\begin{proof}
    \newrefcommand{\ordering}{f}
    \newrefcommand{\biconnectedGraph}{B}

    Note that graph $G$ is $K_{\norm{\F} + \eta}$-minor-free, as otherwise there is a branch in the $\EDF$-elimination forest which leaves a $K_{\norm{\F}}$-minor in a graph induced by the bag of a leaf, meaning that the graph induced by the bag of the leaf is not $\F$-minor-free. Thus, making $K_{\norm{\F} + \eta}$ a minor of $\gadgets(\ell)$ for every $\ell \in X \cup \{\defaultGadgetSymbol\}$ is enough to make all gadgets not be minors of $G$.

    We propose the following construction: take $\abs{X} + 1$ arbitrary non-comparable (under the minor relation) biconnected graphs $\set{\biconnectedGraph_\ell}_{\ell \in X \cup \set{\defaultGadgetSymbol}}$ each of size at most $g(\abs{X})$, that do not have $K_{\norm{\F} + \eta - 2}$ as a minor. To build $\gadgets(\ell)$, connect $\biconnectedGraph_\ell$ with a copy of $K_{\norm{\F} + \eta}$, using two edges between two different vertices of $\biconnectedGraph_\ell$ and two different vertices of the copy of $K_{\norm{\F} + \eta}$. For this to work, it is enough that $\norm{\F} + \eta \geq 5$; if not, we simply take copies of $K_5$ instead of $K_{\norm{\F} + \eta}$\footnote{These requirements could be loosened, but we chose these to simplify the proof.}.

    \newrefcommand{\minorModel}{\varphi}
    \newrefcommand{\firstCopyOfCompleteGraph}{K}
    \newrefcommand{\secondCopyOfCompleteGraph}{K'}
    The resulting graphs will be biconnected because removing any vertex will still leave at least one edge connecting the two still connected graphs $\biconnectedGraph_\ell$ and the copy of $K_{\norm{\F} + \eta}$.
    Let us corroborate that the resulting graphs are not minors of one another. Take $\{\ell,\ell'\}\subseteq X \cup \set{\defaultGadgetSymbol}$ such that $\ell \neq \ell'$, and suppose for a contradiction that $\minorModel$ is a minor model of $\gadgets(\ell)$ in $\gadgets(\ell')$. Let $\firstCopyOfCompleteGraph$ and $\secondCopyOfCompleteGraph$ be the copies of $K_{\norm{\F} + \eta}$ in $\gadgets(\ell)$ and $\gadgets(\ell')$, respectively.

    As there are only two edges between $\biconnectedGraph_{\ell'}$ and $\secondCopyOfCompleteGraph$, there are at most two vertices of $\firstCopyOfCompleteGraph$ whose branch sets in $\minorModel$ contain both vertices from $\biconnectedGraph_{\ell'}$ and $\secondCopyOfCompleteGraph$. All other branch sets of vertices of $\firstCopyOfCompleteGraph$ must belong entirely to either $\biconnectedGraph_{\ell'}$ or $\secondCopyOfCompleteGraph$. As $K_{\norm{\F} + \eta - 2}$ is not a minor of $\biconnectedGraph_{\ell'}$, there must be at least one vertex $v \in \firstCopyOfCompleteGraph$ such that $\minorModel(v) \subseteq \secondCopyOfCompleteGraph$. The edges between $v$ and the other vertices in $\firstCopyOfCompleteGraph$ must be realized in some way in $\minorModel$, which means that every branch set of a vertex in $\firstCopyOfCompleteGraph$ must contain a neighbor of a vertex in $\minorModel(v)$. Therefore, the branch sets of all but at most two vertices in $\firstCopyOfCompleteGraph$ are contained in $\secondCopyOfCompleteGraph$. Additionally, if the branch set for a vertex $w \in \firstCopyOfCompleteGraph$ is contained entirely in $\biconnectedGraph_{\ell'}$, at most two vertices whose branch sets are contained entirely in $\secondCopyOfCompleteGraph$ are neighbors of $w$. We assumed $\norm{\F} + \eta \geq 5$, and hence $w$ is a neighbor of at least three vertices in $\firstCopyOfCompleteGraph$ whose branch sets are contained entirely in $\secondCopyOfCompleteGraph$. Therefore, every branch set of a vertex in $\firstCopyOfCompleteGraph$ must contain a vertex in $\secondCopyOfCompleteGraph$, which means that no vertex of $\secondCopyOfCompleteGraph$ is contained in $\minorModel(V(\biconnectedGraph_\ell))$. Thus, $\minorModel(V(\biconnectedGraph_\ell)) \subseteq V(\biconnectedGraph_{\ell'})$, which contradicts the fact that $\biconnectedGraph_\ell$ is not a minor of $\biconnectedGraph_{\ell'}$.

    We now show a possible construction for the biconnected graphs $\set{\biconnectedGraph_\ell}_{\ell \in X \cup \set{\defaultGadgetSymbol}}$. Fix an ordering $\ordering: X \cup \set{\defaultGadgetSymbol} \rightarrow \set{0,\dots, \abs{X}}$. For each $\ell \in X \cup \{\defaultGadgetSymbol\}$, build $\biconnectedGraph_\ell$ by taking a grid of height two and width $2 \cdot \ordering(\ell)$, and a grid of height three and width $\abs{X} - \ordering(\ell) + 2$, and combining them as in \autoref{fig:gadgets}. It can be easily checked that none of these graphs is a minor of another.
    %\erictodo{Actually showing this could prove difficult, maybe using the fact that the graphs that are smaller have more vertices of degree 4. Is it worth it?}\ig{ok, let's leave it like that, no problem. In can indeed be checked with not much difficulty}.
    The resulting graphs are planar, and thus by the classic result by Kuratowski~\cite{planar-minor-characterization}, they do not contain $K_5$ as a minor. If $\norm{\F} + \eta < 7$, attach a copy of $K_7$ instead of $K_{\norm{\F} + \eta}$ to obtain the gadget $\gadgets(\ell)$. Finally, pick an arbitrary vertex in each of the gadgets to be the boundary vertex.

    \begin{figure}[ht]
        \centering
        \includegraphics[width=.75\textwidth]{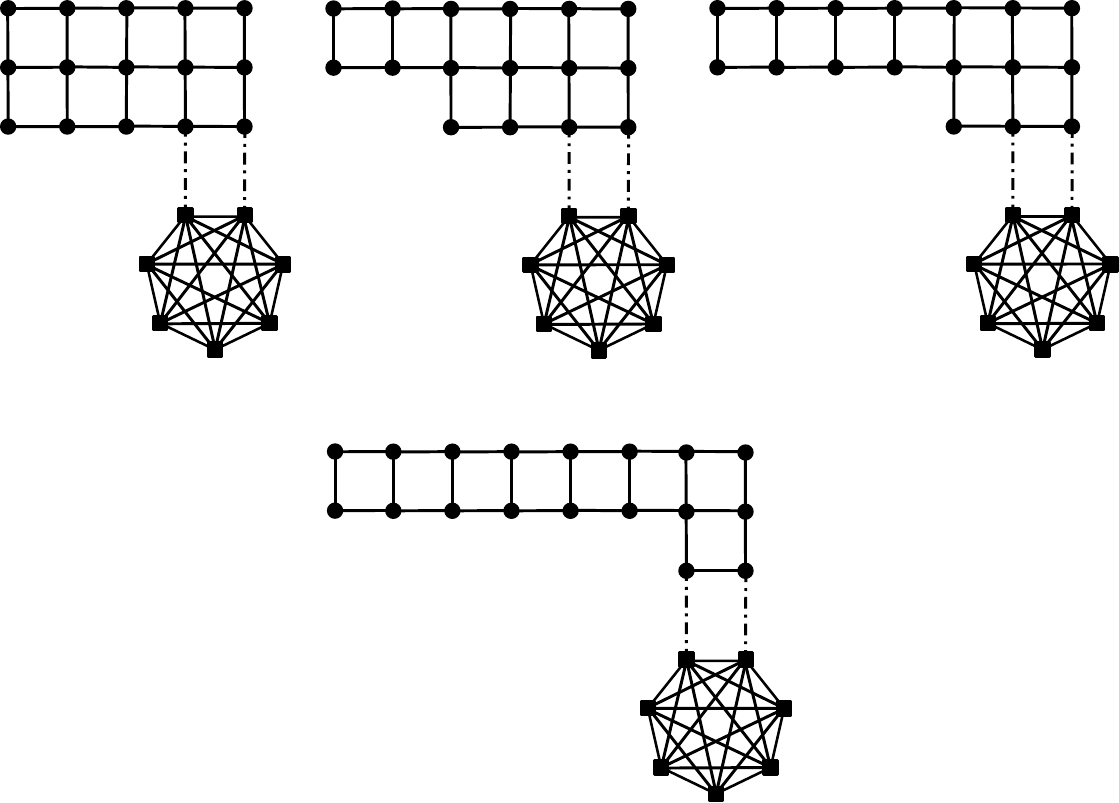}
        \caption{Possible gadgets for $\abs{X} = 3$. No gadget is a minor of another gadget. Each copy of $K_7$ with square vertices is connected to the graph $\biconnectedGraph_\ell$ with circle vertices by two dashed edges.}
        \label{fig:gadgets}
    \end{figure}

    The graph $\biconnectedGraph_\ell$ has size $\bigO(\abs{X})$, while the copy of $K_{\norm{\F} + \eta}$ has size $\bigO((\norm{\F} + \eta)^2)$. Thus, the size of each gadget $\gadgets(\ell)$ is $\bigO(\abs{X} + (\norm{\F} + \eta)^2)$. The total time to construct all gadgets is therefore $\bigO(\abs{X} \cdot (\abs{X} + (\norm{\F} + \eta)^2))$, which is a function of $\abs{X}$, $\norm{\F}$, and $\eta$.
\end{proof}

Extending the graphs with a nice gadgets function will preserve the minor relationship while maintaining the elimination distance of $G$. For this, it is important that we glue gadgets to every vertex, even if it has no labels. Otherwise, a graph $H \in \F$ that is not a minor of $G$ could potentially be a minor of $\extend{G}{\gadgets}$, if for example $H$ is a minor of a gadget in $\gadgets$. Not extending $H$ would therefore not preserve the minor relationship.

The proofs of these properties of graph extension rely on the fact that the branch sets of the biconnected components of a minor of a graph $G$ must intersect with the biconnected components of $G$ in a particular way.

\begin{proposition}[Dekker and Jansen {\fixedspacingcite[Proposition 8]{FVS-via-EDF-DAM}}]
    \label{fact:biconnected-graph-minor-model}
    \newrefcommand{\minorModel}{\varphi}
    Let $H$ be a biconnected graph and let $G$ be a graph which contains $H$ as a minor. Then for any minimal minor model $\minorModel$ of $H$ in $G$, the graph $G[\minorModel(V(H))]$ is biconnected. Furthermore, the graph $G[\minorModel(V(H))]$ is a minor of a biconnected component of $G$.
\end{proposition}

{
\newrefcommand{\minorModel}{\varphi}
\begin{lemma}
    \label{fact:biconnected-components-minor-model}
    Let $\minorModel$ be a minor model of a graph $H$ in a graph $G$. For every biconnected component $B_H$ of $H$, there exists a biconnected component $B_G$ in $G$ such that $B_H$ is a minor of $G[\minorModel(V(B_H)) \cap V(B_G)]$. For an illustration, see \autoref{fig:biconnected-components-minor-model}.
\end{lemma}
\begin{proof}
    \newrefcommand{\minimalMinorModel}{\psi}
    Let $B_H$ be a biconnected component of $H$. We can restrict the domain of $\minorModel$ to the vertices in $B_H$ to obtain a minor model $\minorModel_{B_H}$ of $B_H$. By \autoref{fact:biconnected-graph-minor-model}, this minor model contains a minimal minor model $\minimalMinorModel$ of $B_H$ for which the graph $G[\minimalMinorModel(V(B_H))]$ is biconnected. Therefore, there exists a biconnected component $B_G$ of $G$ that contains $\minimalMinorModel(V(B_H))$. Notice that $\minorModel_{B_H}(V(B_H)) \cap V(B_G)$ contains $\minimalMinorModel(V(B_H))$, and thus $G[\minorModel_{B_H}(V(B_H)) \cap V(B_G)] = G[\minorModel(V(B_H)) \cap V(B_G)]$ contains $B_H$ as a minor.
\end{proof}

\begin{figure}[h]
    \centering
    \includegraphics[width=0.55\textwidth]{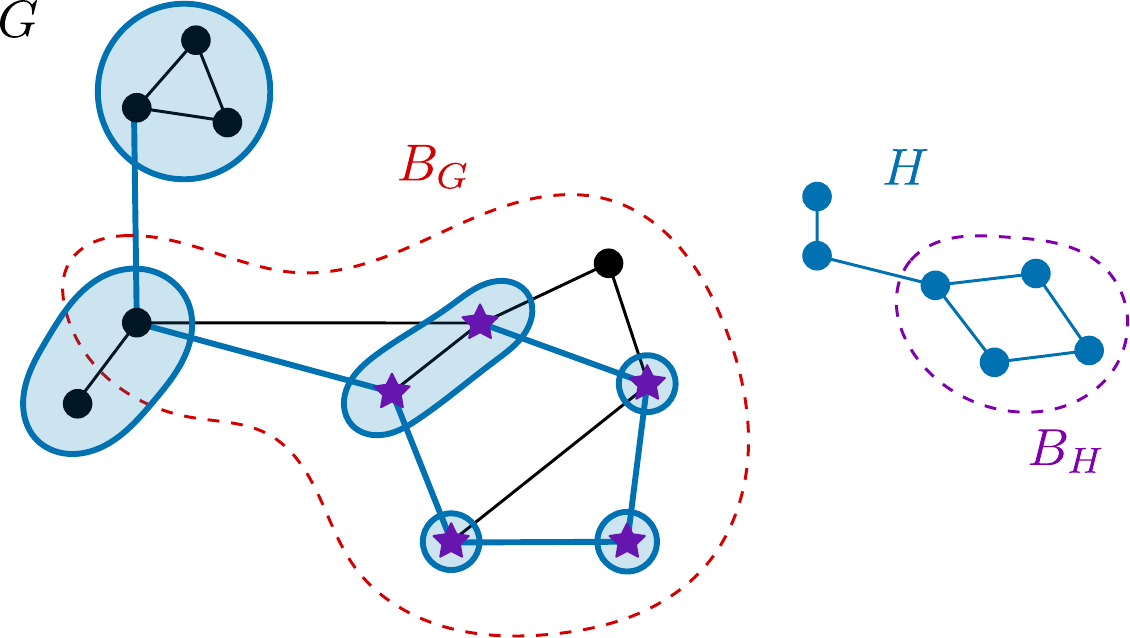}
    \caption{Illustration of \autoref{fact:biconnected-components-minor-model} showing a minor model of a graph $H$ in a graph $G$, in blue. The biconnected component $B_H$ of $H$ can be found as a minor of the graph induced by the purple vertices (represented by stars) of the biconnected component $B_G$ of $G$, which correspond to the intersection of $V(B_G)$ with the branch sets of the vertices in $B_H$.}
    \label{fig:biconnected-components-minor-model}
\end{figure}
}

\renewrefcommand{\gadgets}{\mathsf{Gadgets}}
\newcommand{\extendedG}{\extend{G}{\gadgets}}
\newcommand{\extendedH}{\extend{H}{\gadgets}}
\begin{lemma}[Extending Preserves Minors Lemma]
    \label{fact:extending-preserves-minors}
    \newrefcommand{\minorModel}{\varphi}
    \newrefcommand{\extendedMinorModel}{\noref\extend{\noref\minorModel}{}}
    \newrefcommand{\gadgetsForVertex}[1]{\gadgets_{#1}}
    Let $X$ be a set, and let $G$ be an $X$-labeled graph. Let $\gadgets$ be a nice gadgets function for $G$ and $X$.
    Then an $X$-labeled graph $H$ is a minor of $G$ if and only if $\extendedH$ is a minor of $\extendedG$. Moreover, if $\extendedMinorModel$ is a minor model of $\extendedH$ in $\extendedG$, then the function $\minorModel$ defined as $\minorModel(v) = \extendedMinorModel(\gadgetsForVertex{v}) \cap V(G)$ for every $v \in V(H)$ is a labeled minor model of $H$ in $G$, where $\gadgetsForVertex{v}$ is the union of all copies of gadgets glued to $v$ in $\extendedH$. See \autoref{fig:extending-preserves-minors} for an illustration.
\end{lemma}
\begin{figure}[h]
    \centering
    \includegraphics[width=0.6\textwidth]{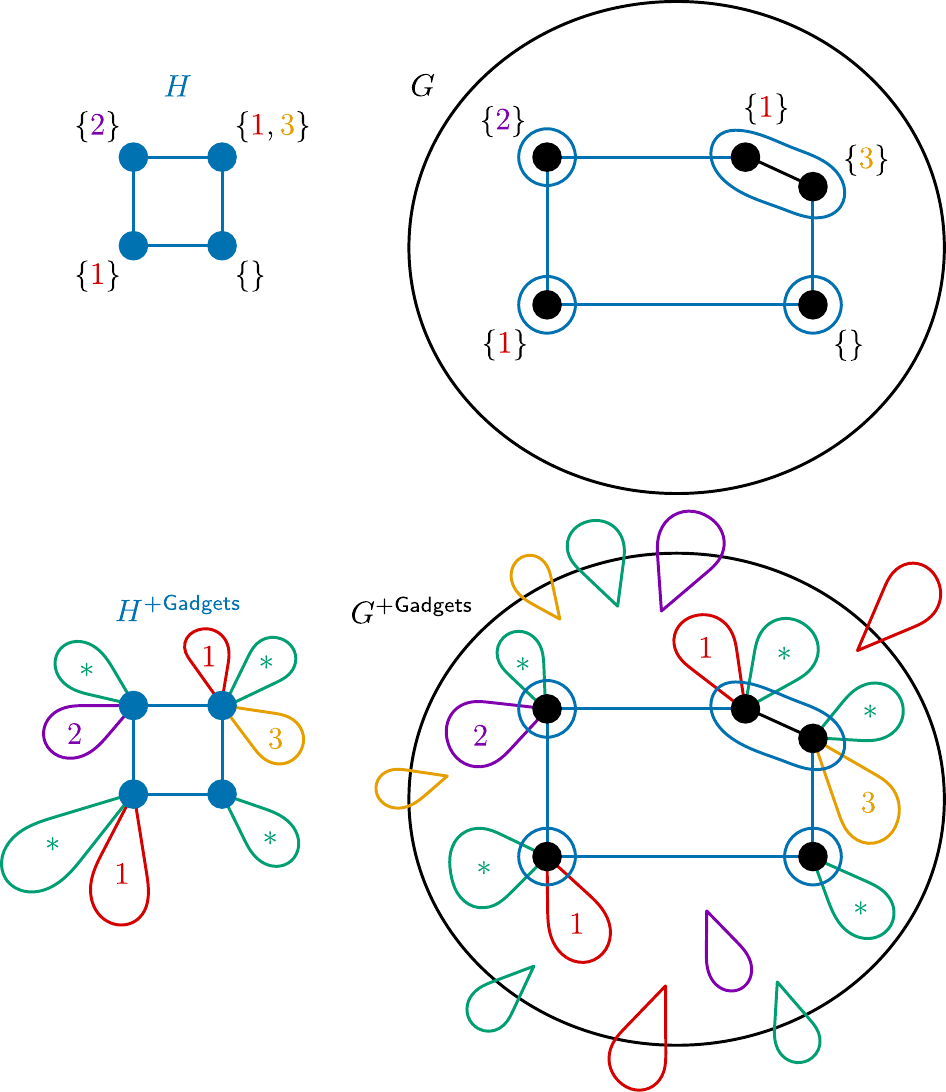}
    \caption{Illustration of \autoref{fact:extending-preserves-minors}. The bubbles in the graphs in the bottom represent the copies of gadgets glued to the vertices of $H$ and $G$. The minor model in blue of $H$ in $G$ is transformed into a minor model of $\extendedH$ in $\extendedG$ by adding the appropriate copies of gadgets to the branch sets. Conversely, the minor model of $\extendedH$ in $\extendedG$ can be transformed into a minor model of $H$ in $G$ by only keeping the vertices in $V(G)$ in the minor model.}
    \label{fig:extending-preserves-minors}
\end{figure}
\begin{proof}
    $\Longrightarrow)$ This implication follows directly from \defref{def:labeled-minor-model} and \defref{def:graph-extension} by adding the appropriate copies of gadgets to the branch sets of a minor model of $H$ in $G$.

    \newrefcommand{\minorModel}{\varphi}
    \newrefcommand{\extendedMinorModel}{\noref\extend{\noref\minorModel}{}}
    \newrefcommand{\branchSet}{\mathsf{BranchSet}}
    \newrefcommand{\gadgetsForVertex}[1]{\gadgets_{#1}}
    $\Longleftarrow)$ Let $\extendedMinorModel$ be a minor model of $\extendedH$ in $\extendedG$.
    Let $\gadgetsForVertex{v}$ denote the union of all the copies of gadgets glued to $v \in V(H)$ in $\extendedH$, as described in the statement of the lemma. Define $\minorModel: V(H) \to \subsets{G}$ such that $\minorModel(v) = \extendedMinorModel(\gadgetsForVertex{v}) \cap V(G)$. We will show that $\minorModel$ is a labeled minor model of $H$ in $G$.

    \newrefcommand{\biconnectedComponent}{B}
    \begin{claim}\label{fact:gadgets-go-to-gadgets}
        Let $v \in V(H)$ and $\ell \in \lab_H(v) \cup \{\defaultGadgetSymbol\}$. Let $\gadgets(\ell)_v$ be the copy of the gadget $\gadgets(\ell)$ glued to $v$ in $\extendedH$. Then the branch set $\extendedMinorModel(\gadgets(\ell)_v)$ contains a copy of a gadget $\gadgets(\ell)$ in $\extendedG$.
    \end{claim}
    \begin{claimproof}
        As $\gadgets(\ell)_v$ is biconnected by \defref{def:nice-gadgets}, by \autoref{fact:biconnected-components-minor-model} there exists a biconnected component $\biconnectedComponent$ of $\extendedG$ such that $G[\extendedMinorModel(V(\gadgets(\ell)_v)) \cap V(\biconnectedComponent)]$ contains $\gadgets(\ell)_v$ as a minor.
        Notice that every time the biconnected graph $\gadgets(\ell)$ is glued to a vertex in $V(G)$, a new biconnected component is created. As $\gadgets(\ell)$ is not a minor of $G$, the biconnected component $\biconnectedComponent$ must correspond to one of the glued gadgets. Additionally, since $\gadgets(\ell)$ is not a minor of $\gadgets(\ell')$ for any distinct labels $\ell$ and $\ell'$, the only gadget that contains $\gadgets(\ell)_v$ as a minor is precisely $\gadgets(\ell)$, and thus $\biconnectedComponent$ must correspond to a copy of $\gadgets(\ell)$ in $\extendedG$.

        Lastly, as the number of vertices in $\gadgets(\ell)_v$ is exactly the number of vertices in $\biconnectedComponent$, the branch set $\extendedMinorModel(V(\gadgets(\ell)_v))$ must contain all vertices in $\biconnectedComponent$.
    \end{claimproof}

    We now follow \defref{def:minor-model} and \defref{def:labeled-minor-model} to show that $\minorModel$ is a labeled minor model of $H$ in $G$.

    Take a vertex $v \in V(H)$. By \autoref{fact:gadgets-go-to-gadgets}, the branch set $\extendedMinorModel(\gadgetsForVertex{v})$ must contain copies in $\extendedG$ of all the gadgets glued to $v$. As each gadget $\gadgets(\ell)$ for a label $\ell \in X$ was glued only to vertices with that label in $G$, the branch set $\extendedMinorModel(\gadgetsForVertex{v}) \cap V(G)$, which is equal to $\minorModel(v)$, must contain all labels in $\lab_H(v)$.

    It remains to show that $\minorModel$ follows the conditions of \defref{def:minor-model} for all $v \in V(H)$.

    \begin{itemize}
        \item \underline{$G[\minorModel(v)]$ is non-empty:} Notice that $\gadgetsForVertex{v}$ contains a copy $\gadgets(\defaultGadgetSymbol)_v$ of $\gadgets(\defaultGadgetSymbol)$. By \autoref{fact:gadgets-go-to-gadgets}, the branch set $\extendedMinorModel(\gadgets(\defaultGadgetSymbol)_v)$ contains a copy of $\gadgets(\defaultGadgetSymbol)$ in $\extendedG$, which contains a vertex in $V(G)$. As $\gadgets(\defaultGadgetSymbol)_v \subseteq \gadgetsForVertex{v}$, the branch set $\extendedMinorModel(\gadgetsForVertex{v})$ contains a vertex in $V(G)$, and thus $\extendedMinorModel(\gadgetsForVertex{v}) \cap V(G)$ is non-empty.
        \item \underline{$G[\minorModel(v)]$ is connected:} Suppose, to the contrary, that $G[\minorModel(v)]$ is disconnected. Observe that $\gadgetsForVertex{v}$ is connected in $\extendedH$, and thus $\extendedMinorModel(\gadgetsForVertex{v})$ is connected in $\extendedG$. Therefore, there exist two vertices $u_1, u_2 \in \extendedMinorModel(\gadgetsForVertex{v}) \cap V(G)$ that are connected in $\extendedG$ but not in $G$. This means that there is a simple path connecting $u_1$ and $u_2$ in $\extendedG$ that uses a vertex not belonging to $V(G)$ in a copy $\gadgets(\ell)_w$ of a gadget $\gadgets(\ell)$. But $\gadgets(\ell)_w$ was glued to the single vertex $w \in V(G)$, and thus the path must contain $w$ twice, contradicting the fact that the path is simple. Hence, $G[\minorModel(v)]$ is connected.
        \item \underline{$\minorModel(v) \cap \minorModel(u) = \emptyset$ for all $u \neq v \in V(H)$:} This follows directly from applying \defref{def:minor-model} to $\extendedMinorModel$.
        \newrefcommand{\Pv}{P_v}
        \newrefcommand{\Pu}{P_u}
        \newrefcommand{\copyOfAGadget}{B}
        \item \underline{If $\{u,v\} \in E(H)$, then there exist $u' \in \minorModel(u)$ and $v' \in \minorModel(v)$ such that $\{u',v'\}\in E(G)$:}
        By \defref{def:minor-model}, there exist $u' \in \extendedMinorModel(u)$ and $v' \in \extendedMinorModel(v)$ such that $\{u',v'\} \in E(\extendedG)$. We will prove that $\{u', v'\} \in E(G)$. To the contrary, suppose that there exists in $\extendedG$ a copy $\copyOfAGadget$ of a gadget glued to a vertex $w \in V(G)$ such that $\{u', v'\} \in E(\extendedG[\copyOfAGadget])$.

        \newrefcommand{\copyOfTheDefaultGadget}{B'}
        Since $\gadgetsForVertex{u}$ is connected, we have that $\extendedMinorModel(\gadgetsForVertex{u})$ is also connected, and by \autoref{fact:gadgets-go-to-gadgets} it contains a copy $\copyOfTheDefaultGadget$ of $\gadgets(\defaultGadgetSymbol)$ in $\extendedG$. This copy $\copyOfTheDefaultGadget$ cannot be exactly $\copyOfAGadget$ because $v'$, which does not belong to $\extendedMinorModel(\gadgetsForVertex{u})$, is in $\copyOfAGadget$. Additionally, $\copyOfTheDefaultGadget$ cannot even be contained in $\copyOfAGadget$, as different gadgets are not minors of one another by \defref{def:nice-gadgets}. There is thus a vertex in $\extendedMinorModel(\gadgetsForVertex{u})$ contained in $\copyOfAGadget$, namely $u'$, and at least one vertex in $\extendedMinorModel(\gadgetsForVertex{u})$ not contained in $\copyOfAGadget$, namely a vertex of $\copyOfTheDefaultGadget$. As $w$ is the only vertex in $\copyOfAGadget$ that is adjacent to vertices outside $\copyOfAGadget$, the connectivity of $\extendedMinorModel(\gadgetsForVertex{u})$ implies that $w \in \extendedMinorModel(\gadgetsForVertex{u})$.
        By a symmetric argument, we get that $w \in \extendedMinorModel(\gadgetsForVertex{v})$, contradicting the fact that $\extendedMinorModel(\gadgetsForVertex{u})$ and $\extendedMinorModel(\gadgetsForVertex{v})$ must be disjoint.
        \qedhere
    \end{itemize}
\end{proof}

\renewrefcommand{\gadgets}{\mathsf{Gadgets}}
\newcommand{\extendedF}{\extend{\F}{\gadgets}}
\begin{lemma}[Extending Preserves Elimination Distance Lemma]
    \label{fact:extending-preserves-edf}
    Let $X$ be a set, let $G$ be an $X$-labeled graph, and let $\F$ be a family of graphs. Let $\gadgets$ be a nice gadgets function for $G$ and $X$. Then $\EDF(G) = \ED{\extendedF}(\extendedG)$.
\end{lemma}
\begin{proof}
    \newrefcommand{\extendedT}{\noref\extend{T}{}}
    \newrefcommand{\leaf}{t}
    \newrefcommand{\extendedLeaf}{\noref\extend{\noref\leaf}{}}
    \newcommand{\bags}{\chi}
    We prove the two inequalities separately.
    \begin{itemize}
        \item\underline{$\EDF(G) \geq \ED{\extendedF}(\extendedG)$:}
        Let $T$ be an $\F$-elimination forest of $G$ with depth $\eta$. We create an $\extendedF$-elimination forest $\extendedT$ of $\extendedG$ with depth $\eta$. The internal nodes of $\extendedT$ will have the same bags as the internal nodes of $T$. Notice that each leaf $\leaf$ of $T$ corresponds with a leaf $\extendedLeaf$ in $\extendedT$ such that $\extendedG[\bags(\extendedLeaf)]$ is the extension of $G[\bags(\leaf)]$. As each bag of a leaf in $T$ induces an $\F$-minor-free graph, by \namedtheoremref{fact:extending-preserves-minors}, the bag of the leaf $\extendedLeaf$ induces an $\extendedF$-minor-free graph.

        \newrefcommand{\newLeaf}{\noref\leaf_{v, \ell}}
        Additionally, a new leaf $\newLeaf$ is created in $\extendedT$ for each copy of a gadget $\gadgets(\ell)$ glued to an internal node $v$ of $T$. The bag of this leaf $\newLeaf$ contains all vertices of the copy of the gadget, except for the single vertex in the boundary, which is identified with $v$.

        As every graph $\noref\extend{H}{} \in \extendedF$ is an extension of a graph $H \in \F$, it contains a copy of $\gadgets(\defaultGadgetSymbol)$ as an induced subgraph. If $\ell = \defaultGadgetSymbol$, the graph induced by $\bags(\newLeaf)$ does not contain $\gadgets(\defaultGadgetSymbol)$ as a minor, as $\bags(\newLeaf)$ does not contain the vertex $v$. On the other hand, by \defref{def:nice-gadgets}, all gadgets are non-comparable according to the minor relation, and so if $\ell \neq \defaultGadgetSymbol$, the graph induced by $\bags(\newLeaf)$ still does not contain $\gadgets(\defaultGadgetSymbol)$ as a minor. This in turn means that the graph induced by $\bags(\newLeaf)$ does not contain any graph in $\extendedF$ as a minor. Thus, $\bags(\newLeaf)$ induces an $\extendedF$-minor-free graph.

        As every bag of a leaf in $\extendedT$ induces an $\extendedF$-minor-free graph, we have that $\extendedT$ is an $\extendedF$-elimination forest of $\extendedG$ with depth $\eta$. Thus, $\ED{\extendedF}(\extendedG) \leq \eta = \EDF(G)$.

        \renewrefcommand{\extendedT}{\noref\extend{T}{}}
        \item\underline{$\EDF(G) \leq \ED{\extendedF}(\extendedG)$:} We proceed by induction on $\eta \coloneq \ED{\extendedF}(\extendedG)$.
        \begin{itemize}
            \item\textbf{Base case ($\eta = 0$):} By \defref{def:elimination-distance}, the graph $\extendedG$ is $\extendedF$-minor-free, and thus by \namedtheoremref{fact:extending-preserves-minors}, $G$ is $\F$-minor-free. Therefore, $\EDF(G) = 0 = \ED{\extendedF}(\extendedG)$.
            \renewrefcommand{\root}{r}
            \item\textbf{Inductive step ($\eta \geq 1$):} Let $\extendedT$ be an $\extendedF$-elimination forest $\extendedT$ of $\extendedG$ with depth $\eta$. Assume $\extendedG$ is connected; otherwise, the same argument can be applied to each of the connected components. Let $\root$ be the root of $\extendedT$.

            Let $v \in V(G)$ be such that the only vertex $w$ in $\bags(\root)$ belongs to a copy of a gadget glued to $v$. By \defref{def:elimination-distance}, the graph $\extendedG \setminus \{w\}$ has $\extendedF$-elimination distance at most $\eta - 1$. Additionally, $\extendedG \setminus \{w\}$ contains the extension of $G \setminus \{v\}$ as an induced subgraph, which must therefore also have $\extendedF$-elimination distance at most $\eta - 1$. By the inductive hypothesis, the graph $G \setminus \{v\}$ must thus have $\F$-elimination distance at most $\eta - 1$. Hence, we have found a vertex $v$ in $G$ whose removal produces a graph with $\F$-elimination distance at most $\eta - 1$, and thus $\EDF(G) \leq \eta$.\qedhere
        \end{itemize}
    \end{itemize}
\end{proof}

The following observation, which follows from \defref{def:labeled-minor-model}, shows that we can in fact keep only the labels appearing in the graphs in $\Q$ when transforming $G$, $\F$, and $\Q$. This will allow us to reduce the number of labels in the input, which will be useful for the running time of the algorithm.
\begin{observation}
    \label{fact:can-keep-only-labels-in-Q}
    \newrefcommand{\XQ}{X_\Q}
    \newrefcommand{\GQ}{G_\Q}
    Let $X$ be a set, $G$ be an $X$-labeled graph, and $\Q$ be a family of $X$-labeled graphs. Let $\XQ \subseteq X$ be the set of labels appearing in at least one vertex in a graph in $\Q$. Define $\GQ$ as the $\XQ$-labeled graph obtained from $G$ by removing all labels that do not appear in $\XQ$ from its vertices. Then the graph $\GQ$ has a $\Q$-minor if and only if $G$ has a $\Q$-minor.
\end{observation}

We are now ready to present the reduction between the \textsc{$\F$-Minor Deletion Hitting Labeled $\Q$} and \textsc{$\F$-Minor Deletion Hitting $\Q$} problems. In our reduction, the elimination distance of the extended graph is the same as the original graph, for the extended and original graph classes respectively. This is stronger than what \autoref{fact:compute-solution-breaking-Q} requires, as it would be enough for the extended graph to have elimination distance not greater than a function of the original elimination distance. However, this stronger result could
be useful for applications of the reduction in future work.

\newrefcommand{\totalLabels}{L}
\renewrefcommand{\extendedG}{\noref\extend{G}{}}
\renewrefcommand{\extendedF}{\noref\extend{\F}{}}
\newrefcommand{\extendedQ}{\noref\extend{\Q}{}}
\let\oldeta\eta
\renewrefcommand{\eta}{\oldeta}
\begin{lemma}[Unlabeling Lemma]
    \label{fact:labeled-to-unlabeled}

    Let
    \begin{itemize}
    \item $X$ be a set;
    \item $\F$ be a finite set of unlabeled graphs;
    \item $\eta$ be a positive integer;
    \item $G$ be an $X$-labeled graph such that $\EDF(G) \leq \eta$;
    \item $\Q$ be a finite set of $X$-labeled graphs; and
    \item $\totalLabels$ be the number of different labels in vertices in graphs in $\Q$.
    \end{itemize}
    There exists an algorithm that runs in time \[O\parens*{f(\totalLabels, \norm{\F}, \eta) \cdot \parens*{\abs{V(G)} + \abs{E(G)} + \sum_{H\in(\F\cup \Q)}(\abs{V(H)} + \abs{E(H)})}}\] for some function $f$ and transforms $G$, $\F$, and $\Q$ into an unlabeled graph $\extendedG$ and finite sets of unlabeled graphs $\extendedF$ and $\extendedQ$, respectively, such that
    \begin{enumerate}
        \item\label{item:labeled-to-unlabeled:ed-is-maintained} $\EDF(G) = \ED{\extendedF}(\extendedG)$;
        \item\label{item:labeled-to-unlabeled:sizes-are-maintained} $\abs{\extendedF} = \abs{\F}$ and $\abs{\extendedQ} = \abs{\Q}$;
        \item\label{item:labeled-to-unlabeled:norm-is-bounded} $\norm{\extendedF} \leq f(\totalLabels, \norm{\F}, \eta) \cdot \norm{\F}$ and $\norm{\extendedQ} \leq f(\totalLabels, \norm{\F}, \eta) \cdot \norm{\Q}$; and
        \item\label{item:labeled-to-unlabeled:reduces-the-problem} $G$ is a \yes-instance of \textsc{$\F$-Minor Deletion Hitting Labeled $\Q$} if and only if $\extendedG$ is a \yes-instance of \textsc{$\extendedF$-Minor Deletion Hitting $\extendedQ$}.
    \end{enumerate}
\end{lemma}
\begin{proof}
    \newrefcommand{\labelsInQ}{\noref\lab(\Q)}
    \renewrefcommand{\gadgets}{\mathsf{Gadgets}}
    Let $\labelsInQ$ be the set of different labels in $\Q$. By \autoref{fact:can-keep-only-labels-in-Q}, we can assume that $X = \labelsInQ$, and thus that $\abs{X} = \totalLabels$. Take $\gadgets$ to be a nice gadgets function for $G$ and $X$ provided by \autoref{fact:existence-of-nice-gadgets}. We extend the graph $G$ and every graph in $\F$ and $\Q$ to create the graph $\extendedG$ and the families $\extendedF$ and $\extendedQ$, respectively. We will now show that $\extendedG$, $\extendedF$, and $\extendedQ$ satisfy the properties of the lemma.

    \hyperref[item:labeled-to-unlabeled:ed-is-maintained]{Item \ref{item:labeled-to-unlabeled:ed-is-maintained}} holds by \namedtheoremref{fact:extending-preserves-edf}.

    Notice that the number of graphs in $\extendedF$ and $\extendedQ$ is the same as in $\F$ and $\Q$, respectively, because we created one new graph for each graph in $\F$ and $\Q$. Thus, \autoref{item:labeled-to-unlabeled:sizes-are-maintained} holds.

    For each vertex $v$ in a graph $H$ in $\F$ or $\Q$, we glued one gadget for each label in $\lab(v)$, and one additional gadget $\gadgets(\defaultGadgetSymbol)$. By \autoref{fact:existence-of-nice-gadgets}, each gadget has size at most $g(\totalLabels, \norm{\F}, \eta)$ for some function $g$. Thus, there are at most $(\totalLabels + 1) \cdot g(\totalLabels, \norm{\F}, \eta)$ vertices in the extension of $H$ for each vertex in $V(H)$. The maximum number of vertices in a graph in $\extendedF$ and $\extendedQ$ is thus at most $(\totalLabels + 1) \cdot g(\totalLabels, \norm{\F}, \eta)\cdot \norm\F$ and $(\totalLabels + 1) \cdot g(\totalLabels, \norm{\F}, \eta)\cdot \norm\Q$, respectively. This proves \autoref{item:labeled-to-unlabeled:norm-is-bounded}.

    \begin{claim}[\autoref{item:labeled-to-unlabeled:reduces-the-problem}]
        $G$ is a \yes-instance of \textsc{$\F$-Minor Deletion Hitting Labeled $\Q$} if and only if $\extendedG$ is a \yes-instance of \textsc{$\extendedF$-Minor Deletion Hitting $\extendedQ$}.
    \end{claim}
    \begin{claimproof}
        \newrefcommand{\solution}{Y}
        \newrefcommand{\connectedComponent}{C}
        $\Longrightarrow)$ Let $\solution$ be a solution to \Fdeletion on $G$ that hits all labeled $\Q$-minors. By \namedtheoremref{fact:extending-preserves-minors}, every minor model in $\extendedG$ of a graph $\extend{H}{} \in \extendedF \cup \extendedQ$ contains vertices that form a labeled minor model in $G$ of $H \in \F \cup \Q$. By the contrapositive statement, the solution $\solution$, which hits all $\F$- and $\Q$-minors in $G$, also hits all $\extendedF$- and $\extendedQ$-minors in $\extendedG$.

        \newrefcommand{\extendedSolution}{\noref\extend{Y}{}}
        $\Longleftarrow)$ Let $\extendedSolution$ be a solution to \textsc{$\extendedF$-Minor Deletion} on $\extendedG$ that hits all $\extendedQ$-minors. Suppose $\extendedSolution$ contains a vertex $v$ in a copy of a gadget glued to a vertex $u \in V(G)$. Recall that every connected component in a graph in $\extendedF$ and $\extendedQ$ contains a copy of a gadget, and that gadgets are not minors of one another. Replacing $v$ with $u$ in $\extendedSolution$ will thus not create any new $\extendedF$- or $\extendedQ$-minors in $\extendedG \setminus \extendedSolution$. We therefore assume that $\extendedSolution \subseteq V(G)$, which means
        by \namedtheoremref{fact:extending-preserves-minors} that $\extendedSolution$ is an $\F$-minor deletion set on $G$ that hits all labeled $\Q$-minors.
    \end{claimproof}

    Finally, the algorithm only creates $f(\totalLabels, \norm\F, \eta)$ vertices for each vertex in a graph in $\F \cup \Q \cup \{G\}$, and thus the time complexity is \[\bigO\parens*{f(\totalLabels, \norm{\F}, \eta) \cdot \parens*{\abs{V(G)} + \abs{E(G)} + \sum_{H\in(\F\cup \Q)}(\abs{V(H)} + \abs{E(H)})}},\] as stated in the lemma.
\end{proof}
\renewcommand{\eta}{\oldeta}

\subsection{Solving the unlabeled version of the problem}
\label{sec:solve-unlabeled-version}
Now, we show how to solve the \FdeletionHittingQ problem parameterized by $\EDF$, which is accomplished in \autoref{fact:F-deletion-hitting-Q-parameterized-by-twF}.
We will be adapting Corollary 5.42 found in the full version of the article by Jansen, de Kroon, and W{\l}odarczyk\cite{vertex-deletion-parameterized-by-elimination-distance-arxiv}, which provides an \FPT algorithm for \Fdeletion when parameterized by the $\EDF$ of the input graph. Many of the lemmas and definitions presented in Section 5 in \cite{vertex-deletion-parameterized-by-elimination-distance-arxiv} will need to be adapted to our setting.

The algorithm by Jansen, de Kroon, and W{\l}odarczyk actually works when parameterizing by the \emph{$\H$-treewidth} of the input graph, which is a parameter that generalizes both treewidth and $\EDF$.

\newrefcommand{\twH}[1]{\noref\tw_{\H}}
\begin{nameddefinition}{$\H$-treewidth}[{\cite[Definition 6]{H-treewidth}}]
\newrefcommand{\bags}{\chi}
For a graph class $\H$, a \emph{tree $\H$-decomposition} of a graph $G$ is a triple $(T, \bags, L)$ where~$L \subseteq V(G)$,~$T$ is a rooted tree, and~$\bags \colon V(T) \to \subsets{V(G)}$, called the \emph{bags of $T$}, such that:
\begin{enumerate}
    \item For each~$v \in V(G)$ the nodes~$\{t \mid v \in \bags(t)\}$ form a {non-empty} connected subtree of~$T$.
    \item For each edge~$\set{u,v} \in E(G)$ there is a node~$t \in V(G)$ with~$\set{u,v} \subseteq \bags(t)$.
    \item For each vertex~$v \in L$, there is a unique~$t \in V(T)$ for which~$v \in \bags(t)$,  with~$t$ being a leaf of~$T$. \label{item:tree:h:decomp:unique}
    \item For each node~$t \in V(T)$, the graph~$G[\bags(t) \cap L]$ belongs to~$\H$.
\end{enumerate}
The \emph{width} of a tree $\H$-decomposition is defined as~$\max(0, \max_{t \in V(T)} |\bags(t) \setminus L| - 1)$. The $\H$-treewidth of a graph~$G$, denoted~$\twH(G)$, is the minimum width of a tree $\H$-decomposition of~$G$.
The connected components of $G[L]$ are called \emph{base components}
and the vertices in $L$ are called \emph{base vertices}.

A pair~$(T, \bags)$ is a (standard) \emph{tree decomposition} if~$(T, \bags, \emptyset)$ satisfies all conditions of an $\H$-decomposition; the choice of~$\H$ is irrelevant.
\end{nameddefinition}

\begin{lemma}[Jansen, de Kroon, and W{\l}odarczyk {\fixedspacingcite[Lemma 2.4]{vertex-deletion-parameterized-by-elimination-distance}}]\label{fact:treedepth-treewidth}
For any class $\H$ defined by a set $\F$ of forbidden minors and graph $G$, we have $\twH(G) \leq \EDF(G)$. Furthermore, given an $\H$-elimination forest of depth $d$ we can construct a tree $\F$-decomposition of width $d$ in polynomial time.
\end{lemma}

\autoref{fact:treedepth-treewidth} implies that an \FPT algorithm parameterized by $\twH(G)$ can be used to construct an \FPT algorithm parameterized by $\EDF(G)$.

Fortunately, the $\H$-treewidth of a graph $G$ can be approximated efficiently when $\H$ is the class of $\F$-minor-free graphs.

\begin{lemma}[Jansen, de Kroon, and W{\l}odarczyk {\fixedspacingcite[Theorem 4.35]{vertex-deletion-parameterized-by-elimination-distance-arxiv}}]
  \label{fact:compute-tree-decomposition}
  Let $\H$ be a class of graphs defined by a finite family of forbidden minors. There exists an algorithm that, given a graph $G$ such that $\twH(G) \leq k$, runs in time $2^{k^{\bigO(1)}} \cdot n^{\bigO(1)}$ and returns a tree $\H$-decomposition of $G$ of width $\bigO(k^5)$.
\end{lemma}

The algorithm by Jansen, de Kroon, and W{\l}odarczyk proceeds inductively on the bags of a tree $\H$-decomposition $T$ of $G$. It computes, for every rooted subtree $T'$ of $T$, a small number of possible subsets of the vertices $A$ that appear exclusively in the bags in $T'$, such that there exists an optimal \Fdeletion solution that takes exactly these vertices in $A$.

Instead of focusing on a subtree of the decomposition, the lemmas focus on the actual subsets of $V(G)$ defined by the subtrees, in the form of \emph{tri-separations} in the graph $G$.

\begin{definition}[tri-separation {\fixedspacingcite[Definition 5.6]{vertex-deletion-parameterized-by-elimination-distance-arxiv}}] \label{def:triseparation}
A \emph{tri-separation} in a graph~$G$ is a partition~$(A,X,B)$ of~$V(G)$ such that no vertex in~$A$ is adjacent to any vertex of~$B$. The set~$X$ is the \emph{separator} corresponding to the tri-separation. The \emph{order} of the tri-separation is defined as~$\abs{X}$.
\end{definition}

To define these subsets of $A$, they first define an equivalence relation between boundaried graphs.
If two different subsets of $A$, when removed, leave two boundaried graphs that are equivalent, then it is enough to keep only one of them.

Two $k$-boundaried graphs $G_1, G_2$ are \emph{compatible} if $\boundaryIndex_{G_1} \circ \boundaryIndex_{G_2}^{-1}$ is a graph isomorphism between $G_1[\boundary(G_1)]$ and $G_2[\boundary(G_2)]$.

\begin{definition}[{cf. \cite[Definition 5.28]{vertex-deletion-parameterized-by-elimination-distance-arxiv}}] \label{def:boundaried:eqv}
  Let $\F$ be a family of graphs. We say that two $k$-boundaried graphs $G_1, G_2$
  are \emph{$(\F,k)$-equivalent} if they are compatible and for every compatible $k$-boundaried graph $G_3$, it holds that $G_1 \oplus G_3$ is $\F$-minor-free if and only if $G_2 \oplus G_3$ is $\F$-minor-free.
\end{definition}

\begin{observation}[{\fixedspacingcite[Observation 5.29]{vertex-deletion-parameterized-by-elimination-distance-arxiv}}]\label{obs:boundaried-deletion-set}
  Consider $k$-boundaried graphs $G_1, G_2$, and $H$, such that
  $G_1$ and $G_2$ are $(\F,k)$-equivalent and compatible with $H$.
  Let $S \subseteq V(H) \setminus \boundary(H)$.
  Then $(H \setminus S) \oplus G_1$ is $\F$-minor-free if and only if $(H \setminus S) \oplus G_2$ is $\F$-minor-free.
\end{observation}

A family of graphs $\R^\F_k$ is called \emph{$(\F, k)$-representative} if it contains a minimal representative from each $(\F, k)$-equivalence class where the underlying graphs are $\F$-minor-free. A family of graphs $\R^{\F}_{\leq k}$ is called \emph{$(\F, \leq k)$-representative} if it is a union of $(\F,t)$-representative families for all $t \in \{1,\dots,k\}$.

Representative families can be computed efficiently.

\begin{lemma}[{\fixedspacingcite[Lemma 5.38]{vertex-deletion-parameterized-by-elimination-distance-arxiv}, cf. \cite[Theorem 31]{minor-representative-families}}]
  \label{fact:compute-representative-family}
Let $\F$ be a family of graphs.
There exists an algorithm that, given an integer $k$, runs in time $2^{O_\F(k\log k)}$
and returns an $(\F, \leq k)$-representative family.
\end{lemma}

The family of subsets of $A$ which are computed by the algorithm by Jansen, de Kroon, and W{\l}odarczyk are called \emph{$A$-exhaustive}.

\begin{nameddefinition}{$A$-exhaustiveness for \Fdeletion}[{\cite[Definition 5.25]{vertex-deletion-parameterized-by-elimination-distance-arxiv}}]\label{def:A-exhaustive-F-deletion}
  Let~$G$ be a graph and let~$A \subseteq V(G)$. We say that a family~$\mathcal{S} \subseteq 2^A$ of subsets of~$A$ is \emph{$A$-exhaustive} for \Fdeletion on~$G$ if for every minimum-size \Fdeletion set~$S \subseteq V(G)$, there exists~$S_A \in \mathcal{S}$ such that for~$S' \coloneqq (S \setminus A) \cup S_A$ we have~$\abs{S'} \leq \abs{S}$ and~$G \setminus S'$ is $\F$-minor-free.
\end{nameddefinition}

We modify slightly this definition for the \FdeletionHittingQ problem as follows.

\begin{figure}[h]
  \centering
  \includegraphics[width=0.3\textwidth]{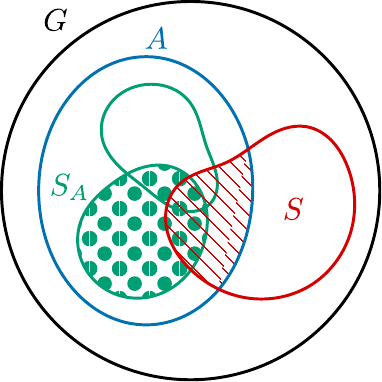}
  \caption{Illustration of \autoref{def:A-exhaustive-F-deletion-hitting-Q}. The set $S$ in red is an optimal solution to \FdeletionHittingQ on $G$. The green sets inside $A$ are the elements of an $A$-exhaustive family $\mathcal{S}$. Replacing the striped part of $S$ with the polka-dotted set $S_A$ produces another optimal solution $S'$ to \FdeletionHittingQ on $G$.}
  \label{fig:A-exhaustive}
\end{figure}

\begin{nameddefinition}{$A$-exhaustiveness for \FdeletionHittingQ} \label{def:A-exhaustive-F-deletion-hitting-Q}
  Let~$G$ be a graph and let~$A \subseteq V(G)$. We say that a family~$\mathcal{S} \subseteq 2^A$ of subsets of~$A$ is \emph{$A$-exhaustive} for \FdeletionHittingQ on~$G$ if for every minimum-size \Fdeletion set~$S \subseteq V(G)$ for which~$G \setminus S$ is $\Q$-minor-free, there exists~$S_A \in \mathcal{S}$ such that for~$S' \coloneqq (S \setminus A) \cup S_A$ we have~$\abs{S'} \leq \abs{S}$ and~$G \setminus S'$ is $(\F \cup \Q)$-minor-free.
\end{nameddefinition}
See \autoref{fig:A-exhaustive} for an illustration of this definition.

Our algorithm, as theirs, will compute small exhaustive families recursively for each subtree of the tree $\H$-decomposition of $G$. We follow their scheme and divide the proof in two lemmas: \autoref{fact:calculate-A-exhaustive-in-leaf} to compute exhaustive families in the leaves of the decomposition, and \autoref{fact:prune-A-exhaustive} to reduce the size of exhaustive families when combining the results in the subtrees.

\autoref{fact:calculate-A-exhaustive-in-leaf} depends on the efficient computation of a variant of \Fdeletion.

\defproblemaparam{Disjoint $\F$-Minor Deletion}
{A graph $G$, integer $s$, and a subset $U \subseteq V(G)$.}{$s.$}
{If a minimum-size \Fdeletion set $S \subseteq V(G) \setminus U$ of $G$ has size at most $s$, return it. Otherwise, conclude that no such set exists.}

\begin{lemma}[Morelle, Sau, Stamoulis, and Thilikos \fixedspacingcite{k-apices-LIPIcs,parameterized-algorithms-vertex-deletion-minor-closed}]
\label{fact:solve-disjoint-F-deletion}
Let $\F$ be a finite family of graphs. Then \textsc{Disjoint \Fdeletion} admits an algorithm with running time $2^{s^{\bigO(1)}} \cdot n^2$, where $s$ is the size of the solution.
\end{lemma}

Additionally, \autoref{fact:calculate-A-exhaustive-in-leaf} uses the fact that any optimal \Fdeletion solution has a small number of vertices in $A \subseteq V(G)$ when $G[A]$ is $\F$-minor-free.

\begin{namedlemma}{Bound on the Size of a Solution in an $\F$-Minor-Free Subgraph}[{\cite[Lemma 5.1]{vertex-deletion-parameterized-by-elimination-distance-arxiv}}]
\label{fact:Y-has-at-most-neighbors-vertices-in-S}
  Let $G$ be a graph and $S \subseteq V(G)$. Let $\F$ be a family of connected graphs. If $G[S]$ is $\F$-minor-free, then every minimum $\F$-minor-free deletion set $Y$ in graph $G$ has at most $\abs{N(S)}$ vertices in $S$.
\end{namedlemma}

\newrefcommand{\vol}{\mathtt{vol}}
We define $\vol(\R^{\F}_{\le k}) = \sum_{{R} \in \R^{\F}_{\leq k}} \abs{V({R})}$.

\begin{lemma}[Adaptation of {\cite[Lemma 5.33]{vertex-deletion-parameterized-by-elimination-distance-arxiv}}]\label{fact:calculate-A-exhaustive-in-leaf}
    Let $\F$ and $\Q$ be sets of connected graphs. There is an algorithm that, given a tri-separation~$(A,X,B)$ of order~$k$ in a graph~$G$ such that $G[A]$ is $\F$-minor-free, and an $(\F \cup \Q,\leq k)$-representative family $\R^{\F\cup\Q}_{\leq k}$, runs in time~$2^{k^{\bigO(1)}} \cdot \vol(\R^{\F\cup\Q}_{\le k})^{\bigO(1)} \cdot n^{\bigO(1)}$ and outputs a family~$\mathcal{S}$ of size at most~$2^k \cdot \abs{\R^{\F\cup\Q}_{\leq k}}$ that is $A$-exhaustive for \FdeletionHittingQ on~$G$.
\end{lemma}
\begin{proof}[Sketch of proof]
The proof is the same as the one for Lemma 5.33 in \cite{vertex-deletion-parameterized-by-elimination-distance-arxiv}, by observing that
\begin{itemize}
  \item the size of an $(\F \cup \Q, k)$-representative family is finite by \autoref{fact:compute-representative-family},
  \item the class of graphs defined by the set of forbidden minors $\F \cup \Q$ is hereditary and union closed,
    \item \textsc{Disjoint ($\F \cup \Q$)-Minor Deletion} admits an algorithm with running time $2^{k^{\bigO(1)}} \cdot n^{\bigO(1)}$ by \autoref{fact:solve-disjoint-F-deletion}, and
  \item even though we are dealing with an additional set of graphs $\Q$ when comparing with the proof of Lemma 5.33 in \cite{vertex-deletion-parameterized-by-elimination-distance-arxiv}, by \namedtheoremref{fact:Y-has-at-most-neighbors-vertices-in-S}, any solution $S$ to \FdeletionHittingQ on $G$ still has at most $k$ vertices in $A$, so the same argument applies.
\end{itemize}

Notice that although our definition of $A$-exhaustive families is different than in the original proof, the sets added to $\mathcal{S}$ belong to $\R^{\F\cup\Q}_{\leq k}$, and thus hit all $(\F \cup \Q)$-minors when replaced in an optimal \Fdeletion solution in $G$ that hits all $Q$-minors.
\end{proof}
\begin{proof}[Adapted copy-pasted proof]
  Initialize $\mathcal{S} = \emptyset$.
  For each subset $X' \subseteq X$, fix an~arbitrary bijection $\lambda \colon \{1,\dots,\abs{X'}\} \to X'$ and consider the graph $G \setminus (X \setminus X')$.
  It admits a tri-separation $(A, X', B)$.
  For each representative $R \in \R^{\F \cup\Q}_{t}$, where $t = \abs{X'}$, which is compatible with $G[B \cup X']$, we perform the gluing operation $G_R \coloneqq G[A \cup X'] \oplus R$
  and execute the algorithm for \textsc{Disjoint ($\F \cup \Q$)-Minor Deletion} on $G_R$ from \autoref{fact:solve-disjoint-F-deletion} with the set of undeletable vertices $U = V(G_R) \setminus A$ and parameter $k$.
  In other words, we seek
  a minimum-size \textsc{$(\F \cup \Q)$-Minor Deletion} set $A' \subseteq A$ of size at most $k$.
  If such a set is found, we add it to $\mathcal{S}$.

  The constructed family clearly has size at most $2^k \cdot |\R^{\F \cup\Q}_{\leq k}|$.
  The running time can thus be upper-bounded by
    $$2^k \cdot \sum_{{R} \in \R^{\F \cup\Q}_{\leq k}} f(k,r_{\F\cup\Q}(k)) (n+\abs{V({R})})^{\bigO(1)} = 2^k \cdot f(k,r_{\F \cup \Q}(k)) \cdot \texttt{vol}(\R^{\F \cup\Q}_{\leq k})^{\bigO(1)} \cdot n^{\bigO(1)}.$$
  It remains to show that $\mathcal{S}$ is indeed $A$-exhaustive.

  Consider a minimum-size solution~$S$ to \FdeletionHittingQ on~$G$. Define the sets~$S_A, S_X, S_B$ as~$S \cap A, S \cap X, S \cap B$, respectively, and let~$X' \coloneqq X \setminus S_X$, $\abs{X'} \eqqcolon t$.
  Fix an~arbitrary bijection $\lambda \colon \{1,\dots,t\} \to X'$.
  Since $G[A]$ is $\F$-minor-free, by \namedtheoremref{fact:Y-has-at-most-neighbors-vertices-in-S} we know that $\abs{S_A} \leq k$. We can indeed apply that lemma because $S$ is a minimum-size $\F$-minor deletion set of $G$.
  The graph $G[B \cup X'] \setminus S_B$ is an induced subgraph of $G\setminus S$ so it is $(\F \cup \Q)$-minor-free.
  The set $\R^{\F \cup\Q}_{t}$ contains a $t$-boundaried graph $R$ that is $(\F\cup\Q,t)$-equivalent to $G[B \cup X'] \setminus S_B$.
  By \autoref{obs:boundaried-deletion-set},
  $S_A$ is an $(\F \cup\Q)$-minor deletion set for $G[A \cup X'] \oplus R$.
  As $G[A \cup X'] \oplus R$ contains an $(\F \cup \Q)$-minor deletion set within $A$ of size at most $k$, some set $S'_A$ with this property has been added to $\mathcal{S}$.
  Furthermore, $S'_A$ is a minimum-size solution, so $\abs{S'_A} \leq \abs{S_A}$.
  Again by \autoref{obs:boundaried-deletion-set}, $S'_A$ is an $(\F\cup\Q)$-minor deletion set for $G[A \cup X'] \oplus G[B \cup X'] \setminus S_B$.
  It means that $S' = (S \setminus A) \cup S'_A = S_B \cup S_X \cup S'_A$ is an $(\F\cup\Q)$-minor deletion set in $G$ and $\abs{S'} \leq \abs{S}$, which finishes the proof.
\end{proof}

The second key part of the algorithm in \autoref{fact:prune-A-exhaustive} tries each set in the $A$-exhaustive family $\mathcal{S}'$ and checks if it hits all $(\F \cup \Q)$-minors in a particular graph. This check can be done efficiently thanks to the following theorem.

\begin{theorem}[\fixedspacingcite{minor-free-recognition-n3,minor-free-recognition-n2,minor-containment-almost-linear-time}]\label{fact:minor-free-recognition}
    There is an algorithm that, given a graph $G$, runs in time $n^{\bigO(1)}$ and decides whether $G$ is $\F$-minor-free for a fixed set of forbidden minors $\F$.
\end{theorem}

\begin{lemma}[Adaptation of {\cite[Lemma 5.34]{vertex-deletion-parameterized-by-elimination-distance-arxiv}}]\label{fact:prune-A-exhaustive}
    There is an algorithm that, given a tri-separation~$(A,X,B)$ of order~$k$ in a graph~$G$, a family~$\mathcal{S}' \subseteq 2^A$ that is $A$-exhaustive for \FdeletionHittingQ on~$G$, and an $(\F \cup \Q,\leq k)$-representative family $\R^{\F\cup\Q}_{\leq k}$, runs in time~$2^k \cdot \abs{\mathcal{S}'} \cdot \vol(\R^{\F\cup\Q}_{\leq k})^{\bigO(1)} \cdot n^{\bigO(1)}$
  and outputs a family~$\mathcal{S} \subseteq \mathcal{S}'$ of size at most~$2^k \cdot \abs{\R^{\F\cup\Q}_{\leq k}}$ that is $A$-exhaustive for \FdeletionHittingQ on~$G$.
\end{lemma}
\begin{proof}[Sketch of proof]
  The proof is the same as the one for Lemma 5.34 in \cite{vertex-deletion-parameterized-by-elimination-distance-arxiv}, by observing that
  \begin{itemize}
    \item the size of an $(\F \cup \Q, k)$-representative family is finite by \autoref{fact:compute-representative-family},
    \item the class of graphs defined by the set of forbidden minors $\F \cup \Q$ can be recognized in polynomial time by \autoref{fact:minor-free-recognition}.\qedhere
  \end{itemize}
\end{proof}
\begin{proof}[Adapted copy-pasted proof]
  Initialize $\mathcal{S} = \emptyset$.
  For each subset $X' \subseteq X$, fix an~arbitrary bijection $\lambda \colon [|X'|] \to X'$ and consider the graph $G \setminus (X \setminus X')$.
  It admits a tri-separation $(A, X', B)$.
  For each graph $R \in \R^{\F \cup \Q}_{t}$, where $t = |X'|$, which is compatible with $(G[B \cup X'], X', \lambda)$,
  we perform the gluing operation
  $G_R = (G[A \cup X'], X', \lambda) \oplus R$.
  %$B$-replacement of $(A, X', B)$ with $(R, X')$; we obtain a graph $(G[A \cup X'], X') \oplus (R, X')$. % with a tri-separation $(A, X',B^H)$.
  Using the polynomial-time recognition algorithm for $(\F\cup\Q)$-minor-free graphs due to \autoref{fact:minor-free-recognition} we choose a minimum-size set $S_A \in \mathcal{S'}$ which is an $(\F\cup\Q)$-minor deletion set for $G_R$, if one exists, and add it to $\mathcal{S}$.

    We construct at most $2^k \cdot |\R^{\F \cup \Q}_{\leq k}|$ graphs~$G_R$. For each graph~$G_R$ we add at most one set to $\mathcal{S}$ and spend~$|\mathcal{S'}| \cdot (n + |R|)^{\bigO(1)}$ time. In total, we perform at most $2^k \cdot |\R^{\F \cup \Q}_{\leq k}| \cdot |\mathcal{S'}| \cdot \sum_{{R} \in \R^{\F \cup \Q}_{\leq k}} (n+|V({R})|)^{\bigO(1)} = 2^k \cdot |\mathcal{S'}| \cdot \texttt{vol}(\R^{\F \cup \Q}_{\leq k})^{\bigO(1)} \cdot n^{\bigO(1)}$
  operations.
  It remains to show that $\mathcal{S}$ is indeed $A$-exhaustive.

  % Consider a minimum-size solution~$S$ to \textsc{$\hh$-deletion} on~$G$. Define~$S_A, S_X, S_B$ as~$S \cap A, S \cap X, S \cap B$, respectively, and let~$X' := X \setminus S_X$, $|X'| = t$.
  % \mic{Fix an~arbitrary bijection $\lambda \colon [|X'|] \to X'$.}
  % The set $\R^{\F \cup \Q}_{t}$ contains a $t$-boundaried graph $R$ that is $(\hh,t)$-equivalent to $(G[B \cup X'] - S_B, X',\lambda)$.
  % By Observation~\ref{obs:boundaried-deletion-set}, a set $A' \subseteq A$ is an $\hh$-deletion set for $(G[A \cup X'], X',\lambda) \oplus (G[B \cup X'] - S_B, X',\lambda)$ if and only if $A'$ is
  % an $\hh$-deletion set for $(G[A \cup X'], X',\lambda) \oplus R$.
  % Since $\mathcal{S'}$ is $A$-exhaustive, it contains such a set and, by the construction above, there \bmp{exists}  $S'_A \in \mathcal{S}$ with this property and minimum size;
  % hence $|S'_A| \leq |S_A|$ and
  % $S' = (S \setminus A) \cup S'_A = S_B \cup S_X \cup S'_A$ is an $\hh$-deletion set in $G$ and $|S'| \leq |S|$. The claim follows.
  Consider a minimum-size solution~$S$ to \FdeletionHittingQ on~$G$. Define the sets~$S_A, S_X, S_B$ as~$S \cap A, S \cap X, S \cap B$, respectively, and let~$X' := X \setminus S_X$, $|X'| = t$. Since $\mathcal{S'}$ is $A$-exhaustive on~$G$, there exists~$\widehat{S}_A \in \mathcal{S}'$ such that~$\widehat{S} \coloneqq (S \setminus A) \cup \widehat{S}_A$ is also a minimum-size solution on~$G$, implying that~$|\widehat{S}_A| \leq |S_A|$.
  Fix an~arbitrary bijection $\lambda \colon [t] \to X'$.
  The set $\R^{\F \cup \Q}_{t}$ contains a $t$-boundaried graph $R$ that is $(\F\cup\Q,t)$-equivalent to $(G[B \cup X'] \setminus S_B, X',\lambda)$.
  By Observation~\ref{obs:boundaried-deletion-set}, a set $A' \subseteq A$ is an $\F\cup\Q$-minor deletion set for $G' \coloneqq (G[A \cup X'], X',\lambda) \oplus (G[B \cup X'] \setminus S_B, X',\lambda)$ if and only if $A'$ is
  an $(\F\cup\Q)$-minor deletion set for $(G[A \cup X'], X',\lambda) \oplus R$.
  Since~$G' \setminus \widehat{S}_A = G \setminus \widehat{S}$ is $\F\cup\Q$-minor-free, we know that $\widehat{S}_A \in \mathcal{S'}$ is such a set. Hence by the construction above, there exists some (possibly different) $S'_A \in \mathcal{S}$ with this property and minimum size;
  hence $|S'_A| \leq |\widehat{S}_A| \leq |S_A|$ and
  $S' = (S \setminus A) \cup S'_A = S_B \cup S_X \cup S'_A$ is an $(\F\cup\Q)$-minor deletion set in $G$ and $|S'| \leq |S|$. The claim follows.
\end{proof}

We can now combine \autoref{fact:calculate-A-exhaustive-in-leaf} and \autoref{fact:prune-A-exhaustive} into the \FPT algorithm for \FdeletionHittingQ given a tree $\H$-decomposition for the class $\H$ of $\F$-minor-free graphs. A key part of this algorithm will consist in the computation of an exhaustive family for a subtree of the tree $\H$-decomposition of $G$ given exhaustive families for its children. We will use the following lemma to do so. Its proof can be obtained from the one of Lemma 5.27 in \cite{vertex-deletion-parameterized-by-elimination-distance-arxiv} simply by replacing \textsc{$\H$-deletion} with \FdeletionHittingQ.

\renewcommand{\S}{\mathcal{S}}
\begin{lemma}[Adaptation of {\cite[Lemma 5.27]{vertex-deletion-parameterized-by-elimination-distance-arxiv}}]
  \label{fact:combine-representative-families}
  Let $\H$ be a graph class and let $G$ be a graph. Let $A_1, A_2 \subseteq V(G)$ be disjoint sets and let $\S_1,\S_2$ be $A_1$-exhaustive (respectively, $A_2$-exhaustive) for \FdeletionHittingQ on $G$. Then for any set $A' \supseteq A_1 \cup A_2$, the family $\S' \subseteq \subsets{A'}$ defined as follows has size at most $\abs{\S_1} \cdot \abs{\S_2} \cdot 2^{\abs{A' \setminus (A_1 \cup A_2)}}$ and is $A'$-exhaustive for \FdeletionHittingQ on $G$:
  \[\S' \coloneqq \{S_1 \cup S_2 \cup S^\star \mid S_1 \in \S_1 \land S_2 \in \S_2 \land S^\star \subseteq A' \setminus (A_1 \cup A_2)\}.\]
\end{lemma}
\begin{proof}[Adapted copy-pasted proof]
  The bound on~$|\mathcal{S}'|$ is clear from the definition. Consider an arbitrary optimal solution~$S \subseteq V(G)$ to~\FdeletionHittingQ on~$G$; we will show that there exists~$\widehat{S} \in \mathcal{S}'$ such that~$(S \setminus A') \cup \widehat{S}$ is an optimal solution. We use a two-step argument.

Since~$\mathcal{S}_1$ is $A_1$-exhaustive, there exists~$S_1 \in \mathcal{S}_1$ such that~$S' \coloneqq (S \setminus A_1) \cup S_1$ is again an optimal solution.

Applying a similar step to~$S'$, as~$\mathcal{S}_2$ is $A_2$-exhaustive there exists~$S_2 \in \mathcal{S}_2$ such that~$S'' \coloneqq (S' \setminus A_2) \cup S_2$ is an optimal solution.

Since~$A_1$ and~$A_2$ are disjoint, we have~$S'' \cap A_1 = S_1$ and~$S'' \cap A_2 = S_2$. Let~$S^\star := S'' \cap (A' \setminus (A_1 \cup A_2))$. It follows that the set~$\widehat{S} = S_1 \cup S_2 \cup S^\star$ belongs to~$\mathcal{S}'$. Now note that~$S \setminus A' = S'' \setminus A'$ as we have only replaced parts of the solution within~$A_1$ and~$A_2$, while~$A' \supseteq A_1 \cup A_2$. Hence~$(S \setminus A') \cup \widehat{S} = S''$ is an optimal solution, which concludes the proof.
\end{proof}

\begin{theorem}[Adaptation of {\cite[Theorem 5.35]{vertex-deletion-parameterized-by-elimination-distance-arxiv}}]
  \label{fact:F-deletion-hitting-Q-given-tree-decomposition}
  Let $\F$ and $\Q$ be finite sets of connected graphs and $\H$ be the class of $\F$-minor-free graphs.
    Then \FdeletionHittingQ can be solved in time~$2^{k^{\bigO(1)}} \cdot n^{\bigO(1)}$
    when given a~tree $\H$-decomposition of width~$k-1$ consisting of~$n^{\bigO(1)}$ nodes.
\end{theorem}
\begin{proof}[Sketch of proof]
Modify the proof of Theorem 5.35 in \cite{vertex-deletion-parameterized-by-elimination-distance-arxiv} as follows:
\begin{enumerate}
    \item Construct an $(\F \cup \Q, \leq k)$-representative family $\R^{\F\cup\Q}_{\leq k}$ using \autoref{fact:compute-representative-family} in time $2^{\bigO(k \log k)}$. This family is of course finite, which is one of the requirements of Theorem 5.35 in \cite{vertex-deletion-parameterized-by-elimination-distance-arxiv}.
  \item Run the algorithm described in the proof, replacing usage of their Lemma 5.33 with \autoref{fact:calculate-A-exhaustive-in-leaf}, usage of their Lemma 5.34 with \autoref{fact:prune-A-exhaustive}, and usage of their Lemma 5.27 with \autoref{fact:combine-representative-families}.
  \item The result of the algorithm is a family $\mathcal{S}$ of subsets of $V(G)$ that is $V(G)$-exhaustive for \FdeletionHittingQ on $G$. If $\mathcal{S}$ is empty, the answer is no, otherwise it is yes.\qedhere
\end{enumerate}
\end{proof}
\begin{proof}[Adapted copy-pasted proof]
First, we construct an $(\F \cup \Q, \leq k)$-representative family $\R^{\F\cup\Q}_{\leq k}$ using \autoref{fact:compute-representative-family} in time $2^{\bigO(k \log k)}$.
Since the output size of the algorithm cannot exceed its running time, we have $|\R^{\F\cup\Q}_{\leq k}| \le \vol(\R^{\F\cup\Q}_{\leq k}) \le 2^{\bigO(k \log k)}$.

The algorithm is based on a variant of dynamic programming in which bounded-size sets of partial solutions are computed, with the guarantee that at least one of the partial solutions which are stored can be completed to an optimal solution. More formally, for each node~$t \in V(T)$ we are going to compute (refer to \cite[Definition 5.7]{vertex-deletion-parameterized-by-elimination-distance-arxiv} for the definitions of $\kappa$ and $\pi$) a set of partial solutions~$\mathcal{S}_t \subseteq 2^{\kappa(t)}$ of size at most~$2^k \cdot |\R^{\F\cup\Q}_{\leq k}|$ which is~$\kappa(t)$-exhaustive for \FdeletionHittingQ in~$G$. As~$\kappa(r) = V(G)$ for the root node~$r$ by Observation 5.8 in \cite{vertex-deletion-parameterized-by-elimination-distance-arxiv}, any minimum-size set~$S \in \mathcal{S}_r$ for which~$G \setminus S$ is $(\F\cup\Q)$-minor-free is an optimal solution to the problem, and the property of $\kappa(r)$-exhaustive families guarantees that one exists.

We do the computation bottom-up in the tree decomposition, using \autoref{fact:prune-A-exhaustive} to prune sets of partial solutions at intermediate steps to prevent them from becoming too large.

Let~$(T,\chi,L)$ be the given tree $\H$-decomposition of width~$k-1$. By Lemma 5.5 in \cite{vertex-deletion-parameterized-by-elimination-distance-arxiv} we may assume that the decomposition is nice and is rooted at some node~$r$.
For~$t \in V(T)$, define~$L_t := L \cap \chi(t)$. Process the nodes of~$T$ from bottom to top. We process a node~$t$ after having computed exhaustive families for all its children, as follows. Let~$X_t := \chi(t) \cap \pi(t)$, let~$A_t := \kappa(t)$ and let~$B_t := V(G) \setminus (A_t \cup X_t)$. By Observation 5.9 in \cite{vertex-deletion-parameterized-by-elimination-distance-arxiv}, the partition~$(A_t,X_t,B_t)$ is a tri-separation of~$G$. The way in which we continue processing~$t$ depends on the number of children it has. As~$T$ is a nice decomposition, node~$t$ has at most two children.

\textbf{Leaf nodes}. For a leaf node~$t \in V(T)$, we construct an exhaustive family of partial solutions~$\mathcal{S}_t \subseteq 2^{\kappa(t)}$ as follows.
By \defref{def:elimination-distance}, vertices of~$L_t$ do not occur {in other bags than~$\chi(t)$}.
Because the decomposition is nice, we have $\chi(t) \setminus L_t = \pi(t)$.
Therefore $\kappa(t) = L_t$ and we have ~$(A_t,X_t,B_t) = (L_t, \chi(t) \setminus L_t, V(G) \setminus \chi(t))$.
Furthermore, $|X_t| \leq k$ since the width of the decomposition is~$k-1$.
As $G[L_t] \in \H$, we can process the tri-separation $(A_t,X_t,B_t)$ with \autoref{fact:calculate-A-exhaustive-in-leaf}
within running time $2^{k^{\bigO(1)}} \cdot \vol(\R^{\F\cup\Q}_{\le k})^{\bigO(1)} \cdot n^{\bigO(1)}$.
We obtain a $\kappa(t)$-exhaustive family of size at most $2^k \cdot |\R^{\F\cup\Q}_{\le k}|$.

\textbf{Nodes with a unique child}. Let~$t$ be a node that has a unique child~$c$, for which a $\kappa(c)$-exhaustive family~$\mathcal{S}_c$ of size~$2^k \cdot |\R^{\F\cup\Q}_{\le k}|$ has already been computed. Recall that vertices of~$L$ only occur in leaf bags, so that~$L_t = \emptyset$ and therefore~$|\chi(t)| \leq k$. Observe that~$\kappa(t) \setminus \kappa(c) \subseteq \chi(t)$, so that~$|\kappa(t) \setminus \kappa(c)| \leq k$. ({A tighter bound is possible by exploiting the niceness property, which we avoid for ease of presentation.}) Compute the following set of partial solutions:
 \begin{equation*}
     \mathcal{S}'_t := \{ S_c \cup S^* \mid S_c \in \mathcal{S}_c, S^* \subseteq \kappa(t) \setminus \kappa(c) \}.
 \end{equation*}
Since the number of choices for~$S_c$ is~$2^k \cdot |\R^{\F\cup\Q}_{\le k}|$, while the number of choices for~$S^*$ is~$2^{k}$, the set~$\mathcal{S}'_t$ has size at most~$2^{2k} \cdot |\R^{\F\cup\Q}_{\le k}|$ and can be computed in time~$2^{2k} \cdot |\R^{\F\cup\Q}_{\le k}| \cdot n^{\bigO(1)}$. Since~$\kappa(c) \subseteq \kappa(t)$ due to Observation 5.8 in \cite{vertex-deletion-parameterized-by-elimination-distance-arxiv} we can invoke Observation 5.26 in \cite{vertex-deletion-parameterized-by-elimination-distance-arxiv} to deduce that the family~$\mathcal{S}'_t$ is $\kappa(t)$-exhaustive for \FdeletionHittingQ on~$G$. As the last step for the computation of this node, we compute the desired exhaustive family~$\mathcal{S}_t$ as the result of applying \autoref{fact:prune-A-exhaustive} to~$\mathcal{S}'_t$ and the tri-separation~$(A_t,X_t,B_t)$ of~$G$, which is done in time~$2^{\bigO(k \log k)} \cdot n^{\bigO(1)}$ because $|\R^{\F\cup\Q}_{\le k}| \le 2^{\bigO(k \log k)}$.

As~$A_t = \kappa(t)$, the lemma guarantees that~$\mathcal{S}_t$ is $\kappa(t)$-exhaustive and it is sufficiently small.

\textbf{Nodes with two children}.
The last type of nodes to handle are those with exactly two children. So let~$t \in V(T)$ have two children~$c_1, c_2$. Since~$t$ is not a leaf we have~$L_t = \emptyset$. Let~$K := \kappa(t) \setminus (\kappa(c_1) \cup \kappa(c_2))$ and observe that~$K \subseteq \chi(t) \setminus L$. Therefore~$|K| \leq k$.

Using the $\kappa(c_1)$-exhaustive set~$\mathcal{S}_{c_1}$ and the~$\kappa(c_2)$-exhaustive set~$\mathcal{S}_{c_2}$ computed earlier in the bottom-up process, we define a set~$\mathcal{S}'_t$ as follows:
\begin{equation*}
    \mathcal{S}'_t := \{ S_1 \cup S_2 \cup S^* \mid S_1 \in \mathcal{S}_{c_1}, S_2 \in \mathcal{S}_{c_2}, S^* \subseteq K \}.
\end{equation*}
As~$\mathcal{S}_{c_1}$ and~$\mathcal{S}_{c_2}$ both have size~$2^k \cdot |\R^{\F\cup\Q}_{\le k}|$, while~$|K| \leq 2^{k}$, we have~$|\mathcal{S}'_t| = 2^{3k} \cdot |\R^{\F\cup\Q}_{\le k}|^2$.
By Observation 5.8 in \cite{vertex-deletion-parameterized-by-elimination-distance-arxiv} we have that $\kappa(c_1) \cap \kappa(c_2) = \emptyset$ and $\kappa(c_1) \cup \kappa(c_2) \subseteq \kappa(t)$, so we can apply \autoref{fact:combine-representative-families} to obtain that the family~$\mathcal{S}'_t$ is $\kappa(t)$-exhaustive for \FdeletionHittingQ on~$G$. The desired exhaustive family~$\mathcal{S}_t$ is obtained by applying \autoref{fact:prune-A-exhaustive} to~$\mathcal{S}'_t$ and the tri-separation~$(A_t,X_t,B_t)$ of~$G$,
which is done in time $2^{\bigO(k \log k)} \cdot n^{\bigO(1)}$

\textbf{Wrapping up}. Using the steps described above we can compute, for each node of~$t \in V(T)$ in a bottom-up fashion, a $\kappa(t)$-exhaustive family~$\mathcal{S}_t$ of size~$2^k \cdot |\R^{\F\cup\Q}_{\le k}|$. Since the number of nodes of~$t$ is~$n^{\bigO(1)}$
the overall running time follows. As discussed in the beginning of the proof, an optimal solution can be found by taking any minimum-size solution from the family~$\mathcal{S}_r$ for the root~$r$.
\end{proof}

\begin{corollary}[Adaptation of {\cite[Corollary 5.42]{vertex-deletion-parameterized-by-elimination-distance-arxiv}}]\label{fact:F-deletion-hitting-Q-parameterized-by-twF}
  Let $\F$ and $\Q$ be finite sets of connected graphs and $\H$ be the class of $\F$-minor-free graphs.
    Then \FdeletionHittingQ on a graph $G$ can be solved in time $2^{k^{\bigO_{\F,\Q}(1)}} \cdot n^{\bigO(1)}$ where $k = \twH(G)$.
\end{corollary}
\begin{proof}
By \autoref{fact:compute-tree-decomposition} we can find a tree $\H$-decomposition of width $\bigO(\twH(G)^5)$ in time $2^{k^{\bigO(1)}} \cdot n^{\bigO(1)}$. We then apply \autoref{fact:F-deletion-hitting-Q-given-tree-decomposition} with parameter $k' = \bigO(\twH(G)^5)$.
\end{proof}

\subsection{Tying everything together}
\label{sec:tying-everything-together}

We finish this section by leveraging \namedtheoremref{fact:labeled-to-unlabeled} and \autoref{fact:F-deletion-hitting-Q-parameterized-by-twF} to prove \autoref{item:compute-opt-that-breaks-Q} of \autoref{fact:compute-solution-breaking-Q}.
\begin{proof}[Proof of \autoref{item:compute-opt-that-breaks-Q} of \autoref{fact:compute-solution-breaking-Q}]
  We first use the algorithm from \namedtheoremref{fact:labeled-to-unlabeled} to transform the labeled graph $C$ into an unlabeled graph $C'$, the set of labeled graphs $\Q$ into a set of unlabeled graphs $\Q'$, and the set of unlabeled graphs $\F$ into a set of unlabeled graphs $\F'$. This step runs in time
  \[O\parens*{f(\totalLabels, \norm{\F}, \eta) \cdot \parens*{\abs{V(C)} + \abs{E(C)} + \sum_{H\in(\F\cup \Q)}(\abs{V(H)} + \abs{E(H)})}}\]
  for some function $f$. The output graph $C'$ has size bounded by the same formula.

    By \namedtheoremref{fact:labeled-to-unlabeled}, graph $C$ admits a solution to the \textsc{$\F$-Minor Deletion Hitting Labeled $\Q$} problem if and only if graph $C'$ admits a solution to the \textsc{$\F'$-Minor Deletion Hitting $\Q'$} problem. We thus apply the algorithm of \autoref{fact:F-deletion-hitting-Q-parameterized-by-twF} to solve \textsc{$\F'$-Minor Deletion Hitting $\Q'$} on the graph $C'$. As $\eta \geq \EDF(C) = \ED{\F'}(C') \geq \tw_{\F'}(C')$ by \namedtheoremref{fact:labeled-to-unlabeled}, this step runs in time $2^{\eta^{\bigO_{\F',\Q'}(1)}} \cdot \abs{V(C')}^{\bigO(1)}$. Notice that by the same lemma the sizes of $\F'$ and $\Q'$ are bounded by a function of $\F, \totalLabels, \eta$, $\abs\Q$, and $\norm\Q$. Therefore, this step takes time \[g\parens*{\F, \totalLabels, \sum_{H \in \Q} \abs{V(H)}, \eta} \cdot \abs{V(C)}^{\bigO(1)}\] for some function $g$, which combined with the previous step gives us the desired time complexity.
\end{proof}

\section{Ingredient 2: Bounding the size of a minimal set \texorpdfstring{$\Q^\star \subseteq \Q$}{Q⋆ ⊆ Q}}
\label{sec:bounding-Q}

This section is devoted to proving \namedtheoremref{fact:main-lemma}, which generalizes Lemma 3 of \cite{minor-hitting} by making the size of the set $\Q^\star$ depend on $\EDF(C)$ instead of the treedepth of $C$. We restate the lemma for convenience.

\newtheorem*{main-lemma}{Lemma \ref{fact:main-lemma}}
\begin{main-lemma}[Main Lemma -- Generalized version of {\cite[Lemma 3]{minor-hitting}}]
    \mainlemma
\end{main-lemma}

%\begin{lemma:main:statement}[Main lemma]
    %\mainlemma
%\end{lemma:main:statement}

Jansen and Pieterse prove their Lemma~3 by stating and proving an inductive version of it~\cite[Lemma 27]{minor-hitting}. We replace a base case in their Lemma 27 so that it also holds for $\EDF$ instead of $\td$. This new base case is described in \namedtheoremref{fact:extra-base-case}. As expected, it deals with the main difference present between treedepth decompositions and $\F$-elimination forests: the bags of the leaves of an $\F$-elimination forest can have many vertices instead of just one. This makes the proof of this base case much more complicated than the one in \cite{minor-hitting}.

\subsection{Inductive version of the main lemma}
To state the inductive version of the lemma, we first need some definitions. A thorough intuitive explanation of these definitions can be found in \cite{minor-hitting}.

\begin{nameddefinition}{\forget}[{\cite[Definition 10]{minor-hitting}}]
    \definedhere{\forget}
    Let $G$ be a $t$-boundaried $X$-labeled graph and let $k \leq t$. Define
    $\forget(G, k)$ as the $k$-boundaried $X$-labeled graph $G'$ obtained from~$G$ by setting $\boundaryIndex_{G'}(v) = \boundaryIndex_G(v)$ for all $v \in V(G)$ for which $\boundaryIndex_G(v) \leq k$, and forgetting the boundary status of the higher-indexed boundary vertices. Define $\forget(G) \coloneqq \forget(G,0)$.

    For a set of graphs $\mathcal{S}$, define $\forget(\mathcal{S},k) \coloneqq \{\forget(G,k) \mid G \in \mathcal{S}\}$ and $\forget(\mathcal{S}) \coloneqq \forget(\mathcal{S},0)$.
\end{nameddefinition}

Observe that the~$\forget$ operation is only used to forget the boundary status of a vertex; it is not used to omit labels from a labelset.

\begin{nameddefinition}{\opt}[{\cite[Definition 11]{minor-hitting}}]
    \definedhere{\opt}
    Let $G$ be a $t$-boundaried graph with boundary set~$S$, let~$\F$ be a family of graphs, and let $\Pi$ (the \emph{prohibitions}) be a set of prohibited $t$-boundaried graphs with boundary~$S$. Define
    \begin{align*}
    \opt(G,\Pi,S) \coloneqq \min \{|Y| \mid \ &Y \subseteq V(G) \wedge Y \cap S = \emptyset, \text{ and}\\
    & \text{$G\setminus Y$ is $\F$-minor-free, and} \\
    & G\setminus Y \text{ has no graph in $\Pi$ as boundaried minor} \}.
    \end{align*}
    Define $\opt(G)\coloneqq \opt(G,\emptyset, \emptyset)$, or simply the size of an optimal $\F$-minor-free deletion set in $G$.
\end{nameddefinition}

\begin{nameddefinition}{$\folio$}[{\cite[Definition 12]{minor-hitting}}]
    \definedhere{\folio}
    For a $t$-boundaried $X$-labeled graph $G$, the $\folio$ of $G$ consists of all $t$-boundaried labeled minors of $G$. In other words, \[\folio(G) \coloneqq \set{G' \mid G' \minorleq G}.\]
    The folio of an unlabeled graph, or an unboundaried graph, is defined analogously.
\end{nameddefinition}

\begin{nameddefinition}{$\ext{t}$}[{\cite[Definition 13]{minor-hitting}}] \label{def:extend}
    \definedhere{\ext}
    Let $H$ be an $X$-labeled $t$-boundaried graph for some~$t \geq 0$ and some set $X$. Let $\ext{1}(H)$ (short for \emph{extend}) be the set of all $(t+1)$-boundaried graphs $H'$ that can be obtained from $H$ by using exactly one of the following steps:
    \begin{itemize}
    \item \textbf{(Do nothing)} Let $H'$ be equal to $H$, thereby forming~$H'$ as a $(t+1)$-boundaried graph in which there is no $(t+1)$'th boundary vertex.
    \item \textbf{(Increase the boundary)} Take a vertex $v \in V(H)\setminus \boundary(H)$, add it to $\boundary(H')$, and set $\boundaryIndex_{H'}(v) \coloneqq t+1$.
    \item \textbf{(Split)} Split a vertex $u \in \boundary(H)$ as follows. Let $V(H') \coloneqq V(H) \cup \{v\}$. Add $v$ to $\boundary(H')$, and let $\boundaryIndex_{H'}(v) \coloneqq t+1$. Add edge $\{u,v\}$ to $H'$.
        For any edge $\{u',u\} \in E(H)$ either keep it in $H'$ or replace it by edge $\{u',v\}$. For each label~$\ell$ on the labelset of~$u$, either keep it on~$u$ or move it to the labelset of~$v$.
    \end{itemize}
    For an integer $t' \geq 1$, define $\ext{t'}(H)$ as the set of~$(t+t')$-boundaried graphs that can be obtained from~$H$ by applying exactly~$t'$ of such operations in a row. Note that for every $H' \in \ext{t'}(H)$, we have that $\abs{V(H')} \leq \abs{V(H)} + t'$ and $\abs{E(H')} \leq \abs{E(H)} + t'$. The extend operation for unlabeled graphs is defined analogously, with the exception that there are no labels to be divided in the \textbf{(Split)} step. For a set of graphs~$\Q$, define~$\ext{1}(\Q) \coloneqq \bigcup _{Q \in \Q} \ext{1}(Q)$, and~$\ext{t}(\Q)$ analogously.
\end{nameddefinition}

For an illustration of \defref{def:extend}, see \cite[Figure 6]{minor-hitting}.

\begin{nameddefinition}{\pieces}[{\cite[Definition 15]{minor-hitting}}] \label{def:pcs}
    \definedhere{\pcs}
    Let $G$ be an $X$-labeled $t$-boundaried graph. Let~$\pcs(G)$ (for \emph{pieces}) contain the following $X$-labeled $t$-boundaried graphs.
    \begin{itemize}
    \item For all  vertices $v \in \boundary(G)$, the set $\pieces(G)$ contains a graph $P$ consisting of a single vertex $u$ with $L_P(u) \coloneqq \emptyset$ and $\boundaryIndex_P(u)\coloneqq \boundaryIndex_G(v)$.
    \item For all $v \in \boundary(G)$, for all $x \in L_G(v)$, the set $\pieces(G)$ contains a graph $P$ consisting of a single vertex $u$ with $L_P(u) \coloneqq \{x\}$ and $\boundaryIndex_P(u)\coloneqq \boundaryIndex_G(v)$.
    \item For every edge $\{u,v\} \in E(G)$ with $u,v\in\boundary(G)$, the set $\pieces(G)$ contains a graph $P$ with vertices $x$ and $y$ and edge $\{x,y\}$. Define $\boundaryIndex_P(x) \coloneqq \boundaryIndex_P(u)$, $\boundaryIndex_P(y)\coloneqq \boundaryIndex_P(v)$, and $L_P(u) \coloneqq L_P(v) = \emptyset$.
    \item For every connected component $C$ of $G \setminus \boundary(G)$, define $C'$ as the vertex set $C$ together with all vertices in $\boundary(G)$ that are adjacent to $C$. Let $\pieces(G)$ contain a graph $P$ that is equal to $G[C']$ after removing all edges between boundary vertices. Set $\boundary(P)$ to be the neighbors of $C$, and remove all labels from the vertices in $\boundary(P)$. Keep all other labels unchanged.
    \end{itemize}
    For unlabeled graphs,~$\pieces(G)$ is defined analogously by treating $G$ as a $\emptyset$-labeled graph.
\end{nameddefinition}

Note that, as pointed out in \cite{minor-hitting}, $\pcs(G)$ can contain the same (labeled boundaried) graph several times if many connected components of $G \setminus \boundary(G)$ are isomorphic and connected the same way with $\boundary(G)$, and that $\bigoplus_{p \in \pcs(G)} p = G$ for every graph $G$.

\begin{nameddefinition}{$\multipieces$}[{\cite[Definition 16]{minor-hitting}}] \label{def:mpcs}
    \definedhere{\mpcs}
    Let $G$ be an $X$-labeled $t$-boundaried graph. For this definition, let two graphs be equal if they are isomorphic, as defined in \defref{def:isomorphism}. Define the \emph{multipieces of $G$}, abbreviated as $\mpcs$ as
    \[\multipieces(G) \coloneqq \braces*{\bigoplus_{p \in P} p  \mid  P \subseteq \pieces(G) \wedge P \neq \emptyset}.\]
    For a set of $X$-labeled $t$-boundaried graphs $\Q$ define
    \[\multipieces(\Q) \coloneqq \bigcup_{Q \in \Q} \multipieces(Q).\]
    Let $\mpcsplus{t}(\Q) \coloneqq \multipieces(\ext{t}(\Q))$.
\end{nameddefinition}

\begin{figure}[h]
    \centering
    \begin{minipage}{0.5\textwidth}
    \includegraphics[width=\textwidth]{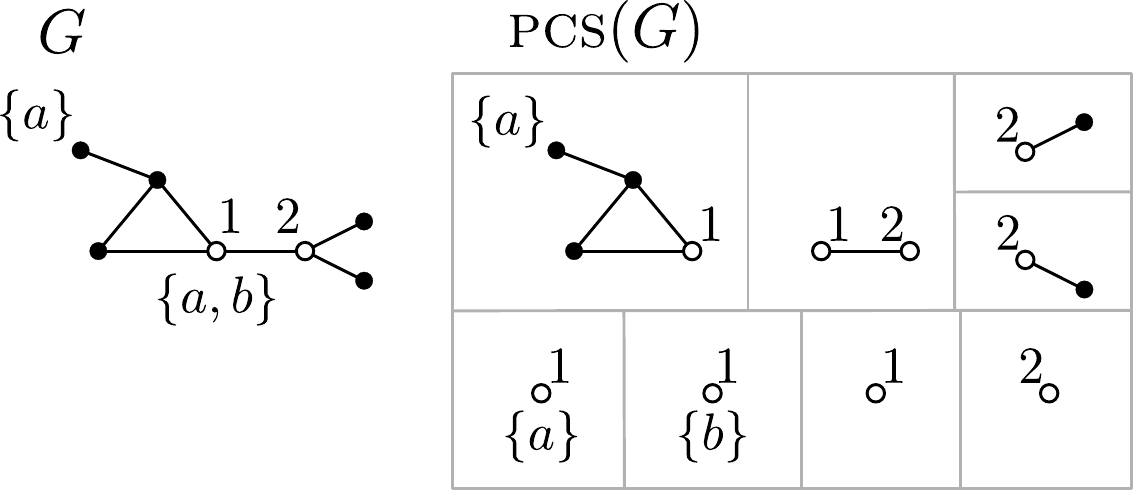}
    \end{minipage}\hspace{.4cm}
    \begin{minipage}{0.17\textwidth}
    \includegraphics[width=\textwidth]{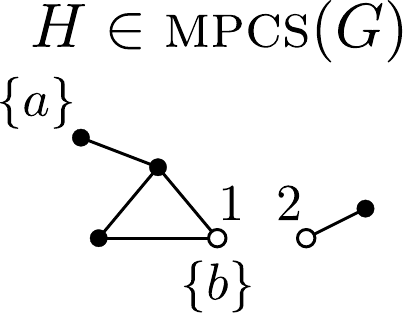}
    \end{minipage}
    \caption{Example of \autoref{def:pcs} and \autoref{def:mpcs} on 2-boundaried $\{a,b\}$-labeled graph $G$. Based off of \cite[Figure 8]{minor-hitting}.}
    \label{fig:pieces}
\end{figure}

Observe that $\pcs(G)$ is a multiset, while $\multipieces(G)$ is a simple set. For an illustration of $\pcs(G)$ and $\multipieces(G)$, see \autoref{fig:pieces}.

The graphs in $\mpcs(G)$ can be also formed by taking any collection of connected components of $G \setminus \boundary(G)$ along with the union $W$ of the neighborhoods in $\boundary(G)$ of the vertices in those connected components, and then adding the edges of an arbitrary subgraph of $W$. The labels of the vertices in connected components are kept, while the labels of a boundary vertex $v$ can be an arbitrary subset of $\lab_G(v)$.

\begin{lemma}[{\fixedspacingcite[Lemma 17]{minor-hitting}}]
    \label{fact:multipieces-of-Q-either-boundaried-or-in-ext}
    Let $t \in \nat$, $\Q$ be a set of $X$-labeled connected graphs, and $H \in \mpcsplus{t}(\Q)$. Then either $H$ contains a boundary vertex or $H \in \ext{t}(\Q)$.
    %\erictodo{Discuss if we want to say $H \in \Q$ or if we want to maintain the statement as is.}\ig{as we already discussed by email, I think it is safer to leave it like that (as otherwise, we would have to add the proof)}\eric{ok!}
\end{lemma}

The following definition is just necessary to state \namedtheoremref{fact:extra-base-case}, but it is not used in the rest of the paper; not even in the proof of \namedtheoremref{fact:extra-base-case}.
\begin{definition}[$\odot$ {\cite[Definition 18]{minor-hitting}}] \label{def:merge}
    Let $\Pi_1$ and $\Pi_2$ be two sets of $t$-boundaried graphs, and let $\F$ be a set of graphs. Define:
    \begin{align*}
    \Pi_1 \odot_\F \Pi_2 := \{G & \in \mpcsplus{t}(\F) \mid\\
    & \forall G_1, G_2 \text{ $t$-boundaried graphs} \colon G_1 \oplus G_2 = G \Rightarrow \\
    & \Pi_1 \text{ contains a minor of $G_1$ or } \Pi_2 \text{ contains a minor of $G_2$}
    \}.
    \end{align*}
    We omit the subscript from~$\odot_\F$ when it is clear from the context.
\end{definition}

\begin{nameddefinition}{$\folioqt$}[{\cite[Definition 19]{minor-hitting}}]
    \definedhere{\folioqt}
    For a set $\Q$ of $X$-labeled connected graphs, an integer~$t$, and an $X$-labeled $t$-boundaried graph $G$, define \[\folioqt(G) \coloneqq \folio(G) \cap \mpcsplus{t}(\Q).\] For a set of graphs $\mathcal{S}$, define $\folioqt(\mathcal{S}) \coloneqq \bigcup_{G \in \mathcal{S}}\folioqt(G)$.
\end{nameddefinition}

\begin{nameddefinition}{optimal solutions such that}[{\cite[Definition 20]{minor-hitting}}]\label{def:minFdelsolwith}
    \definedhere{\optsolst}
    For $X$-labeled $t$-boundaried graphs $G_A,G_B,G_C$ with boundary set~$S\coloneqq \boundary(G_A\oplus G_B \oplus G_C)$, a given family~$\F$ of connected graphs, sets~$\Pi_A, \Pi_B,\Pi_C$ of $t$-boundaried graphs (we will call these \emph{prohibitions}), set $\Q$ of connected $X$-labeled graphs, and set~$R_B$ of $X$-labeled $t$-boundaried graphs, define \optsolst{} (for \emph{opt. solution such that}) as:
    \begin{align*}
    \optsolst(G_A, &G_B, G_C, \Pi_A, \Pi_B, \Pi_C, R_B) \coloneqq\\ \{& Y \in \optsol(G_A \oplus G_B \oplus G_C) \mid \\
        &Y \cap S = \emptyset \text{, and } \\
    &G_i \setminus Y \text{ has no boundaried } \Pi_i\text{-minor for any $i \in \{A,B,C\}$, and}\\
    &R_B = \folioqt(G_B \setminus Y) \}.\qedhere
    \end{align*}
\end{nameddefinition}

Notice we added the $\Q$ subscript to \optsolst{} with respect to Definition 20 in \cite{minor-hitting} even though the definitions are the same, as this set of solutions depends also on $\Q$.

\newrefcommand{\numberof}{\mathsf{numberOf}}
\begin{definition}[{\fixedspacingcite[Definition 21]{minor-hitting}}]
    \definedhere{\numberof}
    Define $\numberof(\ell, t, n, \theta)$ as the number of distinct (meaning not isomorphic; see \defref{def:isomorphism}) $t$-boundaried $\{1,\dots,\ell\}$-labeled graphs on at most~$n$ vertices with at most $\theta$ labels in each vertex.
\end{definition}

Again, we decided to change the notation with respect to \cite{minor-hitting} although the definition is the same.

\begin{observation}[{\fixedspacingcite[Observation 26]{minor-hitting}}]
For $\ell, t, n, \theta \geq 0$, we have $\numberof(\ell, t, n, \theta) \leq n \cdot 2^t \cdot 2^{n^2}\cdot (\ell^\theta + 1)^n$.
\end{observation}

\newrefcommand{\Y}{\mathcal{Y}}
\begin{nameddefinition}{remainder}[{\cite[taken from statement of Lemma 27]{minor-hitting}}]
    Let:
    \begin{itemize}
        \item $X$ be a finite set;
        \item $t \in \nat$;
        \item $\Q$ be a set of $X$-labeled graphs;
        \item $G$ be an $X$-labeled $t$-boundaried graph; and
        \item $\Y$ be a set of subsets of $V(G)$.
    \end{itemize}
    The \emph{remainders of $G$ with respect to $\Y$} are defined as the set $\R$ of inclusion-wise minimal elements of the set $\{\folioqt(G \setminus Y) \mid Y \in \Y\}$. For a remainder $R = \folioqt(G \setminus Y)$ with $Y \in \Y$, we say that $Y$ is a \emph{corresponding solution} for $R$.

    The \emph{remainders of $G$ with respect to $\Y$ that leave a $\Q$-minor} is the subset $\R_\Q$ of $\R$ defined as \[
    \R_\Q \coloneqq \set*{R \in \R \mid \exists q \in \Q, \exists r\in R : q \minorleq \forget(r)}.
    \]
    Lastly, the \emph{remainders of $G$ with respect to $\Y$ that do not leave a $\Q$-minor} is the set $\R_N \coloneqq \R \setminus \R_\Q$.
\end{nameddefinition}

We would like to point out that we defined the remainders as the inclusion-wise minimal elements of the set $\{\folioqt(G \setminus Y) \mid Y \in \Y\}$ because this definition will be applied to the case where $Y \in \defaultoptsolst$. If a solution $Y$ hits more of these fragments in $G$ than another solution $Y'$, then using $Y$ instead of $Y'$ will be preferable for our purposes.

We are now ready to state the inductive version of \namedtheoremref{fact:main-lemma}. This will simply be the result of replacing $\td(G)$ with $\EDF(G)$ in the statement of Lemma 3 in \cite{minor-hitting}. We use $\iscon(G)$ to denote the binary function that indicates if $G$ is connected.

\newrefcommand{\nF}{n_\F}
\begin{namedlemma}{Inductive Version of the Main Lemma}[{-- Adaptation of \cite[Lemma 3]{minor-hitting}}]\label{fact:inductive-main-lemma-ed}
Let:
    \begin{itemize}
        \item $X$ be a finite set;
        \item $t \in \nat$;
        \item $\F$ be a set of connected graphs;
        \item $\Q$ be a set of connected $X$-labeled graphs such that each graph in $\Q$ has at most $\max_{H \in \F}\abs{E(H)}+1$ vertices and $\Q$ is $n_\F$-saturated, with $\nF \coloneqq \min_{H \in \F} \abs{V(H)}$;
        \item $\Pi_A,\Pi_B,\Pi_C\subseteq\mpcsplus{t}(\F)$ such that $\Pi_A \odot \Pi_B \odot \Pi_C \supseteq \ext{t}(\F)$;
        \item $G_A$, $G_B$ and $G_C$ be three $X$-labeled $t$-boundaried graphs;
        \item $G \coloneqq G_A \oplus G_B \oplus G_C$;
        \item $S \coloneqq \boundary(G)$ such that $\EDF(G) \geq \EDF(G_A \setminus S) + \abs{S}$;
        \item $R_B \subseteq \mpcsplus{t}(\Q)$ be a set of isomorphism classes of $X$-labeled $t$-boundaried graphs;
        \item $\Y \coloneqq \defaultoptsolst$;
        \item $\R_\Q$ be the set of remainders of $G_A\oplus G_B$ with respect to $\Y$ that leave a $\Q$-minor;
        \item $\R_N$ be the set of remainders of $G_A\oplus G_B$ with respect to $\Y$ that do not leave a $\Q$-minor;
        \item $\nu(\Pi_A) \coloneqq \abs{\multipieces_{+t}(\F) \setminus \Pi_A}$;
        \item $\xi(R_B) \coloneqq \numberof\parens*{t \cdot \min _{H \in \F} \abs{V(H)}, t, t + \max _{H \in \Q}\abs{V(H)}, \min _{H \in \F} \abs{V(H)}} - \abs{R_B}$; and
        \item $\mu(G_A,\Pi_A, S) \coloneqq \opt(G_A,\Pi_A,S)  - \sum_{C \in \cc(G_A\setminus S)} \opt(C).$
    \end{itemize}

    Then there exist functions $f$ and $g$ such that
    \begin{enumerate}
        \item $\abs{\R_N} \leq f(\EDF(G_A \setminus S), \iscon(G_A \setminus S), \mu(G_A,\Pi_A,S), \nu(\Pi_A),\xi(R_B), \norm{\F}, \abs{S})$, and
        \item there exists $\Q^\star \subseteq \Q$ such that $\abs{\Q^\star} \leq g(\EDF(G_A \setminus S), \iscon(G_A \setminus S), \mu(G_A,\Pi_A,S), \nu(\Pi_A),\xi(R_B), \norm{\F}, \abs{S})$, and for each $R \in \R_\Q$ there exist $q \in \Q^\star$ and $r \in R$ with $q \minorleq \forget(r)$.
    \end{enumerate}
\end{namedlemma}

This statement can be quite daunting, but the main takeaway is that the induction is primarily being made on $\EDF(G_A \setminus S)$. There are some hypotheses that we will be ignoring completely in \namedtheoremref{fact:extra-base-case}; namely, the functions $\nu$, $\xi$, and $\mu$, and the $\Pi$ sets, which we will only use to define the set $\Y$. A thorough intuitive explanation of the rationale behind the statement, including the semantics of the graphs $G_A$, $G_B$, and $G_C$, can be found in \cite{minor-hitting}.

The proof of Lemma 3 in \cite{minor-hitting} proceeds inductively by moving a vertex that reduces the treedepth of $G_A$ to either $G_B$ or $G_C$. The inductive step is split in two parts: one where $G_A \setminus S$ is connected, and another one where it is not. It can be checked that the latter part does not depend on the treedepth of $G_A \setminus S$ other than to apply the inductive hypothesis, while the former actually only uses the fact that there exists a vertex $v\in G_A$ such that $\td(G_A \setminus (S \cup \{v\})) < \td(G_A \setminus S)$, and applies the inductive hypothesis by either removing $v$ from the graph, or adding $v$ to $S$. This means that the inductive step can be applied to any hereditary parameter such that for every connected graph that is not covered by a base case, there exists a vertex that reduces the parameter by at least one when removed. This is true for $\td$, but also for $\EDF$ if we add a base case that deals with the situation where $G_A \setminus S$ is connected and $\EDF(G_A \setminus S) = 0$. This is the content of \namedtheoremref{fact:extra-base-case}, which generalizes the base case BC1 where $G_A \setminus S = \emptyset$ in the proof of Lemma 3 in \cite{minor-hitting}.

\subsection{The \texorpdfstring{$\F$}{F}-minor-free base case}

We now present the new base case for the proof of \namedtheoremref{fact:inductive-main-lemma-ed}. To prove it, we will assume that the following base case is already covered. The proof of this base case can be found in the proof of Lemma 27 of \cite{minor-hitting}.
\begin{itemize}[align=left, labelwidth=1em]
    \item[\textsf{\textbf{\textcolor{lipicsGray}{BC3}}}]\fixedHypertarget{BC3} $R_B \cap \ext{t}(\Q) \neq \emptyset$.
\end{itemize}

The other base cases present in the proof of Lemma 27 in \cite{minor-hitting} are not necessary for our purposes:
\begin{itemize}[align=left, labelwidth=1em]
\item[\textsf{\textbf{\textcolor{lipicsGray}{BC1}}}] will be replaced by our new base case;
\item[\textsf{\textbf{\textcolor{lipicsGray}{BC2}}}] only applies when $G_A \setminus S$ is disconnected; and
\item[\textsf{\textbf{\textcolor{lipicsGray}{BC4}}}] puts restrictions on the size of $R_B$ that we will not be using.
\end{itemize}

This new base case will not be using all the hypotheses of \namedtheoremref{fact:inductive-main-lemma-ed}. In particular, we will be avoiding the sets $\Pi_A$, $\Pi_B$, and $\Pi_C$. Thus, we define a simplified version of \optsolst{} by slightly abusing the notation.

\begin{nameddefinition}{simplified optimal solutions such that}\label{def:simple-opt-solutions-such-that}
    For $X$-labeled $t$-boundaried graphs $G_A,G_B,G_C$ with boundary set~$S\coloneqq \boundary(G_A\oplus G_B \oplus G_C)$, a given family~$\F$ of connected graphs, set $\Q$ of connected $X$-labeled graphs, and set~$R_B$ of $X$-labeled $t$-boundaried graphs, define $\optsolst(G_A, G_B, G_C, R_B)$ as
    \[\set{Y \in \optsol(G_A \oplus G_B \oplus G_C) \mid
        Y \cap S = \emptyset \text{ and } R_B = \folioqt(G_B \setminus Y)}.
    \]
\end{nameddefinition}

Notice that $\defaultoptsolst \subseteq \optsolst(G_A, G_B, G_C,\allowbreak R_B).$ We will thus require that $\Y \subseteq \optsolst(G_A, G_B, G_C, R_B)$.

\renewrefcommand{\Y}{\mathcal{Y}}
\newrefcommand{\Gboundary}{S}
\newrefcommand{\RN}{\R_N}
\newrefcommand{\RQ}{\R_\Q}
\renewcommand{\defaultoptsolst}{\optsolst(G_A, G_B, G_C, R_B)}
\begin{lemma}[$\F$-Minor-Free Base Case]\label{fact:extra-base-case}
    Let:
    \begin{itemize}
        \item $X$ be a finite set;
        \item $t \in \nat$;
        \item $\F$ be a set of connected graphs;
        \item $\Q$ be a set of connected $X$-labeled graphs such that each graph in $\Q$ has at most $\max_{H \in \F}\abs{E(H)}+1$ vertices and $\Q$ is $n_\F$-saturated, with $\nF \coloneqq \min_{H \in \F} \abs{V(H)}$;
        \item $G_A$, $G_B$ and $G_C$ be three $X$-labeled $t$-boundaried graphs;
        \item $G \coloneqq G_A \oplus G_B \oplus G_C$;
        \item $\Gboundary \coloneqq \boundary(G)$ such that $G_A \setminus \Gboundary$ is connected and $\F$-minor-free;
        \item $R_B \subseteq \mpcsplus{t}(\Q)$ be a set of isomorphism classes of $X$-labeled $t$-boundaried graphs;
        \item $\Y \subseteq \defaultoptsolst$;
        \item $\RQ$ be the set of remainders of $G_A\oplus G_B$ with respect to $\Y$ that leave a $\Q$-minor; and
        \item $\RN$ be the set of remainders of $G_A\oplus G_B$ with respect to $\Y$ that do not leave a $\Q$-minor.
    \end{itemize}

    Then there exist functions $f$ and $g$ such that
    \begin{enumerate}
        \item\label{item:extra-base-case:RN} $\abs{\RN} \leq f(\norm{\F}, \abs{\Gboundary})$, and
        \item\label{item:extra-base-case:Q} there exists $\Q^\star \subseteq  \Q$ such that $\abs{\Q^\star} \leq g(\norm{\F}, \abs{\Gboundary})$, and for each $R \in \RQ$ there exist $q \in \Q^\star$ and $r \in R$ with $q \minorleq \forget(r)$.
    \end{enumerate}
\end{lemma}

\subsubsection{Outline of the proof}
We separate the proof in two parts; one for each item in the statement of \namedtheoremref{fact:extra-base-case}.

\newrefcommand{\GAplusRB}{G_{AB'}}
\newrefcommand{\GAplusRBS}{\noref\GAplusRB^\Gboundary}
The proof will heavily rely on upper-bounding the number of different labels that appear in a specific subgraph $H$ of $G_A \oplus G_B$. In the case of \autoref{item:extra-base-case:RN}, this subgraph $H$ -- which will be called $\GAplusRBS$ -- will contain the vertices that potentially belong to some solutions (but not all) in $\Y$ that hit all $\Q$-minors in $G_A \oplus G_B$. Intuitively, if a vertex belongs to all solutions in $\Y$, then it does not affect the number of different remainders in $\R_N$. We will call these vertices \emph{mandatory}, which is a new notion with respect to the proof in \cite{minor-hitting}. The same thing happens if a vertex does not belong to any solution. Thus, the remainder in $G_A \oplus G_B \setminus H$ can be considered as being ``fixed'' for all solutions in $\Y$ that hit all $\Q$-minors. Therefore, the number of different remainders in $\R_N$ will be bounded by the number of possible combinations of different graphs in $\folioqt(H)$, which will in turn be bounded by the number of different labels that appear in $H$, as by hypothesis the maximum number of vertices in a graph in $\Q$ is upper-bounded by $\max_{F\in\F}\abs{E(F)} + 1$.

\newrefcommand{\smallerGAB}{G'_{AB}}
In the case of \autoref{item:extra-base-case:Q}, the subgraph $H$ -- which will be called $\smallerGAB$ -- will contain a small $\Q$-minor-free modulator consisting solely of vertices in $G_A$. The modulator will be big enough to ensure that no solution in $\Y$ that leaves a $\Q$-minor in $G_A \oplus G_B$ takes all the vertices in the modulator. Observe that by \namedtheoremref{fact:Y-has-at-most-neighbors-vertices-in-S}, any solution has at most $\abs{\Gboundary}$ vertices in $G_A$, and thus we just need this modulator to have size at least $\abs{\Gboundary} + 1$. We will then mark labels for each vertex in the modulator, ensuring that if a solution does not take a vertex in the modulator, then it leaves a $\Q$-minor in $H$ that only uses labels in the marked set.
We then build $\Q^\star$ by taking the graphs in $\Q$ that only have labels in the marked set. Again, as the maximum number of vertices in a graph in $\Q$ is upper-bounded by $\max_{F\in\F}\abs{E(F)} + 1$, the size of $\Q^\star$ will be bounded by a function of $\abs{\Gboundary}$ and $\norm\F$.

\subsubsection{Preliminary helpful results}
We now state some results that will be useful in the proof of \namedtheoremref{fact:extra-base-case}. The first is a lemma that will be key in bounding the number of labels in these graphs $H$.

We say a vertex $v$ \emph{reaches} a label $\ell$ in a graph $H$ if there is a path in $H$ from $v$ to a vertex that contains $\ell$ in its labelset. Moreover, $\ell$ is \emph{separated} from $v$ if $v$ does not reach $\ell$.

\begin{figure}[ht]
    \centering
    \includegraphics[width=.4\textwidth]{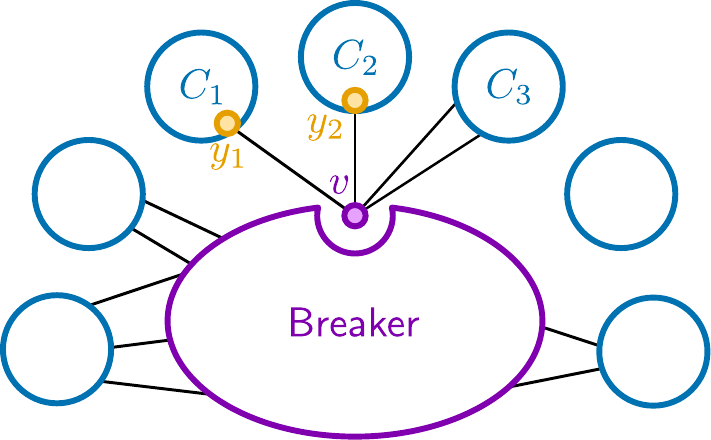}
    \caption{Example of the proof of \autoref{fact:if-v-reaches-too-many-labels-then-mandatory} for $\sizeConstant = 2$, showing the graph $G \setminus (\breaker \setminus \set{v})$. Each connected component $C_i \in \components$ for $i \in \{1,2,3\}$ has less than $\saturationConstant$ different labels, so the set $Y = \{y_1, y_2\}$ separates at most $2 \cdot (\saturationConstant - 1)$ labels from $v$.}
    \label{fig:vertex-in-breaker-reaches-labels}
\end{figure}

\newrefcommand{\breaker}{\mathsf{Breaker}}
\newrefcommand{\saturationConstant}{s}
\newrefcommand{\sizeConstant}{k}
\newrefcommand{\tooManyLabelsReached}{\mathsf{tooManyLabelsReached}}
\newrefcommand{\labelSubset}{L}
\begin{lemma}
\label{fact:if-v-reaches-too-many-labels-then-mandatory}
    Let $X$ be a finite set, $G$ be an $X$-labeled graph, and $\Q$ be a set of $\saturationConstant$-saturated $X$-labeled graphs for a constant $\saturationConstant$. Let $\breaker$ be a subset of $V(G)$ such that $G \setminus \breaker$ is $\Q$-minor-free.
    Take $\sizeConstant \in \nat$, and suppose a vertex $v \in \breaker$ reaches at least $\tooManyLabelsReached \coloneqq \sizeConstant \cdot (\saturationConstant - 1) + \saturationConstant$ different labels in $G \setminus (\breaker \setminus \{v\})$. Define an arbitrary subset $\labelSubset$ of those labels such that $\abs{\labelSubset} = \tooManyLabelsReached$.

    For any set $Y \subseteq V(G) \setminus \{v\}$ of size at most $\sizeConstant$, the graph $G \setminus Y$ contains a $\Q$-minor that only uses labels from $L$.
\end{lemma}
\newrefcommand{\components}{\mathcal{C}}
\begin{proof}
    Let $v \in \breaker$ and $\labelSubset$ be as in the statement of the lemma.
    Let $\components$ be the set of components in $G \setminus \breaker$. Every connected component $C$ in $\components$ is $\Q$-minor-free because of the definition of $\breaker$, and thus, as $\Q$ is $\saturationConstant$-saturated, there are less than $\saturationConstant$ different labels occurring in the vertices of $C$. Take a set $Y \subseteq V(G) \setminus \{v\}$ of size at most $\sizeConstant$. The set $Y$ disconnects at most $\sizeConstant$ components in $\components$ from $v$, and thus separates at most $\sizeConstant \cdot (\saturationConstant - 1)$ different labels in $\labelSubset$ from $v$. \autoref{fig:vertex-in-breaker-reaches-labels} shows an example of this. This leaves at least $\saturationConstant$ different labels in $\labelSubset$ that are reached by $v$ in $G \setminus Y$. As $\Q$ is $\saturationConstant$-saturated, there exists a single-vertex graph $H \in \Q$ that contains $\saturationConstant$ of these labels as its labelset, and hence $H$ is a $\Q$-minor of $G \setminus Y$ that only uses labels from $\labelSubset$.
\end{proof}

The second one is a property of $\folioqt$ that will be useful to reduce the size of the graphs $G_A$ and $G_B$ while maintaining the same remainders with respect to $\Y$.

\begin{lemma}[{\fixedspacingcite[Lemma 24]{minor-hitting}}]
\label{fact:folioqt-of-oplus-is-the-same-if-folioqt-of-right-is-the-same}
    Let $\Q$ be a set of $X$-labeled connected graphs for some set $X$, and let $G_A$, $G_B$, and $G'_B$ be $X$-labeled $t$-boundaried graphs such that $\folioqt(G_B) = \folioqt(G'_B)$. Then $\folioqt(G_A \oplus G_B) = \folioqt(G_A \oplus G'_B).$
\end{lemma}

\begin{corollary}[Sum Consistency of $\folioqt$]
\label{fact:folioqt-of-oplus-is-the-same-if-folioqt-of-both-is-the-same}
    Let $\Q$ be a set of $X$-labeled connected graphs for some set $X$, and let $G_A$, $G'_A$, $G_B$, $G'_B$ be $X$-labeled $t$-boundaried graphs such that $\folioqt(G_A) = \folioqt(G'_A)$ and $\folioqt(G_B) = \folioqt(G'_B)$. Then \[
        \folioqt(G_A \oplus G_B) = \folioqt(G'_A \oplus G'_B).
    \]
    Moreover, if $\folioqt(G_B) = \emptyset$, then $
        \folioqt(G_A \oplus G_B) = \folioqt(G_A).
    $
\end{corollary}
\begin{proof}
    Noting that the $\oplus$ operation is commutative, by \autoref{fact:folioqt-of-oplus-is-the-same-if-folioqt-of-right-is-the-same} and the hypothesis we have that
    \begin{align*}
        \folioqt(G_A \oplus G_B) &= \folioqt(G_A \oplus G'_B) \\
        &= \folioqt(G'_B \oplus G_A) \\
        &= \folioqt(G'_B \oplus G'_A) \\
        &= \folioqt(G'_A \oplus G'_B).
    \end{align*}
    For the second part, it is sufficient to observe that the empty graph has no graph in $\mpcsplus{t}(\Q)$ as a minor, and thus has empty $\folioqt$.
\end{proof}

\subsubsection{Proof of \namedtheoremref{fact:extra-base-case}}

We are now ready to provide the formal proof of our new base case.

\newrefcommand{\totalMarkedLabels}{\mathsf{totalMarkedLabels}}

\begin{proof}
\newrefcommand{\GAB}{G_{AB}}
\newrefcommand{\arbitrarySolution}{Y'}
\newrefcommand{\RBasGraph}{G_{B'}}
Let $\GAB \coloneqq G_A \oplus G_B$. Take an arbitrary solution $\arbitrarySolution \in \Y$, and define $\RBasGraph \coloneqq G_B \setminus \arbitrarySolution$ and $\GAplusRB \coloneqq G_A \oplus \RBasGraph$.

\begin{claim}
\label{fact:we-can-use-GAplusRB-instead-of-GAB}
    For every solution $Y \in \Y$, we have that \[\folioqt(\GAB \setminus Y) = \folioqt(\GAplusRB \setminus (Y \cap V(G_A))).\]
\end{claim}
\begin{claimproof}
    Note that $\folioqt(\RBasGraph) = R_B$ by hypothesis. For every solution $Y \in \Y$ we have that $(G_A \oplus G_B) \setminus Y = (G_A \setminus Y) \oplus (G_B \setminus Y)$. Thus, \namedtheoremref{fact:folioqt-of-oplus-is-the-same-if-folioqt-of-both-is-the-same} gives us
    \begin{align*}
        \folioqt((G_A \oplus G_B) \setminus Y) &= \folioqt((G_A \setminus Y) \oplus (G_B \setminus Y)) \\
        &= \folioqt((G_A \setminus Y) \oplus \RBasGraph) \\
        &= \folioqt(\GAplusRB \setminus (Y \cap V(G_A))).\qedhere
    \end{align*}
\end{claimproof}

\newrefcommand{\YA}{\noref\Y_A}
\autoref{fact:we-can-use-GAplusRB-instead-of-GAB} allows us to restrict ourselves to the graph $\GAplusRB$ instead of $\GAB$, and to the set $\YA \coloneqq \set{Y \cap V(G_A) \mid Y \in \Y}$ instead of $\Y$, as the remainders of $\GAB$ with respect to $\Y$ will be exactly the same as the remainders of $\GAplusRB$ with respect to $\YA$.

In what follows we prove separately the two items in the statement of \autoref{fact:extra-base-case}.

\newrefcommand{\YN}{\noref\Y_N}
\subsubsection*{Proof of \autoref{item:extra-base-case:RN}: $\abs{\RN} \leq f(\norm{\F}, \abs{\Gboundary})$}

We begin by proving \autoref{item:extra-base-case:RN}. Let $\YN$ be the subset of elements in $\YA$ that hit all $\Q$-minors in $\GAplusRB$. We call a vertex $v \in V(G)$ \emph{mandatory} if every element in $\YN$ contains $v$. Note that by definition mandatory vertices belong to $G_A$.

\newrefcommand{\mandatoryVertices}{M}
\renewrefcommand{\components}{\mathcal{C}}
\newrefcommand{\componentsWithoutS}{\noref\components_{\cancel{\Gboundary}}}
The set of all mandatory vertices is denoted by $\mandatoryVertices$. For all $Y \in \YN$, as $\mandatoryVertices \subseteq Y$, we have that $\folioqt(\GAplusRB \setminus Y) = \folioqt((\GAplusRB \setminus \mandatoryVertices) \setminus Y)$. Thus, the number of different sets that $\folioqt(\GAplusRB \setminus Y)$ can attain over all possible elements $Y \in \YN$ will only depend on the vertices taken by $Y$ in $G_A \setminus \mandatoryVertices$, since $Y \subseteq V(G_A)$.

Let $\components$ be the set of connected components of $\GAplusRB \setminus \mandatoryVertices$, and $\componentsWithoutS$ be the subset of components in $\components$ that do not contain a vertex in $\Gboundary$. We will show in the next claim that we can safely ignore the vertices in components in $\componentsWithoutS$ when computing the remainders of elements in $\YN$.

\newrefcommand{\componentsWithoutSinA}{\noref\componentsWithoutS^A}
\newrefcommand{\componentsWithoutSinB}{\noref\componentsWithoutS^{B'}}
As the components in $\componentsWithoutS$ do not contain any boundary vertices, we can partition them into the components $\componentsWithoutSinA$ contained entirely in $G_A$, and the components $\componentsWithoutSinB$ contained entirely in $\RBasGraph$. See \autoref{fig:components-in-GAB-minus-M}.

\begin{figure}[ht]
\centering
\captionsetup{justification=centering}
\includegraphics[width=.48\textwidth]{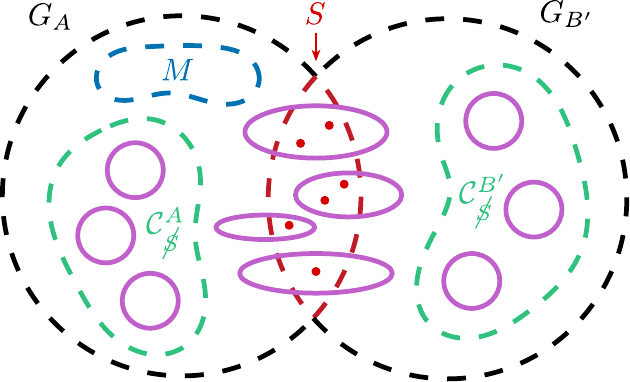}
\caption{Connected components of $\GAplusRB \setminus M$, in purple.}
\label{fig:components-in-GAB-minus-M}
\end{figure}

\begin{claim}
\label{fact:Y-does-not-contain-vertices-in-components-without-S-in-GA}
    No element $Y \in \YN$ contains a vertex in a component in $\componentsWithoutSinA$.
\end{claim}
\begin{claimproof}
    \newrefcommand{\minorModel}{\varphi}
    \newrefcommand{\Ywithoutv}{\hat{Y}}
    Suppose for a contradiction that an element $Y \in \YN$ contains a vertex $v$ in a connected component $C \in \componentsWithoutSinA$, and consider the set $\Ywithoutv \coloneqq Y \setminus \{v\}$. As $Y$ is a subset of an element in $\Y$, which contains only optimal solutions to \Fdeletion in $G$, the graph $G \setminus \Ywithoutv$ must contain an $\F$-minor model $\minorModel$ which contains $v$. The model $\minorModel$ cannot contain a mandatory vertex, as $\mandatoryVertices \subseteq \Ywithoutv$. Thus, the model $\minorModel$ is contained entirely in $G \setminus \mandatoryVertices$. Suppose for a contradiction that $\minorModel$ contains some vertex outside $C$. Then $\minorModel$ must contain a boundary vertex, as $\F$ is a set of connected graphs and $v \in V(G_A)$. But $C$ does not contain any boundary vertices, leading to a contradiction. Therefore, the $\F$-minor model $\minorModel$ is contained entirely in $C$, which contradicts the hypothesis that $G_A\setminus \Gboundary$ is $\F$-minor-free. Therefore, no element in $\YN$ contains a vertex in any connected component in $\componentsWithoutSinA$.
\end{claimproof}

\begin{claim}
    \label{fact:components-without-S-are-Q-minor-free}
    Every component $C$ in $\componentsWithoutS$ is $\Q$-minor-free.
\end{claim}
\begin{claimproof}
    If $\RBasGraph$ contains a $\Q$-minor, then $\folioqt(\RBasGraph) = R_B$ contains a graph in $\Q$, and therefore also in $\ext{t}(\Q)$. By \hyperlink{BC3}{BC3}, we can assume that $R_B \cap \ext{t}(\Q) = \emptyset$, and thus no component in $\componentsWithoutSinB$ contains a $\Q$-minor.

    On the other hand, by \autoref{fact:Y-does-not-contain-vertices-in-components-without-S-in-GA}, no element in $\YN$ contains a vertex in a component in $\componentsWithoutSinA$. In turn, this means that every connected component in $\componentsWithoutSinA$ is $\Q$-minor-free, as every element in $\YN$ hits all $\Q$-minors in $G_A$.
\end{claimproof}

\definedhere{\GAplusRBS}
Let $\GAplusRBS$ be the union of the components in $\components \setminus \componentsWithoutS$.
The graph $\GAplusRBS$ corresponds to the graph induced by the purple connected components in \autoref{fig:components-in-GAB-minus-M} that intersect with the set $\Gboundary$. From now on, we will be working with the graph $\GAplusRBS$ instead of $\GAplusRB$. This decision is supported by the following claim.

\newrefcommand{\componentsWithoutSAsGraph}{H_{\cancel{\Gboundary}}}
\begin{claim}
\label{fact:folioqt-of-GAplusRBS-is-the-same}
    For every element $Y \in \YN$, we have that $\folioqt(\GAplusRB \setminus Y) = \folioqt(\GAplusRBS \setminus Y).$
\end{claim}
\begin{claimproof}
    Let $\componentsWithoutSAsGraph$ be the graph induced by the connected components in $\componentsWithoutS$. By \autoref{fact:components-without-S-are-Q-minor-free}, every component $C$ in $\componentsWithoutS$ is $\Q$-minor-free. As $\Q$ consists of connected graphs, the graph $\componentsWithoutSAsGraph$ is also $\Q$-minor-free.

    By \autoref{fact:multipieces-of-Q-either-boundaried-or-in-ext}, every graph $Q \in \mpcsplus{t}(\Q)$ either has a boundary vertex, or belongs to $\ext{t}(\Q)$. By \defref{def:extend}, if $Q$ has no boundary vertices it must be equal to a graph in $\Q$. As $\componentsWithoutSAsGraph$ has no boundary vertices and is $\Q$-minor-free, we have that $\folioqt(\componentsWithoutSAsGraph) = \emptyset$.

    For every element $Y \in \YN$, as $\mandatoryVertices \subseteq Y$, we can partition the graph $\GAplusRB \setminus Y$ into two disjoint subgraphs: the graph $\GAplusRBS \setminus Y$, and the graph $\componentsWithoutSAsGraph \setminus Y$. The latter is equal to $\componentsWithoutSAsGraph$ by \autoref{fact:Y-does-not-contain-vertices-in-components-without-S-in-GA}. Recall that $S = \boundary(G_A) = \boundary(G_B)$. Thus, $\componentsWithoutSAsGraph$ has no boundary vertices, and so $\GAplusRB \setminus Y = (\GAplusRBS \setminus Y) \oplus \componentsWithoutSAsGraph$. By \namedtheoremref{fact:folioqt-of-oplus-is-the-same-if-folioqt-of-both-is-the-same} and the fact that $\folioqt(\componentsWithoutSAsGraph) = \emptyset$,  we then have that \[\folioqt(\GAplusRB \setminus Y) = \folioqt((\GAplusRBS \setminus Y) \oplus \componentsWithoutSAsGraph) = \folioqt(\GAplusRBS \setminus Y).\claimqedhere\]
\end{claimproof}

In light of \autoref{fact:folioqt-of-GAplusRBS-is-the-same}, to bound the size of $\RN$ we focus on bounding the number of different remainders of $\GAplusRBS$ with respect to $\YN$. For every element $Y \in \YN$, the set $\folioqt(\GAplusRBS \setminus Y)$ is one of the $2^{\abs*{\folioqt(\GAplusRBS)}}$ possible subsets of $\folioqt(\GAplusRBS)$. Thus, we focus on upper-bounding the size of $\folioqt(\GAplusRBS)$. For this, as we will see later, it will be enough to bound the number of different labels in vertices in $\GAplusRBS$. This will be achieved by characterizing some of the mandatory vertices in $\mandatoryVertices$.

\renewrefcommand{\tooManyLabelsReached}{\mathsf{tooManyLabelsReached}}
As $\Q$ is $\nF$-saturated, every vertex $v \in V(G_A)$ that has a labelset of size at least $\nF$ belongs to every element in $\YN$, and thus is mandatory.
On the other hand, take an element $Y \in \YN$. The element $Y$ corresponds to the subset of a solution in $\Y$ in the $\F$-minor-free subgraph $G_A\setminus \Gboundary$. Hence, by \namedtheoremref{fact:Y-has-at-most-neighbors-vertices-in-S}, the element $Y$ has size at most $\abs{\Gboundary}$. Additionally, as $Y \in \YN$, the element $Y$ hits all $\Q$-minors in $\GAplusRB$, and thus also in $\GAplusRBS$. Therefore, by \autoref{fact:if-v-reaches-too-many-labels-then-mandatory}, every vertex $v \in Y$ that reaches at least $\tooManyLabelsReached \coloneqq \abs{\Gboundary} \cdot (\nF - 1) + \nF$ different labels in $\GAplusRBS \setminus (Y \setminus \set{v})$ is mandatory.

Let $Y \in \Y_N$. As by definition $\GAplusRBS$ does not contain any mandatory vertices, we have that:
\begin{enumerate}
    \item Every vertex in $\GAplusRBS$ has less than $\nF$ labels.
    \item\label{item:v-reaches-small-number-of-labels} Every vertex $v \in Y\cap V(\GAplusRBS)$ reaches less than $\tooManyLabelsReached$ different labels in $\GAplusRBS \setminus (Y \setminus \set{v})$.
\end{enumerate}

Each of the at most $\abs{\Gboundary}$ connected components of $\GAplusRBS$ either:
\begin{itemize}
    \item is $\Q$-minor-free, and thus, because $\Q$ is $\nF$-saturated, contains less than $\nF$ different labels; or
    \item has a $\Q$-minor, and thus, as $Y$ hits all $\Q$-minors in $\GAplusRBS$, shares a vertex with $Y$.
\end{itemize}
\newrefcommand{\maxLabels}{\mathsf{maxLabels}}
We can therefore upper bound the number of different labels in $\Q$-minor-free connected components of $\GAplusRBS$ by $\abs{\Gboundary} \cdot (\nF - 1)$, and the number of different labels in components that have a $\Q$-minor by the number of different labels reached by vertices in $Y$. Recall that by \namedtheoremref{fact:Y-has-at-most-neighbors-vertices-in-S} there are at most $\abs{\Gboundary}$ vertices in $Y$, and thus by \autoref{item:v-reaches-small-number-of-labels} this number of labels can be upper bounded by $\abs{\Gboundary} \cdot (\tooManyLabelsReached - 1)$. In total, the number of different labels in vertices in $\GAplusRBS$ is at most
\begin{align*}
    \maxLabels &\coloneqq \abs{\Gboundary} \cdot(\nF - 1) + \abs{\Gboundary}\cdot(\tooManyLabelsReached - 1) \\
    &= \abs{\Gboundary} \cdot(\nF + \tooManyLabelsReached - 2).
\end{align*}

\newcommand{\maxSizeOfFolioOfComponentsWithSAsGraph}{\mathtt{maxSizeOfFolioOf}\componentsWithSAsGraph}
The graphs in $\folioqt(\GAplusRBS)$ therefore satisfy the following properties:
\begin{itemize}
    \item Their vertices have labels of a set of at most $\maxLabels$ labels.
    \item They have boundary with indices in a set of at most $\abs{\Gboundary}$ elements.
    \item They have at most $\max_{H\in \F} \abs{E(H)} + 1 + \abs{\Gboundary}$ vertices, as they are in $\mpcsplus{t}(\Q)$, and each vertex added when extending a graph in $\Q$ must have a boundary index in $\Gboundary$.
\end{itemize}
The number of different graphs that meet all of these properties is a function of $\abs{\Gboundary}$ and $\norm{\F}$, and thus $\abs*{\folioqt(\GAplusRBS)}$ is bounded by a function of $\abs{\Gboundary}$ and $\norm{\F}$.

As mentioned earlier, for every element $Y \in \YN$, the set $\folioqt(\GAplusRBS \setminus Y)$ is one of the $2^{\abs*{\folioqt(\GAplusRBS)}}$ possible subsets of $\folioqt(\GAplusRBS)$. By \autoref{fact:folioqt-of-GAplusRBS-is-the-same}, there are therefore at most $2^{\abs*{\folioqt(\GAplusRBS)}}$ remainders in $\RN$.

\subsubsection*{Proof of \autoref{item:extra-base-case:Q}: there exists $\Q^\star \subseteq  \Q$ such that $\abs{\Q^\star} \leq g(\norm{\F}, \abs{\Gboundary})$}

We now turn our attention to constructing a set $\Q^\star \subseteq \Q$ with size upper bounded by a function of $\norm{\F}$ and $\abs{\Gboundary}$ such that every remainder in $\RQ$ contains a graph that contains a $\Q^\star$-minor.

\newrefcommand{\YQ}{\mathcal{Y}_\Q}
Following \autoref{fact:we-can-use-GAplusRB-instead-of-GAB} we focus on $\GAplusRB$ instead of $G_A \oplus G_B$. Define $\YQ \coloneqq \YA \setminus \YN$, meaning, the subset of elements in $\YA$ that leave a $\Q$-minor in $\GAplusRB$. We thus want to prove that there exists a set $\Q^\star \subseteq \Q$ of size not greater than $g(\norm{\F}, \abs{\Gboundary})$ such that every element $Y \in \YQ$ leaves a $\Q^\star$-minor in $\GAplusRB$. Assume that $\YQ$ is not empty, as otherwise taking $\Q^\star \coloneqq \emptyset$ would satisfy the statement.

We proceed to mark labels ensuring that every set $Y \in \YQ$ leaves a $\Q$-minor in $\GAplusRB$ that only uses those labels. We will then build $\Q^\star$ by taking the graphs in $\Q$ that only use labels that are marked.

By \hyperlink{BC3}{BC3}, we can assume that $R_B \cap \ext{t}(\Q) = \emptyset$. As $\folioqt(\RBasGraph) = R_B$, this implies that $\RBasGraph$ has no $\Q$-minors.

\renewrefcommand{\breaker}{\mathsf{Breaker}}
\newrefcommand{\smallerGA}{G_{A'}}
\begin{claim}
\label{fact:Z-can-be-small}
    There exists an induced subgraph $\smallerGA$ of $G_A$ and a set $\breaker \subseteq V(\smallerGA)$ such that:
    \begin{enumerate}
        \item Every set $Y \in \YQ$ leaves a $\Q$-minor in $\smallerGA \oplus \RBasGraph$.
        \item $\breaker$ hits all $\Q$-minors in $\smallerGA \oplus \RBasGraph$.
        \item\label{item:Z-is-small} $\breaker$ has size at most $\abs{\Gboundary} + 1$.
    \end{enumerate}
\end{claim}
\begin{claimproof}
    See \autoref{fig:breaker-in-GA} for an illustration of $\smallerGA$ and the set $\breaker$.
    For a graph $H$ and an induced subgraph $H'$ of $H$, define $\optin{\Q}(H, H')$ to be the minimum size of a subset of $V(H')$ that hits all $\Q$-minors in $H$, and $\optsolin{\Q}(H, H')$ to be the set of such subsets that have minimum size. As $\RBasGraph$ has no $\Q$-minors, there exists a set in $\optsolin{\Q}(\GAplusRB, G_A)$.

    If $\optin{\Q}(\GAplusRB, G_A) \leq \abs{\Gboundary} + 1$, we simply take any set in $\optsolin{\Q}(\GAplusRB, G_A)$ as $\breaker$, and define $\smallerGA \coloneqq G_A$. By definition, $\breaker$ hits all $\Q$-minors in $\smallerGA \oplus \RBasGraph$. Also, by definition, every set $Y \in \YQ$ leaves a $\Q$-minor in $\smallerGA \oplus \RBasGraph$.

    Suppose then that $\optin{\Q}(\GAplusRB, G_A) > \abs{\Gboundary} + 1$. Removing any vertex $v$ from $G_A$ reduces $\optin{\Q}(\GAplusRB, G_A)$ by at most one, as otherwise there would exist a set of size lower than $\optin{\Q}(\GAplusRB, G_A)$ that contains $v$ and hits all $\Q$-minors in $\GAplusRB$. In other words, we have that \[
    \optin{\Q}(\GAplusRB, G_A) \leq \optin{\Q}(\GAplusRB \setminus \{v\}, G_A \setminus \{v\}) + 1.\]
    Thus, we can obtain a graph $\smallerGA$ that is an induced subgraph of $G_A$ such that $\optin{\Q}(\smallerGA \oplus \RBasGraph, \smallerGA) = \abs{\Gboundary} + 1$
    by repeatedly removing an arbitrary vertex from $G_A$,
    and take any set in $\optsolin{\Q}(\smallerGA \oplus \RBasGraph, \smallerGA)$ to be $\breaker$. Notice that $Y \in \YQ$ still leaves a $\Q$-minor in $\smallerGA \oplus \RBasGraph$, as otherwise it would need to take more than $\abs{\Gboundary}$ vertices in $\smallerGA$, contradicting \namedtheoremref{fact:Y-has-at-most-neighbors-vertices-in-S}.
\end{claimproof}

\begin{figure}[ht]
\centering
\includegraphics[width=.5\textwidth]{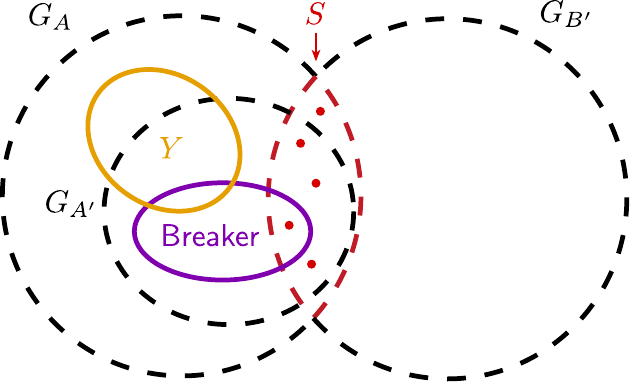}
\caption{Every element $Y \in \YQ$ must miss at least one vertex in the $\Q$-minor-free deletion set $\breaker$ of $\smallerGA \oplus \RBasGraph$.}
\label{fig:breaker-in-GA}
\end{figure}

\newrefcommand{\minorModel}{\varphi}
\definedhere{\smallerGAB} Take $\smallerGA$ and $\breaker$ to be as in \autoref{fact:Z-can-be-small}, and define $\smallerGAB \coloneqq \smallerGA \oplus \RBasGraph$. Observe that it is not necessary to show that there exists an efficient algorithm to construct the set $\breaker$; we are only interested on showing that such a vertex set exists. In the same vein, we also do not need to construct efficiently the set $\Q^\star$.

Note that every set $Y \in \YQ$ must miss a vertex $v \in \breaker$ that belongs to a $\Q$-minor model in $\smallerGAB \setminus Y$, as otherwise $Y$ would hit all $\Q$-minors in $\smallerGAB$. We will mark a limited number of labels for every vertex $v \in \breaker$, such that if a set $Y \in \YQ$ misses $v$, and $v$ belongs to a $\Q$-minor model in $\smallerGAB \setminus Y$, then $Y$ leaves a $\Q$-minor model $\minorModel$ that contains only marked labels. As $\smallerGA$ is an induced subgraph of $G_A$, the $\Q$-minor model $\minorModel$ will also be present in $\GAplusRB \setminus Y$. This usage of set $\breaker$ corresponds to the usage of set $Z$ in the proof of Lemma 4 in \cite{FVS-via-EDF-DAM}.

\newrefcommand{\labelsReachedBy}[1]{\mathsf{LabelsReachedBy}(#1)}
\newrefcommand{\tooManylabelsReached}{\mathsf{tooManyLabelsReached}}
For every $v \in \breaker$, consider the set $\labelsReachedBy{v}$ of labels reached by $v$ in $\smallerGAB \setminus (\breaker \setminus \set{v})$, which includes the labels in $v$. If $\labelsReachedBy{v}$ has less than $\tooManylabelsReached\coloneqq \abs{\Gboundary} \cdot (\nF - 1) + \nF$ labels, we mark all of them. Otherwise, we mark $\tooManylabelsReached$ labels in $\labelsReachedBy{v}$. See \autoref{fig:labels-reached-by-vertex-in-breaker} for an example.

\begin{figure}[ht]
\centering
\includegraphics[width=.45\textwidth]{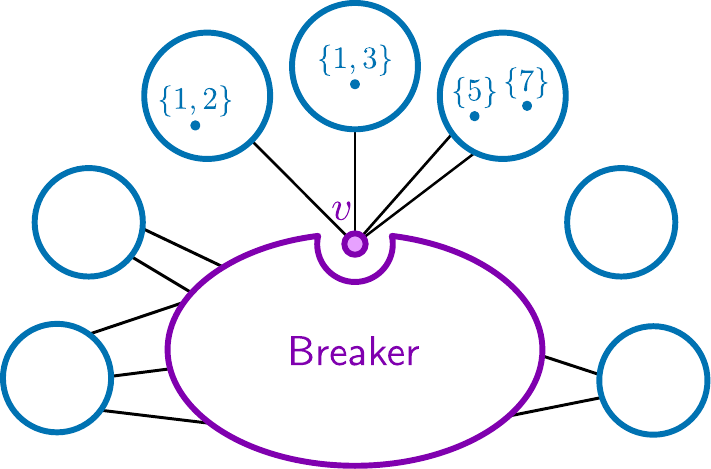}
\caption{Example of the labels reached by a vertex $v \in \breaker$ in the connected components of $\smallerGAB \setminus (\breaker \setminus \set{v})$. Here, $v$ reaches the set of labels $\set{1,2,3,5,7}$. All the connected components in blue are $\Q$-minor-free.}
\label{fig:labels-reached-by-vertex-in-breaker}
\end{figure}

\begin{claim}
    \label{fact:Y-leaves-a-Q-minor-with-marked-labels}
    Every set $Y \in \YQ$ leaves a $\Q$-minor in $\smallerGAB$ that contains only marked labels.
\end{claim}
\begin{claimproof}
    Consider the set $\breaker_Y \coloneqq \breaker \setminus Y$ of vertices of $\breaker$ left by $Y$, which is non-empty by \autoref{fact:Z-can-be-small}. First, suppose there exists a vertex $v \in \breaker_Y$ such that $\abs{\labelsReachedBy{v}} \geq \tooManylabelsReached$. Then by \autoref{fact:if-v-reaches-too-many-labels-then-mandatory} and \namedtheoremref{fact:Y-has-at-most-neighbors-vertices-in-S}, the graph $\smallerGAB \setminus Y$ contains a $\Q$-minor that only uses labels from $\labelsReachedBy{v}$ that were marked.

    Otherwise, suppose that for every vertex $v \in \breaker_Y$, we have that \[\abs{\labelsReachedBy{v}} < \tooManylabelsReached,\] and thus for every vertex $v \in \breaker_Y$ we mark all labels in $\labelsReachedBy{v}$.

    Suppose a label $\ell$ is reached by $v \in \breaker_Y$ in $\smallerGAB \setminus Y$, but not necessarily in $\smallerGAB \setminus (\breaker \setminus \set{v})$. There exists a path $P$ in $\smallerGAB \setminus Y$ from $v$ to a vertex that contains $\ell$ in its labelset. Let $v'$ be the last vertex in $P$ that belongs to $\breaker_Y$, and notice that $\ell$ must belong to $\labelsReachedBy{v'}$. See \autoref{fig:path-to-label-in-breakerY} for a diagram of this situation. Thus, $\ell$ is marked. This means that every label reached by $v \in \breaker_Y$ in $\smallerGAB \setminus Y$ is marked.

    \begin{figure}[ht]
        \centering
        \includegraphics[width=.45\textwidth]{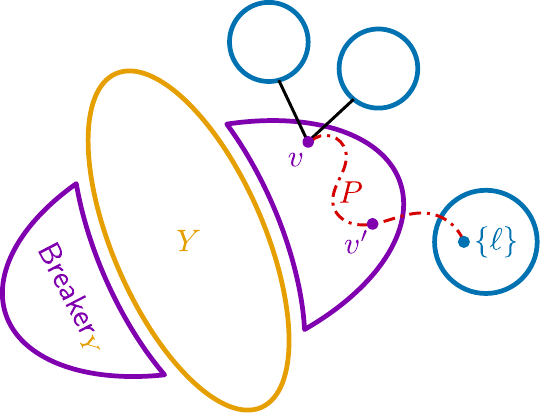}
        \caption{Label $\ell$ is reached by $v \in \breaker_Y$ in $\smallerGAB \setminus Y$ through a path $P$. The last vertex in $P$ that belongs to $\breaker_Y$ is $v'$, which reaches $\ell$ in $\smallerGAB \setminus (\breaker \setminus \set{v'})$.}
        \label{fig:path-to-label-in-breakerY}
    \end{figure}

    As $\breaker$ hits all $\Q$-minors in $\smallerGAB$ while $Y \in \YQ$ leaves a $\Q$-minor in $\smallerGAB \setminus Y$, there exists a vertex $v \in \breaker_Y$
    that belongs to a $\Q$-minor model $\minorModel$ in $\smallerGAB \setminus Y$. As $\Q$ consists of connected graphs, all labels in $\minorModel$ are reached by $v$ in $\smallerGAB \setminus Y$. Additionally, by the previous paragraph, all labels reached by $v$ in $\smallerGAB \setminus Y$ are marked. Thus, all labels in $\minorModel$ are marked.
\end{claimproof}

In total, we have marked at most $\tooManylabelsReached\cdot\abs{\breaker}$ labels in $\smallerGAB$, which by \autoref{item:Z-is-small} of \autoref{fact:Z-can-be-small} equals at most
\[
    \totalMarkedLabels \coloneqq (\abs{\Gboundary}(\nF - 1) + \nF)\cdot(\abs{\Gboundary} + 1)
\]
marked labels. We define the set $\Q^\star$ as the subset of graphs in $\Q$ that have only marked labels, which by \autoref{fact:Y-leaves-a-Q-minor-with-marked-labels} will contain a minor of $\smallerGAB \setminus Y$ for every $Y \in \YQ$. As each graph in $\Q$ has at most $\max_{H\in \F} \abs{E(H)} + 1$ vertices, the size of $\Q^\star$ is bounded by the number of different graphs with at most $\max_{H\in \F} \abs{E(H)} + 1$ vertices and with labels from a set of size at most $\totalMarkedLabels$, which is a function of $\abs{\Gboundary}$ and $\norm{\F}$.
\end{proof}

\subsection{Proof of \namedtheoremref{fact:main-lemma}}

\Namedtheoremref{fact:extra-base-case} combined with the rest of the proof of Lemma 27 of \cite{minor-hitting} completes the proof of \namedtheoremref{fact:inductive-main-lemma-ed}. We now prove \namedtheoremref{fact:main-lemma}.

\begin{lemma:main:statement}[Main Lemma for $\EDF$]
    \mainlemma
\end{lemma:main:statement}
\begin{proof}
Let $\F$, $\Q$, and $C$ be given. Apply \namedtheoremref{fact:inductive-main-lemma-ed} with $\Q$, $\F$, $G_A := C$, $G_B$ and $G_C$ empty, $\Pi_A  := \F$, $\Pi_B = \Pi_C = \emptyset$ and $R_B = \emptyset$. The lemma gives a set $\Q^\star\subseteq \Q$ that satisfies the required size bound, by the fact that $\EDF(G_A \setminus S) = \EDF(C)$ and the same reasons as in the proof of Lemma 27 in \cite{minor-hitting}. The rest of the proof follows exactly as the proof of Lemma 27 in \cite{minor-hitting}.
\end{proof}

\section{Lower bounds}
\label{sec:lower-bounds}

In this section we present two different types of lower bounds. We start in \autoref{sec:lower-bound-Bart} by describing the reduction of Dekker and Jansen~\cite{FVS-via-EDF-DAM} for families $\F$ of biconnected graphs containing a planar graph, since our reductions consist in appropriate modifications of theirs.  Then, in \autoref{sec:lower-bounds-not-biconnected} we show that the same dichotomy still holds for infinitely many families $\F$ that contain graphs that are {\sl not} biconnected. Finally, in \autoref{sec:lower-bounds-degree} we present a lower bound on the exponent of $\abs{X}$ of our kernels. We also discuss how the elimination distance and the collection $\F$ appear in the exponent, by relating our positive and negative results to the ones by Giannopoulou et al.~\cite{GiannopoulouJLS17}
(parameterized by the solution size, which corresponds to elimination distance zero).

\subsection{Reduction of Dekker and Jansen for biconnected families containing a planar graph}
\label{sec:lower-bound-Bart}

In this section we give a sketch of the proof of~\cite[Theorem 2]{FVS-via-EDF-DAM} by Dekker and Jansen, formally stated below. This reduction was in part inspired by other previous work~\cite{vertex-cover-bridge-depth,HolsKP22}. Recall that, as discussed in \autoref{sec:consequences}, we can relax the hypothesis of~\cite[Theorem 2]{FVS-via-EDF-DAM} and assume only that at least one graph in $\F$ is planar, instead of all of them.

\begin{proposition}[Dekker and Jansen~\cite{FVS-via-EDF-DAM}]\label{fact:lower-bound-bart}
Let ${\mathcal C}$ be a minor-closed class of graphs and let $\F$ be a finite set of biconnected graphs on at least three vertices containing at least one planar graph. If ${\mathcal C}$ has unbounded elimination distance to an $\F$-minor-free graph, then \Fdeletion does not admit a polynomial kernel in the size of a ${\mathcal C}$-modulator, unless $\NP \subseteq \coNP/\poly$.
\end{proposition}

To prove \autoref{fact:lower-bound-bart}, Dekker and Jansen~\cite{FVS-via-EDF-DAM} present a polynomial parameter transformation from the \textsc{CNF Satisfiability} problem parameterized by the number of variables $n$ to the \Fdeletion problem parameterized by the size of a $\C$-modulator, where $\C$ is a minor-closed graph family with unbounded elimination distance to an $\F$-minor-free graph, and $\F$ is a graph family of biconnected graphs in at least three vertices with at least one planar graph. The size of the modulator in the reduction turns out to be linear in $n$ when $\F$ is fixed.

The proof is then completed through the use of the following proposition.

\begin{proposition}[{\fixedspacingcite[Theorem 4]{FVS-via-EDF-DAM}}]\label{fact:cnfsat-no-poly-kernel}
There is no polynomial-time algorithm that, given a {\sc CNF} formula $F$ on $n$ variables, constructs an instance $I$ of a fixed decision problem $Q$ such that
\begin{enumerate}
    \item $F$ is satisfiable if and only if $I$ is a {\sf yes}-instance of $Q$,
    \item $\abs I = \bigO(n^c)$ for some constant $c$,
\end{enumerate}
unless $\NP \subseteq \coNP / \poly$.
In other words, there is no \emph{polynomial compression} from {\sc CNF Satisfiability} to any fixed decision problem $Q$, unless $\NP \subseteq \coNP / \poly$.
\end{proposition}

We introduce the notion of a necklace, which is a crucial structure in the proof of \autoref{fact:lower-bound-bart}.
\begin{definition}[$\F$-necklace {\fixedspacingcite[Definition 7]{FVS-via-EDF-DAM}}]
    Let $G$ be a graph and let $\mathcal F$ be a collection of connected graphs.
    $G$ is an $\mathcal F$-\emph{necklace} of length $t$ if there exists a partition of $V(G)$ into $S_1, \dots, S_t$ such that
    \begin{itemize}
        \item $G[S_i] \in \mathcal F$ for each $i \in \set{1,\dots,t}$ (these subgraphs are the \emph{beads} of the necklace),
        \item $G$ has precisely one edge between $S_i$ and $S_{i+1}$ for each $i \in \set{1,\dots,t-1}$,
        \item $G$ has no edges between any other pair of sets $S_i$ and $S_j$.
    \end{itemize}
\end{definition}

When the length of the necklace is not relevant, we simply speak of an $\mathcal F$-necklace.
The following definition specifies a special type of necklace.

\newrefcommand{\leftEndpoint}{\mathsf{left}}
\newrefcommand{\rightEndpoint}{\mathsf{right}}
\begin{definition}[uniform $\F$-necklace {\fixedspacingcite[Definition 8]{FVS-via-EDF-DAM}}]
    \label{def:uniform-necklace}
    Let $\mathcal F$ be a collection of connected graphs.
    Let $G$ be an $\mathcal F$-necklace of length $t$, and $H$ be a graph in $\F$ with two (not necessarily distinct) vertices $\leftEndpoint, \rightEndpoint \in V(H)$.
    We say that $G$ is a \emph{uniform necklace with structure} $(H, \leftEndpoint, \rightEndpoint)$ if it satisfies two additional conditions.
    \begin{itemize}
        \item Each bead $G[S_i]$ is isomorphic to $H$.
        \item There exist graph isomorphisms $f_i \colon V(H) \rightarrow V(G[S_i])$ for each bead $G[S_i]$, such that for each $i \in \set{1,\dots,t-1}$, the edge between $G[S_i]$ and $G[S_{i+1}]$ has precisely the endpoints $f_i(\leftEndpoint)$ and $f_{i+1}(\rightEndpoint)$.
    \end{itemize}
\end{definition}

Notice that for any $H \in \F$ we can assume no proper minor of $H$ is contained in $\F$ in the context of \Fdeletion. Otherwise, we can remove $H$ from $\F$, as the deletion of all $H$-minors follows from the deletion of the proper minors of $H$.

When $\F$ consists of biconnected graphs and is minor-minimal, every optimal $\F$-minor deletion set for an $\F$-necklace takes exactly one vertex in each bead, as the following proposition states.

\begin{proposition}[{\fixedspacingcite[Lemma 6]{FVS-via-EDF-DAM}}]
    \label{fact:necklace-minors}
    Let $\F$ be a collection of connected graphs and let $G$ be an $\F$-necklace.
    Let $\mathcal H$ be a collection of biconnected graphs on at least three vertices.
    If no graph in $\F$ contains an $\mathcal H$-minor, then $G$ contains no $\mathcal H$-minor.
\end{proposition}

One of the key results to prove \autoref{fact:lower-bound-bart} is the following. %\ig{important: they assume all the graphs to be planar, but we don't need to, with one of them being planar it is enough. We have to mention that (as we already did before)}\eric{Done!}.

\begin{proposition}[cf. {\cite[Lemma 7]{FVS-via-EDF-DAM}}]\label{prop:long-uniform-necklaces}
Let $\F$ be a finite collection of connected graphs containing a planar graph. Any minor-closed graph family $\C$ with unbounded elimination
distance to an $\F$-minor-free graph contains arbitrarily long uniform $\F$-necklaces.
\end{proposition}

In their original Lemma 7, Dekker and Jansen assume that all graphs in $\F$ are planar. Fortunately, the same proof works if only one of them is planar, as explained in \autoref{sec:consequences}. The statement of \autoref{prop:long-uniform-necklaces} reflects this adaptation.

\newcommand{\ClauseGadget}{\mathsf{ClauseGadget}}
\newcommand{\VariableGadget}{\mathsf{VariableGadget}}

The reduction by Dekker and Jansen creates a graph $G$ with a $\C$-modulator $X$, and defines a \emph{budget} $k$, corresponding to the number of vertices that can be taken by the sought $\F$-minor deletion set. The resulting graph $G$ is depicted in \autoref{fig:reduction-Bart}.

Let $C_1,\dots,C_m$ and $x_1,\dots,x_n$ be the clauses and variables of the \textsc{CNF Satisfiability} instance, respectively. By \autoref{prop:long-uniform-necklaces}, the graph class $\C$ contains arbitrarily long uniform $\F$-necklaces; let $(H, \leftEndpoint, \rightEndpoint)$ be the structure of such necklaces.

\newrefcommand{\bead}[2]{\mathsf{Bead}_{#1,#2}}
\subsubsection{Clause gadgets}
For each clause $C_i$, we create a clause gadget $\ClauseGadget_i$ that is a uniform $\F$-necklace with structure $(H, \leftEndpoint, \rightEndpoint)$ of length $\abs{C_i}$. For each literal $\ell$ in $C_i$, we designate a corresponding bead $\bead{i}{\ell}$ of $\ClauseGadget_i$. Let $\leftEndpoint_{i, \ell}$ and $\rightEndpoint_{i, \ell}$ be the copies of $\leftEndpoint$ and $\rightEndpoint$, respectively, in $\bead{i}{\ell}$. In each $\bead{i}{\ell}$, we choose an arbitrary vertex different from $\leftEndpoint_{i, \ell}$ and $\rightEndpoint_{i, \ell}$ and designate it as the \emph{literal} vertex $v_{i, \ell}$. Note that this is always possible as $H$ has at least three vertices. Also note that $H$ is a biconnected graph.

\subsubsection{Variable gadgets}
\begin{figure}[ht]
    \centering
    \includegraphics[width=0.35 \textwidth]{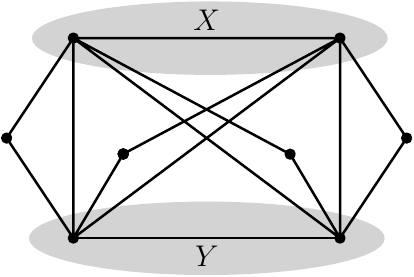}
    \caption{Illustration of $\VariableGadget$ with $H = {K_3}$.
    Observe that the subgraph induced by $X$ is indeed isomorphic to $K_3 \setminus \set{v}$. Figure by Dekker and Jansen~\cite[Figure 2]{FVS-via-EDF-arxiv}.}\label{fig:VariableGadget}
\end{figure}

\begin{figure}[ht]
    \centering
    \includegraphics[width=0.7 \textwidth]{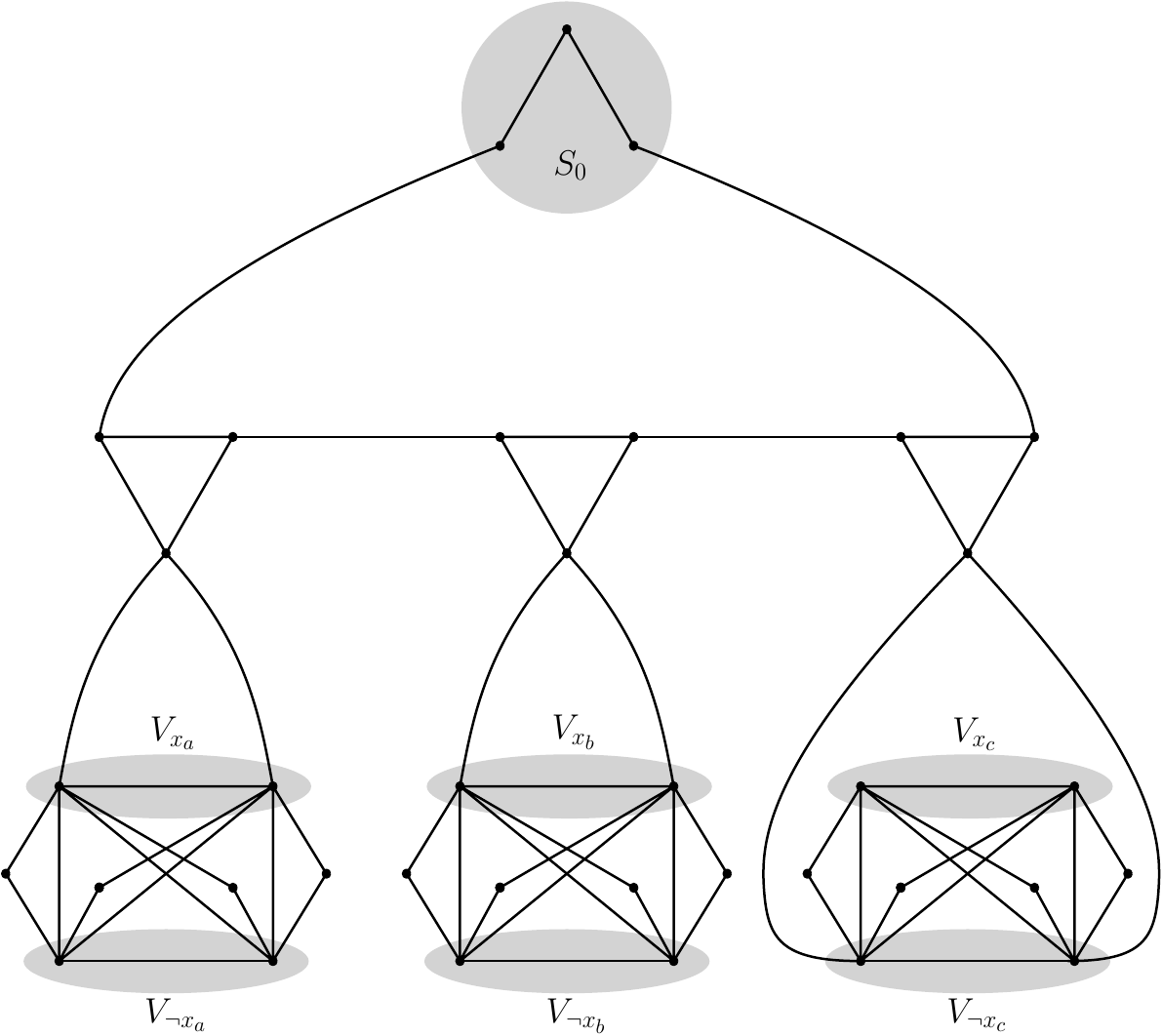}
    \caption{Illustration of the reduction for a clause $C_i \coloneqq (x_a \vee x_b \vee \neg x_c)$.
    Let $\ell$ denote a literal of this clause.
    We pick $H = K_3$ and we pick distinct $\leftEndpoint, \rightEndpoint \in V(H)$, so the uniform necklace structure becomes $(K_3, \leftEndpoint, \rightEndpoint)$.
    We add such a uniform necklace of length $3$.
    The first and last bead are connected to the extra gadget $S_0$.
    We then connect each bead to a variable gadget.
    For the bead corresponding to $\ell$, we pick a literal vertex $v_{i, \ell}$ in this bead which is not adjacent to the next or previous bead, and neither to $S_0$.
    We then ensure that the subgraph induced by $v_{i, \ell}$ and $V_{\ell}$ equals $K_3$.
    Observe that, since the last literal of the clause is a negated variable, the last bead is connected to $V_{\neg x_c}$. Figure by Dekker and Jansen~\cite[Figure 3]{FVS-via-EDF-arxiv}.
    }\label{fig:reduction-Bart}
\end{figure}

For these gadgets we have to define a special graph, depicted in \autoref{fig:VariableGadget}.

\begin{definition}[$\VariableGadget$ (cf. {\cite[Definition 9]{FVS-via-EDF-DAM}})]
    \label{def:VariableGadget}
    Let $v^+, v^-, u \in V(H)$ with $v^+ \neq v^-$, and let $c = \abs{V(H)}$.
    Define $\VariableGadget$ to be the graph which can be constructed with the following two steps.
    \begin{enumerate}
        \item Create $c-1$ vertices $v^+_1, \dots, v^+_{c-1}$ and define the set $V^+ \coloneqq \set{v^+_1, \dots, v^+_{c-1}}$.
        Then, add precisely those edges such that $\VariableGadget[V^+]$ is isomorphic to $H \setminus \set{u}$.
        Repeat this construction for a disjoint set of vertices $V^- \coloneqq \set{v^-_1, \dots, v^-_{c-1}}$ such that $\VariableGadget[V^-]$ is also isomorphic to $H\setminus\set{u}$.
        \item Then, for every pair $v^+_i \in V^+$ and $v^-_j \in V^-$, add $c-2$ extra vertices to the graph and let $H_{i, j}$ be the set containing these vertices.
        Add edges to the graph such that $\VariableGadget[H_{i, j} \cup \set{v^+_i, v^-_j}]$ is isomorphic to $H$.
    \end{enumerate}
    In total, the graph will have $2(c-1) + (c-1)^2(c-2)$ vertices.
\end{definition}

Note that we slightly modify the definition by Dekker and Jansen~\cite[Definition 9]{FVS-via-EDF-DAM} to avoid the clash of variables $x,y$ in their reduction, confirmed by the authors via private email.

For each variable $x_i$, we create a variable gadget $\VariableGadget_i$ that is a copy of $\VariableGadget$, and add the sets $V^+$ and $V^-$ to the $\C$-modulator $X$. For the sake of simplicity, the vertices $v^+$, $v^-$, and $u$ are taken to be the same between different instances of the gadget, although this is not strictly necessary. We designate the set $V^+$ of $\VariableGadget_i$ as $V_{x_i}$, and the set $V^-$ as $V_{\neg x_i}$. If a solution in $G$ takes all vertices in $V_{x_i}$, we interpret this as setting variable $x_i$ to true, and if it takes all vertices in $V_{\neg x_i}$, we interpret this as setting variable $x_i$ to false. This decision is motivated by the following two propositions, which are key to the reduction.

\begin{proposition}[{\fixedspacingcite[Lemma 13]{FVS-via-EDF-DAM}}]
    \label{fact:VariableGadget-no-minor-then-X-or-Y}
    Let $\VariableGadget$ be the graph defined in Definition~\ref{def:VariableGadget} with $V^+$, $V^-$ and $c$ defined accordingly.
    Let $Z \subseteq V(\VariableGadget)$ be a vertex set with $\abs Z \le c-1$.
    If $\VariableGadget \setminus Z$ contains no $H$-minor, then $Z$ is equal to $V^+$ or $V^-$.
\end{proposition}

The implication in \autoref{fact:VariableGadget-no-minor-then-X-or-Y} is actually an equivalence.
For the other direction, Dekker and Jansen prove a slightly stronger statement.

\begin{proposition}[{\fixedspacingcite[Lemma 14]{FVS-via-EDF-DAM}}]
    \label{fact:VariableGadget-X-or-Y-then-no-minor}
    Let $\VariableGadget$ be the graph defined in Definition~\ref{def:VariableGadget} with $V^+$, $V^-$ and $c$ defined accordingly.
    If $Z \subseteq V(\VariableGadget)$ is equal to $V^+$ or $V^-$, then any biconnected graph that $\VariableGadget \setminus Z$ contains as a minor is a proper minor of $H$.
\end{proposition}

Thus, if a solution for $G$ is forced to take an optimal solution in $\VariableGadget_i$, the vertices taken in $\VariableGadget_i$ must either be \textsl{exactly} $V_{x_i}$ or $V_{\lnot x_i}$, which corresponds to a truth assignment for the variable $x_i$.

\subsubsection{Extra gadget}
Finally, the last vertices added to $G$ are part of a copy of $H - e$ for some edge $e \in E(H)$, where $e \coloneq \set{a,b}$. This copy is also added to $X$.

\subsubsection{Edges between gadgets}
For each clause $C_i$ and each literal $\ell$ in $C_i$, we connect the literal vertex $v_{i, \ell}$ of the clause gadget $\ClauseGadget_i$ to the vertices in $V_{\ell}$ of the variable gadget corresponding to $\ell$ such that $G[V_\ell \cup \set{v_{i,\ell}}]$ is isomorphic to $H$. Notice that if a solution does not take any vertex in $V_{\ell}$, it is forced to take the literal vertex $v_{i, \ell}$ to hit the $H$-minor model $V_{\ell} \cup \{v_{i, \ell}\}$.

Additionally, for each clause gadget $\ClauseGadget_i$, we connect vertex $a$ of the extra gadget to the copy of $\leftEndpoint$ in the first bead of $\ClauseGadget_i$, and vertex $b$ of the extra gadget to the copy of $\rightEndpoint$ in the last bead of $\ClauseGadget_i$. As the extra gadget is only missing an edge to create an $H$-minor model, if a solution does not take any vertex in the extra gadget, then it must disconnect the first and last bead of $\ClauseGadget_i$. Because $H$ is biconnected, the only way to accomplish this is by taking at least one copy of $\leftEndpoint$ or $\rightEndpoint$ in a bead---which \textsl{are not} literal vertices.

Finally, the budget $k$ is set to the sum of $\opt(\ClauseGadget_i)$ over all clause gadgets $\ClauseGadget_i$, plus $n \cdot \abs{V_{\ell}}$ for any literal $\ell$. By \autoref{fact:necklace-minors}, an optimal solution for a clause gadget must take one vertex in each bead. On the other hand, by \cref{fact:VariableGadget-no-minor-then-X-or-Y,fact:VariableGadget-X-or-Y-then-no-minor},
an optimal solution for $\VariableGadget_i$ takes either all vertices in $V_{x_i}$ or all vertices in $V_{\neg x_i}$ for each variable $x_i$. Thus, if there exists a solution for $G$ of size at most $k$, it must take an optimal solution in each clause gadget and an optimal solution in each variable gadget, and no vertices in the extra gadget.

Let $Y$ be an $\F$-minor deletion set for $G$ with at most $k$ vertices. The construction ensures the following:
\begin{enumerate}[label=(\roman*)]
    \item\label{item:taking-Vell-frees-vell} For each clause $C$ and each literal $\ell$ in $C$ (which is $x_i$ or $\neg x_i$ for some $i \in \{1,\dots,n\}$), if $Y$ does not take the vertices in $V_{\ell}$, it is forced to take the literal vertex $v_\ell$ in the clause gadget corresponding to $C$.
    \item\label{item:one-vertex-must-be-free} In each clause gadget, the solution $Y$ takes at least one vertex that is not a literal vertex. This is ensured by the adjacencies with the extra gadget in $X$.
\end{enumerate}

If a solution for $G$ does not take any vertex in the extra gadget and does not take any vertex in $V_\ell$ for any literal $\ell$ in $C_i$, it must take every literal vertex in $\ClauseGadget_i$ and at least one non-literal vertex in $\ClauseGadget_i$, thus taking more than $\opt(\ClauseGadget_i)$ vertices in $\ClauseGadget_i$. On the other hand, if a solution does indeed take all vertices in $V_\ell$ for some literal $\ell$ in $C_i$, it can ``skip'' the literal vertex $v_{i, \ell}$, taking instead a non-literal vertex in the corresponding bead, and hitting the $H$-minor model formed with the vertices of the extra gadget. Dekker and Jansen show that this is enough to hit all $H$-minor models in $G$. Moreover, they show that hitting all $H$-minor models in $G$ is equivalent to hitting all $\F$-minor models in $G$, using the fact that $\F$ is assumed to be minor-minimal and that $H$ is biconnected.

Therefore, if $Y$ satisfies every clause by taking the correct vertices in the variable gadgets, it can take $k$ vertices in total, while if $Y$ does not satisfy every clause, it must take more than $k$ vertices. This means that the original \textsc{CNF Satisfiability} instance is satisfiable if and only if there exists an $\F$-minor deletion set for $G$ of size at most $k$.

Finally, observe that, for fixed $\F$, the creation and connection of these gadgets can be executed in time polynomial in the size of the original formula.

\subsection{Dichotomy for some graphs $H$ that are not biconnected}
\label{sec:lower-bounds-not-biconnected}

In the kernelization dichotomy given in \autoref{fact:dichotomy}, we crucially assume that all graphs in $\F$ are biconnected so that we can apply the lower bound by Dekker and Jansen~\cite{FVS-via-EDF-DAM}, as stated in \autoref{fact:lower-bound-bart}.

On the other hand, note that our kernel (cf. \autoref{fact:poly-kernel}) only assumes the graphs in $\F$ to be connected (but not necessarily biconnected). The goal of this section is to show that there exist infinitely many families $\F$ containing graphs that are {\sl not} biconnected such that the lower bound of \autoref{fact:lower-bound-bart} still holds, hence yielding a dichotomy for these families such as the one in \autoref{fact:dichotomy}. We point out that the relevance of the following result is not given by the actual families $\F$ to which it applies, but as a proof of concept that such families $\F$ do exist, hence opening the door for eventually obtaining a dichotomy for all families $\F$ containing connected graphs.

The families presented in this section are not the only non-biconnected families for which the lower bound of \autoref{fact:lower-bound-bart} holds. In \autoref{sec:lower-bounds-treelike} we give another class of families $\F$ for which we can obtain the same dichotomy.

%\ig{IMPORTANT: Eric, I stated the following theorem just for a family containing a single graph $H$ that satisfied the required property. Probably we can obtain lower bounds for larger families with little more effort. I didn't have time to think about that, Eric, please do so}\eric{Done!}

 %\ig{Also, we need to say very clearly, before stating the following theorem and also in the introduction, that the reduction of \autoref{fact:lower-bound-not-biconnected} is strongly based on that of Dekker and Jansen~\cite{FVS-via-EDF-DAM}, which is also inspired by previous reductions, such as~\cite{vertex-cover-bridge-depth,HolsKP22}.}\eric{Done! We need to remember this when we change the introduction though.}

\newcommand{\conditionTitle}[1]{\textup{\textsf{\textbf{\textcolor{lipicsGray}{#1}}}}}
\newcommand{\bigComponentsCondition}{\hyperref[item:big-components]{\conditionTitle{(Big components)}}}
\newcommand{\planarCondition}{\hyperref[item:planar]{\conditionTitle{(Planar)}}}
\newcommand{\nonMinorBlocksCondition}{\hyperref[item:non-minor-blocks]{\conditionTitle{(Non-minor components)}}}

\begin{restatable}[Non-biconnected lower bound]{theorem}{lowerboundnonbiconnected}
\label{fact:lower-bound-not-biconnected}
Let ${\mathcal C}$ be a minor-closed class of graphs and let $\F$ be a finite graph family meeting the following conditions:

\begin{enumerate}[label=(\roman*), align=left, leftmargin=!]
    \item[\conditionTitle{(Connected)}] Every graph $H \in \F$ is connected.
    \item[\conditionTitle{(Planar)}]\label{item:planar} $\F$ contains at least one planar graph.
    \item[\conditionTitle{(Big components)}]\label{item:big-components} Every biconnected component of every graph $H \in \F$ has at least three vertices.
    \item[\conditionTitle{(Non-minor components)}]\label{item:non-minor-blocks} For every two (not necessarily distinct) graphs $H,H' \in \F$ and every biconnected component $B$ of $H$, there exists a corresponding biconnected component $B'$ of $H'$ such that for every edge $e \in E(B)$, the biconnected component $B'$ is not a minor of $H - e$.
\end{enumerate}

If ${\mathcal C}$ has unbounded elimination distance to an $\F$-minor-free graph, then \Fdeletion does not admit a polynomial kernel in the size of a ${\mathcal C}$-modulator, unless $\NP \subseteq \coNP/\poly$.
\end{restatable}

Notice that the original families $\F$ considered in \autoref{fact:lower-bound-bart} satisfy the conditions of \namedtheoremref{fact:lower-bound-not-biconnected}. Thus, our result can be seen as a strict generalization of the lower bound by Dekker and Jansen~\cite{FVS-via-EDF-DAM}.

\begin{figure}[ht]
\centering
\includegraphics[width=0.8\textwidth]{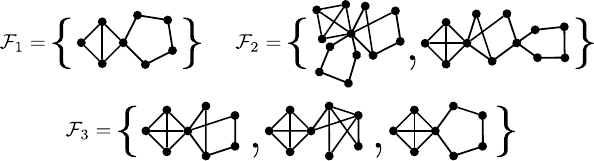}
\caption{Examples of families $\F$ satisfying the conditions of \namedtheoremref{fact:lower-bound-not-biconnected}. }
\label{fig:examples-non-biconnected-families}
\end{figure}

Before proving \namedtheoremref{fact:lower-bound-not-biconnected}, we show some examples---depicted in \autoref{fig:examples-non-biconnected-families}---of families $\F$ that satisfy the conditions of the theorem. To build such examples, it will be useful to have in mind the following observation.

\begin{observation}
\label{fact:blocks-minor-incomparable}
Let $\F$ be a family of graphs meeting the conditions of \namedtheoremref{fact:lower-bound-not-biconnected}, and let $H$ be a graph in $\F$. If $B_1$ and $B_2$ are two different biconnected components of $H$, then $B_1$ is not a minor of $B_2$ and $B_2$ is not a minor of $B_1$.
\end{observation}
\begin{proof}
Suppose, to the contrary, that there exist two different biconnected components $B_1$ and $B_2$ of a graph $H \in \F$ such that $B_1$ is a minor of $B_2$. Take an edge $e \in E(B_1)$. As every biconnected component is a minor of itself, we have that every biconnected component of $H$ different from $B_1$ is a minor of some biconnected component in $H - e$. Additionally, $B_1$ is a minor of $B_2$, the latter being also a biconnected component of $H - e$. Thus, every biconnected component of $H$ is a minor of some biconnected component in $H - e$, contradicting condition \nonMinorBlocksCondition.
\end{proof}

In fact, for families $\F$ with only one graph, condition \nonMinorBlocksCondition\ is equivalent to saying that all biconnected components of that graph are pairwise minor-incomparable. Thus, our first example is a family $\F_1$ containing a single graph $H$ meeting exactly this condition.

For families with more than one graph, instead, \nonMinorBlocksCondition\ is weaker than asking for every possible pair of different biconnected components of graphs in $\F$ to be minor-incomparable. For our second example, we take a graph $H$ whose biconnected components are pairwise minor-incomparable, and we build a family $\F_2$ containing graphs obtained by arranging the same biconnected components of $H$ in different ways. For every two graphs $H, H' \in \F_2$, biconnected component $B$ of $H$, and edge $e \in E(B)$, the same biconnected component $B$ in $H'$ is not a minor of $H - e$ by \autoref{fact:biconnected-components-minor-model}, thus meeting condition \nonMinorBlocksCondition.

Finally, for our third example, we take a family $\F_3$ such that every graph $H' \in \F_3$ has a biconnected component $B'$ that no other graph in $\F_3$ has as a minor. For every biconnected component $B$ of any graph $H \in \F_3$, the corresponding biconnected component when considering $H'$ in condition \nonMinorBlocksCondition\ will be exactly $B'$.

The proof of \namedtheoremref{fact:lower-bound-not-biconnected} is strongly based on the reduction by Dekker and Jansen~\cite{FVS-via-EDF-DAM} described in \autoref{sec:lower-bound-Bart}. In their reduction, when building the gadgets, some vertices are left as arbitrary choices; our reduction instead restricts these to be vertices of a specific biconnected component $B$ of the graph $H$.

Specifically, we define the variable gadgets as follows.

\newrefcommand{\NonBiconnectedVariableGadget}{\mathsf{NonBiconnectedVariableGadget}}

\begin{definition}[$\NonBiconnectedVariableGadget_{H, B}$]
\label{def:NonBiconnectedVariableGadget}
Let $H$ be a graph and $B$ be a biconnected component of $H$ with at least two vertices. Define $\NonBiconnectedVariableGadget_{H, B}$ to be the graph which can be constructed with the following two steps.
\begin{enumerate}
    \item Let $u \in B$. Create $|V(H)|-1$ vertices $V^+ \coloneqq \set{v^+_1, \dots, v^+_{|V(H)|-1}}$ and add edges to the graph such that $\NonBiconnectedVariableGadget_{H, B}[V^+]$ is isomorphic to $H \setminus \set{u}$.
    Then, repeat this construction for a set of vertices $V^- \coloneqq \set{v^-_1, \dots, v^-_{|V(H)|-1}}$ such that $\NonBiconnectedVariableGadget_{H, B}[V^-]$ is also isomorphic to $H \setminus \set{u}$.
    \item For every pair $v^+_i \in V^+$ and $v^-_j \in V^-$, add $|V(H)|-2$ extra vertices to the graph and let $H_{i, j}$ be the set containing these vertices.
    Add edges to the graph such that there exists an isomorphism $\varphi$ from $\NonBiconnectedVariableGadget_{H, B}[H_{i, j} \cup \set{v^+_i, v^-_j}]$ to $H$ where $\varphi(v^+_i)$ and $\varphi(v^-_j)$ belong to $B$.
\end{enumerate}
\end{definition}

\begin{figure}[ht]
    \centering
    \includegraphics{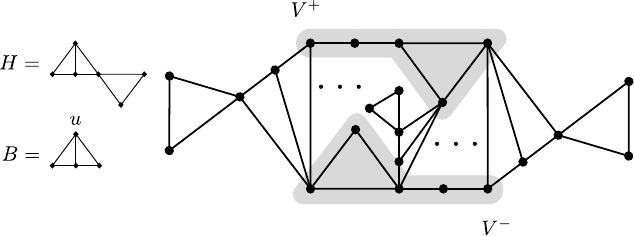}
    \caption{An example of $\NonBiconnectedVariableGadget_{H, B}$. Each vertex $v^+ \in V^+$ is connected to each vertex $v^- \in V^-$ through a copy of $H$ where the images of $v^+$ and $v^-$ belong to the biconnected component $B$ of $H$.}
    \label{fig:NonBiconnectedVariableGadget}
\end{figure}

An example of \defref{def:NonBiconnectedVariableGadget} is depicted in \autoref{fig:NonBiconnectedVariableGadget}.

Note that the graph $\VariableGadget$ from \autoref{def:VariableGadget} is a particular case of the graph $\NonBiconnectedVariableGadget_{H, B}$ where $B = H$.
This more general construction still maintains properties that are analogous to those in \autoref{fact:VariableGadget-no-minor-then-X-or-Y} and \autoref{fact:VariableGadget-X-or-Y-then-no-minor}. In fact, the proof of Lemma 13 in~\cite{FVS-via-EDF-DAM} does not actually require the graph $H$ to be biconnected, and thus we obtain the following statement.

\begin{lemma}[cf. \autoref{fact:VariableGadget-no-minor-then-X-or-Y}]
    \label{fact:NonBiconnectedVariableGadget-no-minor-then-X-or-Y}
    Let $H$ be a graph and $B$ be a biconnected component of $H$, with $|V(H)| \geq 3$ and $|V(B)| \geq 2$. Consider the graph $\NonBiconnectedVariableGadget_{H, B}$ as in \autoref{def:NonBiconnectedVariableGadget}, with $V^+$ and $V^-$ defined accordingly.
    Let $Z \subseteq V(\NonBiconnectedVariableGadget_{H, B})$ be a vertex set with $\abs Z \leq \abs{V(H)}-1$.
    If $\NonBiconnectedVariableGadget_{H, B} \setminus Z$ contains no $H$-minor, then $Z$ is equal to $V^+$ or $V^-$.
\end{lemma}

\begin{lemma}[cf. \autoref{fact:VariableGadget-X-or-Y-then-no-minor}]
\label{fact:NonBiconnectedVariableGadget-X-or-Y-then-no-B'}
Let $H$ and $H'$ be graphs and $B$ and $B'$ be biconnected components of $H$ and $H'$, respectively. Consider the graph $\NonBiconnectedVariableGadget_{H, B}$ as in \autoref{def:NonBiconnectedVariableGadget}, with $V^+$ and $V^-$ defined accordingly. Assume that $B'$ is not a minor of $H - e$ for any $e \in E(B)$. If $Z \subseteq V(\NonBiconnectedVariableGadget_{H, B})$ is equal to $V^+$ or $V^-$, then $B'$ is not a minor of any biconnected component in $\NonBiconnectedVariableGadget_{H, B} \setminus Z$. Moreover, $H'$ is not a minor of $\NonBiconnectedVariableGadget_{H, B} \setminus Z$.
\end{lemma}
\begin{proof}
Notice that the biconnected components of $\NonBiconnectedVariableGadget_{H, B}$ are copies of the biconnected components of $H$ or, in the case of $V^+$ and $V^-$, copies of the biconnected components of $H \setminus \set{v}$ for some $v \in V(B)$. Notice that every biconnected component of $H$ is a minor of $H - e$, except possibly $B$. As $B'$ is not a minor of $H - e$ for any $e \in E(B)$, we have that $B'$ is not a minor of any biconnected component in $\NonBiconnectedVariableGadget_{H, B}$, except possibly copies of $B$. Additionally, as $Z$ is equal to $V^+$ or $V^-$, it contains at least one vertex in each copy of $B$ in $\NonBiconnectedVariableGadget_{H, B}$. Hence, $B'$ is not a minor of any biconnected component in $\NonBiconnectedVariableGadget_{H, B} \setminus Z$, and thus by \autoref{fact:biconnected-components-minor-model}, $H'$ is not a minor of $\NonBiconnectedVariableGadget_{H, B} \setminus Z$.
\end{proof}

We are now ready of prove \namedtheoremref{fact:lower-bound-not-biconnected}. We restate the theorem for convenience.

\lowerboundnonbiconnected*

\begin{proof}
We will give a reduction from \textsc{CNF Satisfiability} to $\F$-\textsc{Minor Deletion} similar to the one by Dekker and Jansen presented in \autoref{sec:lower-bound-Bart}. This, along with \autoref{fact:cnfsat-no-poly-kernel}, will give us the statement of the theorem.

Given a formula $F$ with $n$ variables $x_1, \dots, x_n$ and $m$ clauses $C_1,\dots,C_m$, we will construct a graph $G$ with a $\C$-modulator $X$ and a budget $k$ such that $F$ is satisfiable if and only if there exists an $\F$-minor deletion set for $G$ of size at most $k$. The size of $X$ will be linear in $n$ when $\F$ is fixed.

The graph $G$ will be constructed by connecting some gadgets, which we describe below. The budget $k$ will be equal to the sum of the number of literals $\abs{C_i}$ in each clause $C_i$ plus $(\abs{V(H)} - 1) \cdot n$.

As mentioned earlier, this construction is almost the same as the one by Dekker and Jansen. The difference lies in that we restrict some vertices to belong to a specific biconnected component $B$ of a graph in $H$ instead of being arbitrary. Nothing else is changed in the construction.

By \autoref{prop:long-uniform-necklaces} and condition \planarCondition, the graph class $\C$ contains arbitrarily long uniform $\F$-necklaces; let $(H,\leftEndpoint,\rightEndpoint)$ be the structure of such necklaces. Denote the biconnected component of $H$ containing $\leftEndpoint$ as $B$.
We now describe the different gadgets of the reduction and how they connect to each other in $G$.

\textbf{Variable gadgets.} For each variable $x_i$, we create a variable gadget $\VariableGadget_i$ that is a copy of $\NonBiconnectedVariableGadget_{H, B}$ as described in \autoref{def:NonBiconnectedVariableGadget}. We designate the set $V^+$ of $\VariableGadget_i$ as $V_{x_i}$, and the set $V^-$ as $V_{\neg x_i}$. We add all vertices in $V^+$ and $V^-$ to the $\C$-modulator $X$, but \textsl{not} the vertices we added connecting $V^+$ and $V^-$. This is supported by the fact that the connections between $V^+$ and $V^-$ induce minors of $H$, and thus $\VariableGadget_i \setminus (V^+ \cup V^-)$ consists of connected components that are minors of a uniform necklace with structure $(H,\leftEndpoint,\rightEndpoint)$, and hence belong to $\C$.

\textbf{Clause gadgets.} For each clause $C_i$, we create a clause gadget $\ClauseGadget_i$ which is a uniform $\F$-necklace with structure $(H,\leftEndpoint,\rightEndpoint)$ and length equal to the number of literals in $C_i$. For each literal $\ell$ in $C_i$, let $\bead{i}{\ell}$ be the bead corresponding to $\ell$, and let $\leftEndpoint_{i, \ell}$ and $\rightEndpoint_{i, \ell}$ be the copies of $\leftEndpoint$ and $\rightEndpoint$, respectively, in $\bead{i}{\ell}$. We pick one vertex $v_{i, \ell} \in V(B)$ in the bead $\bead{i}{\ell}$ different from $\leftEndpoint_{i, \ell}$ and $\rightEndpoint_{i, \ell}$ that does not disconnect $\leftEndpoint_{i, \ell}$ from $\rightEndpoint_{i, \ell}$ in the bead. This vertex exists because each biconnected component of $H$ has at least three vertices. These will be the \emph{literal vertices} of the clause gadget. We add edges between each literal vertex $v_{i, \ell}$ and the vertices in $V_\ell$ such that $G[V_\ell \cup \set{v_{i, \ell}}]$ is isomorphic to $H$.

\textbf{Extra gadget.} We create an extra gadget which is a copy of $H - e$ for some edge $e \in E(B)$, where $e = \set{a,b}$. This edge exists because $B$ has at least two vertices. We add all vertices of the extra gadget to the $\C$-modulator $X$. Then, for every clause gadget $\ClauseGadget_i$, we add an edge between $a$ and vertex $\leftEndpoint$ in the first bead of $\ClauseGadget_i$, and an edge between $b$ and vertex $\rightEndpoint$ in the last bead of $\ClauseGadget_i$.

An illustration of this construction is given in \autoref{fig:non-biconnected-reduction}.

\begin{figure}[ht]
\centering
\includegraphics{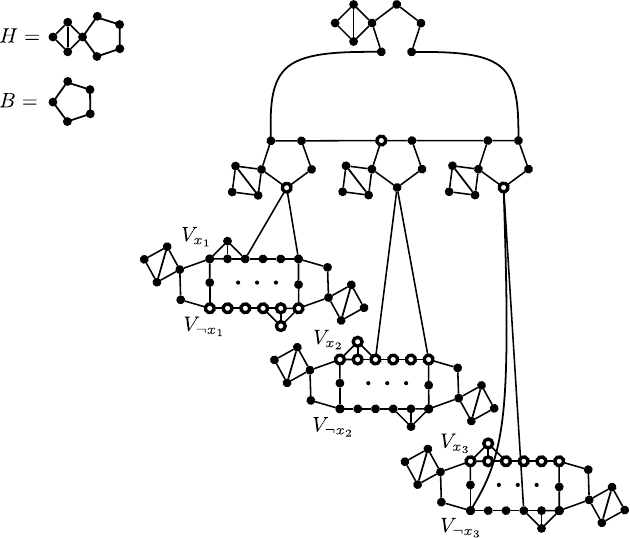}
\caption{An example construction for a clause $C_i \coloneqq (x_1 \lor x_2 \lor \neg x_3)$ in the reduction of \autoref{fact:lower-bound-not-biconnected} for $\F \coloneqq \set{H}$. Each bead of the clause gadget $\ClauseGadget_i$ is connected through the literal vertex $v_{i, \ell}$ to the corresponding variable gadget $\VariableGadget_j$ for the variable $x_j$ of the literal $\ell$. The first and last bead of $\ClauseGadget_i$ are connected to the extra gadget. The hollowed-out vertices from an optimal $\F$-deletion solution in the graph.}
\label{fig:non-biconnected-reduction}
\end{figure}

Each connected component in $G \setminus X$ is a minor of a uniform necklace with structure $(H,\leftEndpoint,\rightEndpoint)$: the clause gadgets are themselves uniform $\F$-necklaces with that structure, and for each variable $x_i$, the connections between $V_{x_i}$ and $V_{\neg x_i}$ in $\VariableGadget_i$ are minors of $H$. Notice that, as $\C$ is minor-closed and contains arbitrarily long $\F$-necklaces, $\C$ also contains the unions of arbitrarily many copies of such necklaces, which are simply minors of a sufficiently long $\F$-necklace. Therefore, $G\setminus X \in \C$.

The modulator $X$ consists of the $\abs{V(H)}$ vertices in the extra gadget, and the $2 \cdot (\abs{V(H)} - 1) \cdot n$ vertices belonging to $V^+ \cup V^-$ in each variable gadget. Thus, the size of $X$ is linear in $n$ when $\F$ is fixed.

It remains to show that $F$ is satisfiable if and only if there exists an $\F$-minor deletion set for $G$ of size at most $k$.

\textcolor{lipicsGray}{$\implies)$} Suppose $F$ is satisfiable. We will build an $\F$-minor deletion set $Y$ for $G$ of size at most $k$ as follows. For each variable $x_i$, if $x_i$ is assigned true, we add all vertices in $V_{x_i}$ to $Y$, and if $x_i$ is assigned false, we add all vertices in $V_{\neg x_i}$ to $Y$. For each clause $C_i$ there must exist a literal $\ell$ in $C_i$ that is satisfied by the truth assignment. We add to $Y$ the vertex $\leftEndpoint_{i, \ell}$, and the literal vertices $v_{i, \ell'}$ corresponding to all the literals $\ell'$ different from $\ell$ in $C_i$. Finally, we do not take any vertex in the extra gadget. An illustration of the set $Y$ can be seen in \autoref{fig:non-biconnected-reduction}.

Observe that $Y$ has $\abs{V(H)} - 1$ vertices for each of the $n$ variable gadgets, and $\abs{Y \cap \ClauseGadget_i}$ is equal to the number of literals in $C_i$ for each $i \in \set{1, \ldots, m}$. Thus, the size of $Y$ is exactly $k$.

We now show that $Y$ hits all $\F$-minors in $G$, completing the proof of the implication.

Let $H' \in \F$, and $B'$ be a biconnected component of $H'$ corresponding to $B$ as stated in condition \nonMinorBlocksCondition. We will show that there is no $B'$-minor in any biconnected component of $G \setminus Y$, thus proving by \autoref{fact:biconnected-components-minor-model} that there is no $H'$-minor in $G \setminus Y$.

\cref{fact:NonBiconnectedVariableGadget-X-or-Y-then-no-B'} implies that there is no $B'$-minor in any biconnected component of the variable gadgets in $G \setminus Y$. On the other hand, consider the connected components of the clause gadgets remaining in $G \setminus Y$. As $Y$ contains one vertex in each
copy of $B$ in the clause gadgets, these connected components are $\F'$-necklaces, where $\F'$ is the family of graphs resulting from removing a vertex $v \in V(B)$ from $H$. Additionally, notice that \nonMinorBlocksCondition\ implies that $B'$ is not a minor of $H
\setminus v$ for any $v \in V(B)$. Thus, by \autoref{fact:necklace-minors}, there is no $B'$-minor in any biconnected component of a bead of a clause gadget in $G \setminus Y$. Moreover, the set $Y$ completely disconnects the variable gadgets and the clause gadgets: for each literal $\ell$, the set $Y$ either takes all of $V_\ell$, or it takes all the neighbors of $V_\ell$ in the clause gadgets, namely, the literal vertices $v_{i, \ell}$. Hence, there is no biconnected component containing a $B'$-minor that uses vertices both from the clause gadgets and from the variable gadgets.

It remains to show that there is no $B'$-minor model in $G \setminus Y$ containing vertices of the extra gadget. Recall that $Y$ has at least one vertex $\leftEndpoint_{i, \ell}$ in each clause gadget $\ClauseGadget_i$. This ensures that in $G \setminus Y$ there is no path from the vertex $\leftEndpoint$ in the first bead of $\ClauseGadget_i$ and the vertex $\rightEndpoint$ in the last bead of $\ClauseGadget_i$. Hence, for every vertex $v$ of a clause gadget, every path between $v$ and the extra gadget passes through the same edge---either between $\leftEndpoint$ in the first bead and $a$, or $\rightEndpoint$ in the last bead and $b$---, and thus there is no biconnected component that uses vertices both from a clause gadget and from the extra gadget. Lastly, the extra gadget is exactly $H - \set{a,b}$ for some $\set{a,b} \in E(H)$, and thus by condition \nonMinorBlocksCondition\ there is no $B'$-minor in the extra gadget. This concludes the proof of this implication.

\textcolor{lipicsGray}{$\impliedby)$} Let $Y$ be an $\F$-minor deletion set for $G$ of size at most $k$. We will show that $F$ is satisfiable by assigning truth values to the variables according to the vertices taken in the variable gadgets, and showing that at least one literal vertex in each clause gadget is not taken by $Y$, thus ensuring that at least one literal in each clause is satisfied.

Note that each clause gadget $\ClauseGadget_i$ contains a number of beads equal to the number of literals in $C_i$. Each bead induces a graph in $\F$, and thus any optimal $\F$-hitting set for $G$ must take at least one vertex in each bead. Therefore, $Y$ has at most $k - \sum_{i=1}^m \abs{C_i} = (\abs{V(H)} - 1) \cdot n$ vertices in the variable and extra gadgets. As there are $n$ variable gadgets, by the pigeonhole principle, there is a variable gadget $\VariableGadget_i$ such that $Y$ takes at most $\abs{V(H)} - 1$ vertices in $\VariableGadget_i$. By \autoref{fact:NonBiconnectedVariableGadget-no-minor-then-X-or-Y}, we have that $Y$ takes either $V_{x_i}$ or $V_{\neg x_i}$ in $\VariableGadget_i$; both consisting of exactly $\abs{V(H)} - 1$ vertices. Following this same logic for every variable gadget, we obtain that $Y$ takes either $V_{x_i}$ or $V_{\neg x_i}$ for every variable $x_i$. Thus, $\abs{V(H)} - 1$ vertices are taken in each variable gadget. This means that exactly $\abs{C_i}$ vertices are taken by $Y$ in each clause gadget, one in each bead. Moreover, $Y$ takes no vertices in the extra gadget.

We assign truth values to the variables as follows: for every variable $x_j$, if $V_{x_j}$ is taken, then $x_j$ is assigned true; otherwise, $V_{\lnot x_j}$ is taken, and $x_j$ is assigned false. We will see that this assignment satisfies every clause.

\begin{claim}
\label{fact:if-V-l-not-taken-then-v-l-taken}
For every literal $\ell$ in $C_i$, if $V_{\ell}$ is not taken by $Y$, then the literal vertex $v_{i, \ell}$ is taken by $Y$.
\end{claim}
\begin{claimproof}
If $V_{\ell}$ and $v_{i, \ell}$ are both not taken by $Y$, then $V_{\ell} \cup \set{v_{i, \ell}}$ is an $H$-minor model in $G \setminus Y$, contradicting the fact that $Y$ is an $\F$-minor deletion set.
\end{claimproof}

\begin{claim}
\label{fact:at-least-one-literal-vertex-not-taken}
There exists at least one literal vertex in $\ClauseGadget_i$ that is not taken by $Y$.
\end{claim}
\begin{claimproof}
Suppose that $Y$ takes all literal vertices in $\ClauseGadget_i$. Then, as there are $\abs{C_i}$ literal vertices in $\ClauseGadget_i$, $Y$ takes no more vertices in $\ClauseGadget_i$. As the literal vertex does not separate $\leftEndpoint_{i, \ell}$ and $\rightEndpoint_{i, \ell}$ in $\bead{i}{\ell}$, there exists a path in $\ClauseGadget_i$ from $\leftEndpoint$ in the first bead to $\rightEndpoint$ in the last bead that does not use any vertex of $Y$. This path, together with the extra gadget, forms an $H$-minor model in $G \setminus Y$, contradicting the fact that $Y$ is an $\F$-minor deletion set.
\end{claimproof}

Combining \autoref{fact:if-V-l-not-taken-then-v-l-taken} and \autoref{fact:at-least-one-literal-vertex-not-taken}, we see that for each clause $C_i$, there exists a literal $\ell$ in $C_i$ such that $V_{\ell}$ is taken by $Y$. This implies that $\ell$ is assigned true in the truth assignment, and thus $C_i$ is satisfied. As this holds for every clause, we conclude that $F$ is satisfiable.
%\ig{Eric, please try to be very clear about which stuff is taken from Bart (most!), and what is new}\eric{Done! I think. I said that the construction is the same and that we only changed which specific vertices we use when connecting the gadgets.}
\end{proof}

% \begin{theorem}[Lower bound when treewidth is unbounded]
%     Let ${\mathcal C}$ be a minor-closed class of graphs, and let $\F$ be a family of connected graphs with a planar graph $H \in \F$. Suppose there exists a block $B$ of $H$ with at least three vertices and an edge $e \in E(B)$ such that for every graph $H' \in \F$, there exists a corresponding block $B'$ of $H'$ that is not a minor of $H - e$.

%     If ${\mathcal C}$ has unbounded treewidth, then \Fdeletion does not admit a polynomial kernel in the size of a ${\mathcal C}$-modulator, unless $\NP \subseteq \coNP/\poly$.
% \end{theorem}
% \begin{proof}
% Big treewidth implies big arbitrary planar necklaces, in particular, big $H$-necklaces where $H$ is connected to the rest of the necklace via a single vertex that is an endpoint of the edge $e$. If we plug this necklace into the reduction and create a graph $G$, every solution for $G$ will take an endpoint of $e$ in each copy of $B$, including in the extra gadget if we connect it correctly. Thus, there will be no other $\F$-minor after removing the solution, and thus the reduction will work as always.
% \end{proof}

\subsection{On the degree of the polynomial kernels}
\label{sec:lower-bounds-degree}

For the upper bound, we show in \autoref{fact:explicit-bound-kernel-size} in \autoref{sec:explicit-upper-bounds} that the size of the kernel presented in \autoref{fact:poly-kernel} for the graph family $\F$ and the graph class $\C$ is
\[
    \bigO\parens*{\abs{X}^{2^{2^{\polynomial(\EDF(\C) + \norm\F)}}}},
\]
where $\norm\F = \max_{H \in \F} \abs{V(H)}$. The exponent of $\abs{X}$ is mainly driven by the size of the sets $\Q^\star$ given by \autoref{fact:main-lemma}. These sets $\Q^\star$ play the role of minimal blocking sets in previous work~\cite{vertex-cover-treedepth-poly-kernel-Algorithmica, vertex-cover-bridge-depth, HolsKP22}.

% \ig{IMPORTANT: This is my proposal for the presentation of this subsection (it is just a draft, needs to be polished):} I would \textbf{NOT} state the lower and upper bounds in terms of blocking sets (which are actually not defined in this paper), but in terms of the parameter that governs the dichotomy of \autoref{fact:dichotomy}, which is the elimination distance to the class of $\F$-minor-free graphs, denoted by $\EDF$.

% For the \textbf{upper bound}, that is, our kernel presented in \autoref{fact:poly-kernel}, it is easy to verify that, by the marking algorithm of Jansen and Pieterse~\cite{minor-hitting} that we reuse, the size of the obtained kernel is roughly \ig{please verify that! doesn't the induction step add some extra stuff to the exponent?}  $|X|^{\bigO(q)}$, where $q$ is the upper bound on the set $\Q^\star$ given by \autoref{fact:main-lemma}, which only depends on $\F$ and $\EDF(\C)$, with $\C$ begin the minor-closed graph class such that $X \subseteq V(G)$ is a $\C$-modulator. \igm{shall we give a rough upper bound on $q$?? I think we SHOULD}More precisely, as discussed in \autoref{sec:bounding-Q}, the dependency on $\F$ in the degree of the kernelization algorithm is given by $\norm\F =\max_{H\in \F} \abs{V(H)}$.

% (We should mention (again) that these sets $\Q^\star$  play the role of minimal blocking sets in previous papers on this topic~\cite{vertex-cover-treedepth-poly-kernel-Algorithmica, vertex-cover-bridge-depth, HolsKP22}.)

\medskip

On the other hand, for the lower bound, Jansen and Pieterse~\cite{FVS-via-EDF-DAM} use \autoref{prop:long-uniform-necklaces}---which corresponds to their Lemma 7---to obtain an arbitrarily long uniform $\F$-necklace, assuming that the elimination
distance of $\C$ to an $\F$-minor-free graph is large. They later use this necklace for their reduction in \autoref{fact:lower-bound-bart}.

What if the class $\C$, instead, does \emph{not} contain arbitrarily long uniform $\F$-necklaces? In this case, we cannot use the same reduction as in \autoref{fact:lower-bound-bart}. Nonetheless, we show that we can still give a lower bound,  depending on $\EDF(\C)$, on the degree of a polynomial kernel for \Fdeletion. This will be done in this subsection via a reduction from $d$-{\sc CNF-SAT}.

We will need the following result of Dell and van Melkebeek~\cite{DellM14}.

\begin{proposition}[Dell and van Melkebeek~\cite{DellM14}]\label{prop:lower-bound-d-SAT}
Let $d \geq 3$ be an integer. For any $\varepsilon > 0$,
the $d$-{\sc CNF-SAT} problem parameterized by the number of variables, denoted by $n$, does not admit a polynomial compression with bitsize $\bigO(n^{d-\varepsilon})$ unless $\NP \subseteq \coNP/\poly$.
\end{proposition}

\newrefcommand{\nm}{\textsc{nm}_\F}
We will also need to show an explicit lower bound on the size of a uniform $\F$-necklace in the class $\C$, when $\C$ has large elimination distance to $\F$-minor free graphs. Dekker and Jansen~\cite{FVS-via-EDF-DAM} already state that there exists a function to this effect by a combination of their Lemma 12 and Proposition 7. Let $\nm(G)$ be the length of the longest $\F$-necklace in $G$.

\begin{proposition}[{\fixedspacingcite[Lemma 12]{FVS-via-EDF-DAM}}]
    \label{fact:ed-bounded-by-g-of-nm}
    For any finite collection $\mathcal F$ of connected planar graphs, there exists a polynomial function $g$ such that for any graph $G$,
    \[
        \EDF(G) \leq g(\nm(G)).
    \]
\end{proposition}

\begin{proposition}[{\fixedspacingcite[Proposition 7]{FVS-via-EDF-DAM}}]
    \label{fact:long-necklaces-imply-long-uniform-necklaces}
    Let $\F$ be a finite collection of connected graphs.
    Let $\C$ be a minor-closed graph family that contains arbitrarily long $\F$-necklaces.
    Then $\C$ also contains arbitrarily long uniform $\F$-necklaces.
    Moreover, there exists a graph $H \in \F$ and vertices $\leftEndpoint, \rightEndpoint \in V(H)$ such that all uniform necklaces with structure $(H, \leftEndpoint, \rightEndpoint)$ are contained in $\C$.
\end{proposition}

We analyze their proofs of \autoref{fact:ed-bounded-by-g-of-nm} and \autoref{fact:long-necklaces-imply-long-uniform-necklaces} in depth to obtain an explicit lower bound for the size of uniform $\F$-necklaces in $\C$.

\newrefcommand{\necklaceLength}{\mathsf{necklaceLength}_\F}
\begin{lemma}
    \label{fact:big-edf-big-necklaces}
    Let $\F$ be a finite collection of connected graphs, and let $\C$ be a minor-closed graph class such that $\EDF(\C) \geq d$. There exists a function $\necklaceLength: \mathds{N} \to \mathds{N}$ with
    \[
    \necklaceLength(t) \in \Omega\parens*{
        \frac{
            \sqrt[11]{t}
        }{
            \norm{\F}
        }
    }
    \]
    such that $\C$ contains an $\F$-necklace of length at least $\necklaceLength(d)$.
\end{lemma}
\begin{proof}
    \newrefcommand{\twBound}{p}
    Let $G$ be a graph in $\C$ such that $\EDF(G) \geq d$. In the proof of \autoref{fact:ed-bounded-by-g-of-nm}, Dekker and Jansen define $g$ as
    \[
        g(t) \coloneq \sum_{i=1}^t f_{\F}(t),
    \]
    where $f_{\F}(\nm(G))$ is an upper bound for the size of a set $X$ such that $\nm(G \setminus X) < \nm(G)$. This function $f_{\F}$ is given in their Lemma 9 \cite{FVS-via-EDF-DAM}, and is defined as a bound on the treewidth of $G$ as a function of $\nm(G)$. They give a bound on the treewidth of $G$ as a function of $\nm(G)$ \textsl{and} $\norm{\F}$ on their Lemma 10, where they state that there exists a polynomial $\twBound: \nat \to \nat$ with $\twBound(n) \in \bigO(n^9 \polylog n)$ such that \[\tw(G) < \twBound(4 \cdot \norm\F (\nm(G) + 1)).\]
    Tracing back the steps, we have that
    \begin{align*}
        \tw(G) &\in \bigO([4 \cdot \norm\F (\nm(G) + 1)]^9 \cdot \polylog [4 \cdot \norm\F (\nm(G) + 1)])\\
        &\subseteq \bigO(\norm{\F}^{10} \cdot \nm(G)^{10}).
    \end{align*}
    Then, by \autoref{fact:ed-bounded-by-g-of-nm} we obtain that
    \begin{align*}
        \EDF(G) &\leq g(\nm(G))\\
        &\leq \nm(G) \cdot f_{\F}(\nm(G))\\
        &\in \bigO(\nm(G)^{11} \cdot \norm{\F}^{10}).
    \end{align*}
    Thus, we have that $G$ contains as a minor an $\F$-necklace of size
    \begin{align*}
    \necklaceLength(\EDF(G)) &\in \Omega\parens*{\sqrt[11]{\frac{\EDF(G)}{\norm{\F}^{10}}}}\\
    &\subseteq \Omega\parens*{
        \frac{\sqrt[11]{\EDF(G)}}{\norm{\F}}
    }.\qedhere
    \end{align*}
\end{proof}

\begin{lemma}
    \label{fact:big-edf-big-uniform-necklaces}
    Let $\F$ be a finite collection of connected graphs, and let $\C$ be a minor-closed graph class such that $\EDF(\C) \geq d$. There exists a function $h_\F: \mathds{N} \to \mathds{N}$ with
    \[
    h_\F(t) \in \Omega\parens*{
        \frac{
            \sqrt[11]{t}
        }{
            \abs\F\cdot \norm{\F}^3
        }
    }
    \]
    such that $\C$ contains a uniform $\F$-necklace of length at least $h_\F(d)$.
\end{lemma}
\begin{proof}
    Let $\necklaceLength$ be the function defined in \autoref{fact:big-edf-big-uniform-necklaces}.  Let $G$ be a graph in $\C$ such that $\EDF(G) \geq d$. By \autoref{fact:big-edf-big-necklaces}, $G$ contains as a minor an $\F$-necklace of size at least $\necklaceLength(\EDF(G))$. We now need to analyze the proof of \autoref{fact:long-necklaces-imply-long-uniform-necklaces} to obtain a lower bound on the size of a uniform $\F$-necklace that is a minor of $G$.

    The proof begins by showing that there exists a graph $H \in \F$ such that $G$ contains an $\set{H}$-necklace of size at least
    \begin{equation}\label{eq:H-necklace-length}
        \frac{\nm(G)}{\abs{\F}}
    \end{equation}
    as a minor.

    % Note that the number of non-isomorphic graphs in $\F$ is bounded by a function of the size of its biggest graph. Therefore, we have that $G$ contains an $\set{H}$-necklace-minor of size at least
    % \begin{equation}\label{eq:H-necklace-length}
    %     \frac{\nm(G)}{\numberof(0, 0, \norm{\F}, 0)}.
    % \end{equation}

    Then, having an $\set{H}$-necklace of size $t$, the proof shows that there exists a uniform $\set{H}$-necklace that is a minor of $G$ of size at least
    \begin{equation}\label{eq:uniform-necklace-from-necklace}
        \ceil*{\frac{t-2}{\abs{V(H)}^2}}.
    \end{equation}

    Replacing $t$ in \autoref{eq:uniform-necklace-from-necklace} with \autoref{eq:H-necklace-length} and upper bounding $\abs{V(H)}$ by $\norm\F$ we obtain that $G$ contains as a minor a uniform $\set{H}$-necklace of size at least
    \begin{equation}\label{eq:uniform-necklace-length}
      \ceil*{
        \frac{
            \frac{\nm(G)}{\abs{\F}}-2
        }{
            \norm{\F}^2
        }
      } = \ceil*{
        \frac{
          \nm(G)
        }{
          \abs\F \cdot \norm{\F}^2
        } - \frac{2}{\norm{\F}^2}
      }.
    \end{equation}

    Finally, inserting $\necklaceLength(\EDF(G))$ into \autoref{eq:uniform-necklace-length} we conclude that $G$ has a uniform $\F$-necklace-minor of size at least
    \begin{align*}
        h_\F(\EDF(G)) &\coloneq \ceil*{
            \frac{
                \necklaceLength(\EDF(G))
            }{
                \abs\F \cdot \norm{\F}^2
            } - \frac{2}{\norm{\F}^2}
        }\\
        &\in \Omega\parens*{
            \frac{
                \frac{\sqrt[11]{\EDF(G)}}{\norm{\F}}
            }{
                \abs\F\cdot \norm{\F}^2
            }
        }\\
        &= \Omega\parens*{
            \frac{
                \sqrt[11]{\EDF(G)}
            }{
                \abs\F\cdot \norm{\F}^3
            }
        }.
    \end{align*}

    As $\C$ is minor closed and contains $G$, it also contains a uniform $\F$-necklace of size at least $h_\F(\EDF(G))$, which is what we wanted to show.
\end{proof}

With this, we prove the following result.

\begin{theorem}[Kernel degree lower bound]\label{fact:degree-lower-bound}
Let $\C$ be a minor-closed class of graphs closed under disjoint union%\igm{as discussed below, probably we need to add the hypothesis that ${\mathcal C}$ is CLOSED UNDER DISJOINT UNION}\ericm{Done!}
and let $\F$ be a finite set of biconnected graphs on at least three vertices containing at least one planar graph. Define $\eta \coloneqq \EDF({\mathcal C})$, and let $h_\F: \mathds{N} \to \mathds{N}$ be the function defined in \autoref{fact:big-edf-big-uniform-necklaces}.

If $\NP \not\subseteq \coNP/\poly$, then \Fdeletion does not admit a polynomial kernel of bitsize
$
    \bigO_\F(
        \abs{X}^{h_\F(\eta)-\varepsilon}
    )
$
for any $\varepsilon > 0$, where $X$ is a given ${\mathcal C}$-modulator of the input graph.
%\ig{this function $f$ comes from the function given by \autoref{prop:long-uniform-necklaces}, we just have to make the calculations going through the proof of Bart. Please state a concrete upper bound on this function $f$ in the statement of the theorem.  I think that the function of the elimination distance is polynomial in the length of the desired necklace, mainly given by the polynomial Grid Theorem of Chuzhoy and Tan~\cite{tighter-excluded-grid-theorem} (then, $f$ will be just the inverse of that polynomial of degree 9). Also, please make clear what we mean by ``size'' in this statement (number of vertices, bit size?}\eric{Done!}
\end{theorem}

\begin{proof}
We will describe a linear parameter transformation from the $d$-\textsc{CNF-SAT} problem parameterized by the number of variables $n$ to the \Fdeletion problem parameterized by the size of a given $\C$-modulator $X$, where $h_\F(\EDF(\C)) = d$. By \autoref{prop:lower-bound-d-SAT}, this will be enough to prove the statement.

Following the framework of the previous reductions, we will be using copies of a uniform $\F$-necklace as the connected components of $G \setminus X$.
In contrast to them, we cannot simply assume that $\C$ contains an arbitrarily big uniform $\F$-necklace and take a minor of it to form $G \setminus X$. This is why we need the condition that $\C$ be closed under disjoint union.

Let $F$ be a $d$-\textsc{CNF-SAT} formula with $n$ variables. Consider a graph $H \in \F$ such that $\C$ contains a uniform $\set{H}$-necklace $N$ of size at least $d$, ensured to exist by \autoref{fact:big-edf-big-uniform-necklaces}. As $\C$ is closed under disjoint union, any disjoint union of copies of $N$ is contained in $\C$. Note that taking the disjoint union of copies of $N$ does not increase the elimination distance to $\F$-minor free graphs, and thus this does not contradict the existence of such a class $\C$.

We then can follow the same reduction as in \autoref{sec:lower-bound-Bart} using $N$ as the connected components of $G \setminus X$---or minors of $N$, if a clause has less than $d$ literals---to obtain a linear parameter transformation, as the size of $X$ is linear on $n$ when $\F$ is fixed.
\end{proof}

The reduction in \autoref{fact:degree-lower-bound} also applies to the families $\F$ considered in \autoref{fact:lower-bound-not-biconnected} and \autoref{fact:lower-bound-treelike}, since essentially the only change is to reduce from $d$-{\sc CNF-SAT} instead of {\sc CNF-SAT}. Thus, we also obtain a lower bound on the degree of the polynomial kernels for these families $\F$.

If we analyze the reduction in \autoref{fact:degree-lower-bound} thoroughly, we can see that the $\F$-necklaces present in $G \setminus X$ do not actually need to be uniform for the proof to go through. In fact, in a non-uniform $\F$-necklace, assuming that the graphs in $\F$ are biconnected and minor-minimal, taking one vertex from each bead is enough to hit all $\F$-minors in the necklace. Thus, we could improve the result replacing the function $h_\F$ given by \autoref{fact:big-edf-big-uniform-necklaces} with the function $\necklaceLength$ given by \autoref{fact:big-edf-big-necklaces}, therefore showing for these families $\F$ that \Fdeletion does not admit a polynomial kernel of bitsize
\[
    \bigO\parens*{|X|^{\necklaceLength(\EDF(\C)) - \varepsilon}}
\]
for any $\varepsilon > 0$. This change also works for the families $\F$ defined in \autoref{fact:lower-bound-treelike}; we omit the details for brevity. However, it does not necessarily work for the ones considered in \autoref{fact:lower-bound-not-biconnected}---the proof of the reduction for these families relies on the fact that the same biconnected component $B$ contains the left endpoint in each bead of the necklace, which is not necessarily the case for non-uniform necklaces.

Summarizing, we have the following:
\begin{itemize}
  \item Our upper bound is of the form $\bigO(|X|^{g(\EDF(\C), \norm\F)})$ for some function $g$ that \emph{increases} when $\norm\F$ increases and $\EDF(\C)$ is fixed, as proven in \autoref{fact:explicit-bound-kernel-size}.
  \item Our lower bound is of the form $\bigO(|X|^{f(\EDF(\C), \norm\F)-\varepsilon})$ for some function $f$ that \emph{decreases} when $\norm\F$ increases and $\EDF(\C)$ is fixed, as proven in \autoref{fact:degree-lower-bound}.
\end{itemize}

One could wonder whether this gap regarding $\norm\F$ between $g$ and $f$ could be closed. In fact, when considering families $\F$ that are not necessarily all biconnected but that do contain a planar graph, it cannot. Indeed, consider the case where $\EDF(\C) = 0$. Here, the parameter is just the size of a given modulator to an $\F$-minor-free graph. As mentioned in \autoref{sec:consequences}, when $\F$ contains a planar graph, this problem is equivalent to \Fdeletion parameterized by the solution size $k$. Giannopoulou et al.~\cite{GiannopoulouJLS17} proved that, for this problem, there exist classes of families $\F$ for which a uniform kernel exists, and other classes for which it does not. Specifically, they prove the following:
\begin{enumerate}
    \item\label{item:treewidth} \textsc{Treewidth-$t$ Vertex Deletion} parameterized by the solution size $k$ does not admit a kernel of size $\bigO(k^{\frac{t}{4} -\varepsilon})$ for any $\varepsilon > 0$, unless $\NP \subseteq \coNP/\poly$ \cite[Theorem 1.1]{GiannopoulouJLS17}.
    \item\label{item:treedepth} \textsc{Treedepth-$t$ Vertex Deletion} parameterized by the solution size $k$ admits a kernel of size $\bigO(k^6)$ for every fixed $t$ \cite[Theorem 1.2]{GiannopoulouJLS17}.
\end{enumerate}

The first item implies that the degree of our kernel cannot depend solely on $\EDF(\C)$, as there is no constant upper bound on the degree of the kernel that holds for every \textsc{Treewidth-$t$ Vertex Deletion} problem. On the other hand, \autoref{item:treedepth} hints at the fact that the degree of the kernel in the lower bound cannot be an increasing function on $\norm\F$, as the obstructions for graphs of treedepth $t$ grow with $t$. Nonetheless, contrary to the case of treewidth-$t$ graphs, the obstructions for graphs of treedepth $t$ are not all biconnected, as they contain a path for every $t$. It may thus be the case that for all biconnected families $\F$ there does exist a lower bound on the degree of the kernel that also grows with $\norm\F$. In this case, the gap between the upper and lower bounds may be closed.

Notice that, contrary to the case of $\norm\F$, the degree of the polynomial kernels \emph{must} grow with $\EDF(\C)$ when all the graphs in $\F$ are biconnected and at least one is a planar graph, or when considering the families $\F$ defined in \autoref{fact:lower-bound-not-biconnected} or \autoref{fact:lower-bound-treelike}. This means, interestingly, that the degree of the kernel behaves very differently with respect to $\norm\F$ and $\EDF(\C)$:
\begin{itemize}
\item For some classes of families $\F$, the degree grows with $\norm\F$, and for others it does not.
\item For every one of the families $\F$ for which we give a lower bound, the degree grows with $\EDF(\C)$.
\end{itemize}

\newpage
\section{Discussion and further research}
\label{sec:conclusions}

The most natural open problem that pops up from our work is to unveil the right kernelization dichotomies for the (connected) families $\F$ that are neither covered by \autoref{fact:dichotomy} nor by \autoref{fact:lower-bound-not-biconnected}. For instance, the panorama is already hazy for $\F = \{P_3\}$, the path on three vertices, which is not biconnected. Does bridge-depth play a role in the dichotomy for this problem?

Another apparently challenging problem is to get rid of the connectivity assumption about the graphs in the collection $\F$ in \autoref{fact:poly-kernel}. This is crucially exploited several times in our approach, which strongly builds on the one by Jansen and Pieterse for treedepth~\cite{minor-hitting}. Note that the case where $\F$ contains disconnected graphs in indeed relevant, as for example it is well-known that the minor obstruction set of any surface of positive genus contains disconnected graphs~\cite{book-surfaces}.

Our main result (\autoref{fact:poly-kernel}) fits within the active line of work of using {\em hybrid parameterizations}, which
simultaneously capture the connectivity structure of the
input instance (typically, via a width parameter) and properties of its optimal solutions; see~\cite{vertex-deletion-parameterized-by-elimination-distance} and the references therein for a complete account. One of the main takeaways of the breakthrough article by Jansen, de Kroon, and W{\l}odarczyk~\cite{vertex-deletion-parameterized-by-elimination-distance} is that, for many natural vertex-deletion problems to a graph class ${\mathcal C}$, including \Fdeletion, \FPT algorithms parameterized by the solution size can be lifted to \FPT algorithms parameterized by the elimination distance to ${\mathcal C}$. With this viewpoint, \autoref{fact:poly-kernel} can be seen as an analog to this lifting result of~\cite{vertex-deletion-parameterized-by-elimination-distance} with respect to the existence of polynomial kernels. In fact, the result of~\cite{vertex-deletion-parameterized-by-elimination-distance} even holds for a more general parameter recently defined by Eiben et al.~\cite{H-treewidth} called \emph{${\mathcal C}$-treewidth}, which is a common generalization of treewidth and elimination distance. Thus, for the families $\F$ not covered by \autoref{fact:dichotomy}, we may hope to generalize \autoref{fact:poly-kernel} to the parameterization by the size of a modulator to graphs with bounded ${\mathcal H}$-treewidth, where ${\mathcal H}$ is the class of $\F$-minor-free graphs. However, this looks implausible, as Cygan et al.~\cite{treewidth-modulator-parameterized-by-treewidth-modulator} showed that already the ``simplest'' case of \textsc{Vertex Cover} ($\F = \{K_2\}$) does not admit a polynomial kernel when parameterized by the size of a modulator to graphs with treewidth two, assuming $\NP \not\subseteq \coNP/\poly$.

Another way we may try to go beyond \autoref{fact:poly-kernel} is by replacing the \Fdeletion problem with its (induced) subgraph counterpart, where we want to delete vertices to make the graph $\F$-subgraph-free (resp. $\F$-induced-subgraph-free). However, Bougeret, Jansen, and Sau~\cite{F-subgraph-deletion-kernel-dichotomies-LIPIcs} recently showed that, for some choices of $\F$, these problems do not admit polynomial kernels when parameterized by the size of a modulator to graphs with bounded treedepth, unless $\NP \subseteq \coNP/\poly$. As the elimination distance to any graph class is not greater than the treedepth, these problems also do not admit polynomial kernels when parameterized by the size of a modulator to graphs with bounded $\EDF$.

The gap between our upper and lower bounds on the degree of the polynomial kernels could also be narrowed. As discussed in \autoref{sec:lower-bounds-degree}, we already know that the degree of the kernel in the upper bound must depend on $\norm\F$, while in the lower bound it cannot. Nonetheless, we could try to close the gap with respect to $\EDF(\C)$, as it appears in both bounds.
Additionally, recall that the obstructions for treedepth-$t$ graphs are not all biconnected. Could we maybe find a class of biconnected families $\F_t$ for which there exists a uniform polynomial kernel of degree independent of $\norm{\F_t}$, as is the case for the families of treedepth-$t$ graphs? Or is every polynomial kernel for every class of families $\F_t$ forced to have a degree that depends on $\norm{\F_t}$?

\newpage
\bibliography{references}

\begin{thebibliography}{10}

\bibitem{AgrawalKLPRSZ22delet}
Akanksha Agrawal, Lawqueen Kanesh, Daniel Lokshtanov, Fahad Panolan, M.~S.
  Ramanujan, Saket Saurabh, and Meirav Zehavi.
\newblock {Deleting, Eliminating and Decomposing to Hereditary Classes Are All
  FPT-Equivalent}.
\newblock In {\em Proc. of the 32st Annual ACM-SIAM Symposium on Discrete
  Algorithms (SODA)}, pages 1976--2004, 2022.
\newblock \href {https://doi.org/10.1137/1.9781611977073.79}
  {\path{doi:10.1137/1.9781611977073.79}}.

\bibitem{AoikeGHKKKKO22}
Yuuki Aoike, Tatsuya Gima, Tesshu Hanaka, Masashi Kiyomi, Yasuaki Kobayashi,
  Yusuke Kobayashi, Kazuhiro Kurita, and Yota Otachi.
\newblock An improved deterministic parameterized algorithm for cactus vertex
  deletion.
\newblock {\em Theory of Computing Systems}, 66(2):502--515, 2022.
\newblock \href {https://doi.org/10.1007/S00224-022-10076-X}
  {\path{doi:10.1007/S00224-022-10076-X}}.

\bibitem{minor-representative-families}
Julien Baste, Ignasi Sau, and Dimitrios~M. Thilikos.
\newblock Hitting minors on bounded treewidth graphs. iv. an optimal algorithm.
\newblock {\em SIAM Journal on Computing}, 52(4):865–912, July 2023.
\newblock \href {https://doi.org/10.1137/21m140482x}
  {\path{doi:10.1137/21m140482x}}.

\bibitem{FPT-equals-kernel}
Hans~L. Bodlaender.
\newblock Kernelization: New upper and lower bound techniques.
\newblock In {\em Proc. 4th {IWPEC}}, pages 17--37, 2009.
\newblock \href {https://doi.org/10.1007/978-3-642-11269-0_2}
  {\path{doi:10.1007/978-3-642-11269-0_2}}.

\bibitem{BodlaenderTY11}
Hans~L. Bodlaender, St{\'{e}}phan Thomass{\'{e}}, and Anders Yeo.
\newblock Kernel bounds for disjoint cycles and disjoint paths.
\newblock {\em Theoretical Computer Science}, 412(35):4570--4578, 2011.
\newblock \href {https://doi.org/10.1016/j.tcs.2011.04.039}
  {\path{doi:10.1016/j.tcs.2011.04.039}}.

\bibitem{vertex-cover-bridge-depth}
Marin Bougeret, Bart M.~P. Jansen, and Ignasi Sau.
\newblock Bridge-depth characterizes which minor-closed structural
  parameterizations of vertex cover admit a polynomial kernel.
\newblock {\em {SIAM} Journal on Discrete Mathematics}, 36(4):2737--2773, 2022.
\newblock \href {https://doi.org/10.1137/21m1400766}
  {\path{doi:10.1137/21m1400766}}.

\bibitem{F-subgraph-deletion-kernel-dichotomies-LIPIcs}
Marin Bougeret, Bart M.~P. Jansen, and Ignasi Sau.
\newblock Kernelization dichotomies for hitting subgraphs under structural
  parameterizations.
\newblock In {\em Proc. of the 51st International Colloquium on Automata,
  Languages, and Programming (ICALP 2024)}, volume 297 of {\em LIPIcs}, pages
  33:1--33:20, 2024.
\newblock \href {https://doi.org/10.4230/LIPIcs.ICALP.2024.33}
  {\path{doi:10.4230/LIPIcs.ICALP.2024.33}}.

\bibitem{BougeretS17}
Marin Bougeret and Ignasi Sau.
\newblock How much does a treedepth modulator help to obtain polynomial kernels
  beyond sparse graphs?
\newblock In Daniel Lokshtanov and Naomi Nishimura, editors, {\em Proc. 12th
  IPEC}, volume~89 of {\em LIPIcs}, pages 10:1--10:13. Schloss Dagstuhl -
  Leibniz-Zentrum fuer Informatik, 2017.
\newblock \href {https://doi.org/10.4230/LIPIcs.IPEC.2017.10}
  {\path{doi:10.4230/LIPIcs.IPEC.2017.10}}.

\bibitem{vertex-cover-treedepth-poly-kernel-Algorithmica}
Marin Bougeret and Ignasi Sau.
\newblock How much does a treedepth modulator help to obtain polynomial kernels
  beyond sparse graphs?
\newblock {\em Algorithmica}, 81(10):4043--4068, 2019.
\newblock \href {https://doi.org/10.1007/s00453-018-0468-8}
  {\path{doi:10.1007/s00453-018-0468-8}}.

\bibitem{isomorphism-elimination-distance}
Jannis Bulian and Anuj Dawar.
\newblock Graph isomorphism parameterized by elimination distance to bounded
  degree.
\newblock {\em Algorithmica}, 75(2):363--382, 2016.
\newblock \href {https://doi.org/10.1007/s00453-015-0045-3}
  {\path{doi:10.1007/s00453-015-0045-3}}.

\bibitem{elimination-distance}
Jannis Bulian and Anuj Dawar.
\newblock Fixed-parameter tractable distances to sparse graph classes.
\newblock {\em Algorithmica}, 79(1):139--158, 2017.
\newblock \href {https://doi.org/10.1007/s00453-016-0235-7}
  {\path{doi:10.1007/s00453-016-0235-7}}.

\bibitem{elimination-distance-fpt}
Jannis Bulian and Anuj Dawar.
\newblock Fixed-parameter tractable distances to sparse graph classes.
\newblock {\em Algorithmica}, 79(1):139--158, 2017.
\newblock URL: \url{https://doi.org/10.1007/s00453-016-0235-7}, \href
  {https://doi.org/10.1007/S00453-016-0235-7}
  {\path{doi:10.1007/S00453-016-0235-7}}.

\bibitem{tighter-excluded-grid-theorem}
Julia Chuzhoy and Zihan Tan.
\newblock Towards tight(er) bounds for the excluded grid theorem.
\newblock {\em Journal of Combinatorial Theory, Series B}, 146:219–265,
  January 2021.
\newblock \href {https://doi.org/10.1016/j.jctb.2020.09.010}
  {\path{doi:10.1016/j.jctb.2020.09.010}}.

\bibitem{CyganFKLMPPS15}
Marek Cygan, Fedor~V. Fomin, Lukasz Kowalik, Daniel Lokshtanov, D{\'{a}}niel
  Marx, Marcin Pilipczuk, Michal Pilipczuk, and Saket Saurabh.
\newblock {\em Parameterized Algorithms}.
\newblock Springer, 2015.
\newblock \href {https://doi.org/10.1007/978-3-319-21275-3}
  {\path{doi:10.1007/978-3-319-21275-3}}.

\bibitem{treewidth-modulator-parameterized-by-treewidth-modulator}
Marek Cygan, Daniel Lokshtanov, Marcin Pilipczuk, Michal Pilipczuk, and Saket
  Saurabh.
\newblock On the hardness of losing width.
\newblock {\em Theory of Computing Systems}, 54(1):73--82, 2014.
\newblock \href {https://doi.org/10.1007/s00224-013-9480-1}
  {\path{doi:10.1007/s00224-013-9480-1}}.

\bibitem{FVS-via-EDF-WG}
David Dekker and Bart M.~P. Jansen.
\newblock {Kernelization for Feedback Vertex Set via Elimination Distance to a
  Forest}.
\newblock In {\em Proc. of the 48th International Workshop Graph-Theoretic
  Concepts in Computer Science (WG)}, volume 13453 of {\em LNCS}, pages
  158--172, 2022.
\newblock \href {https://doi.org/10.1007/978-3-031-15914-5\_12}
  {\path{doi:10.1007/978-3-031-15914-5\_12}}.

\bibitem{FVS-via-EDF-arxiv}
David Dekker and Bart M.~P. Jansen.
\newblock Kernelization for feedback vertex set via elimination distance to a
  forest, 2022.
\newblock URL: \url{https://arxiv.org/abs/2206.04387}, \href
  {https://arxiv.org/abs/2206.04387} {\path{arXiv:2206.04387}}.

\bibitem{FVS-via-EDF-DAM}
David J.~C. Dekker and Bart M.~P. Jansen.
\newblock Kernelization for feedback vertex set via elimination distance to a
  forest.
\newblock {\em Discrete Applied Mathematics}, 346:192--214, 2024.
\newblock \href {https://doi.org/10.1016/j.dam.2023.12.016}
  {\path{doi:10.1016/j.dam.2023.12.016}}.

\bibitem{DellM14}
Holger Dell and Dieter van Melkebeek.
\newblock Satisfiability allows no nontrivial sparsification unless the
  polynomial-time hierarchy collapses.
\newblock {\em Journal of the {ACM}}, 61(4):23:1--23:27, 2014.
\newblock \href {https://doi.org/10.1145/2629620} {\path{doi:10.1145/2629620}}.

\bibitem{Diestel16}
Reinhard Diestel.
\newblock {\em Graph Theory}.
\newblock Springer-Verlag, Heidelberg, 5th edition, 2016.

\bibitem{DonkersJ21}
Huib Donkers and Bart M.~P. Jansen.
\newblock {A Turing kernelization dichotomy for structural parameterizations of
  ${\mathcal F}$-Minor-Free Deletion}.
\newblock {\em Journal of Computer and System Sciences}, 119:164--182, 2021.
\newblock \href {https://doi.org/10.1016/J.JCSS.2021.02.005}
  {\path{doi:10.1016/J.JCSS.2021.02.005}}.

\bibitem{DonkersJW22}
Huib Donkers, Bart M.~P. Jansen, and Michal Wlodarczyk.
\newblock Preprocessing for outerplanar vertex deletion: An elementary kernel
  of quartic size.
\newblock {\em Algorithmica}, 84(11):3407--3458, 2022.
\newblock \href {https://doi.org/10.1007/S00453-022-00984-2}
  {\path{doi:10.1007/S00453-022-00984-2}}.

\bibitem{DowneyF13}
Rodney~G. Downey and Michael~R. Fellows.
\newblock {\em Fundamentals of Parameterized Complexity}.
\newblock Texts in Computer Science. Springer, 2013.
\newblock \href {https://doi.org/10.1007/978-1-4471-5559-1}
  {\path{doi:10.1007/978-1-4471-5559-1}}.

\bibitem{H-treewidth}
Eduard Eiben, Robert Ganian, Thekla Hamm, and O~joung Kwon.
\newblock Measuring what matters: A hybrid approach to dynamic programming with
  treewidth.
\newblock {\em Journal of Computer and System Sciences}, 121:57--75, 2021.
\newblock \href {https://doi.org/10.1016/j.jcss.2021.04.005}
  {\path{doi:10.1016/j.jcss.2021.04.005}}.

\bibitem{FioriniJP10}
Samuel Fiorini, Gwena{\"{e}}l Joret, and Ugo Pietropaoli.
\newblock Hitting diamonds and growing cacti.
\newblock In {\em Proc. of the 14th International Conference on Integer
  Programming and Combinatorial Optimization (IPCO)}, volume 6080 of {\em
  LNCS}, pages 191--204, 2010.
\newblock \href {https://doi.org/10.1007/978-3-642-13036-6\_15}
  {\path{doi:10.1007/978-3-642-13036-6\_15}}.

\bibitem{FlumG06}
J{\"{o}}rg Flum and Martin Grohe.
\newblock {\em Parameterized Complexity Theory}.
\newblock Springer-Verlag, 2006.
\newblock \href {https://doi.org/10.1007/3-540-29953-X}
  {\path{doi:10.1007/3-540-29953-X}}.

\bibitem{planar-F-deletion-kernel}
Fedor~V. Fomin, Daniel Lokshtanov, Neeldhara Misra, and Saket Saurabh.
\newblock {Planar ${\cal F}$-Deletion: Approximation, Kernelization and Optimal
  {FPT} Algorithms}.
\newblock In {\em Proc. of the 53rd Annual {IEEE} Symposium on Foundations of
  Computer Science (FOCS)}, 2012.
\newblock \href {https://doi.org/10.1109/FOCS.2012.62}
  {\path{doi:10.1109/FOCS.2012.62}}.

\bibitem{FominLSZ19}
Fedor~V. Fomin, Daniel Lokshtanov, Saket Saurabh, and Meirav Zehavi.
\newblock {\em Kernelization: Theory of Parameterized Preprocessing}.
\newblock Cambridge University Press, 2019.
\newblock \href {https://doi.org/10.1017/9781107415157}
  {\path{doi:10.1017/9781107415157}}.

\bibitem{FominS16}
Fedor~V. Fomin and Torstein J.~F. Str{\o}mme.
\newblock Vertex cover structural parameterization revisited.
\newblock In {\em Proc. of the 42nd International Workshop on Graph-Theoretic
  Concepts in Computer Science (WG)}, volume 9941 of {\em LNCS}, pages
  171--182, 2016.
\newblock \href {https://doi.org/10.1007/978-3-662-53536-3_15}
  {\path{doi:10.1007/978-3-662-53536-3_15}}.

\bibitem{GiannopoulouJLS17}
Archontia~C. Giannopoulou, Bart M.~P. Jansen, Daniel Lokshtanov, and Saket
  Saurabh.
\newblock Uniform kernelization complexity of hitting forbidden minors.
\newblock {\em {ACM} Transactions on Algorithms}, 13(3):35:1--35:35, 2017.
\newblock \href {https://doi.org/10.1145/3029051} {\path{doi:10.1145/3029051}}.

\bibitem{GokeMM20}
Alexander G{\"{o}}ke, D{\'{a}}niel Marx, and Matthias Mnich.
\newblock {Hitting Long Directed Cycles Is Fixed-Parameter Tractable}.
\newblock In {\em Proc. of the 47th International Colloquium on Automata,
  Languages, and Programming (ICALP)}, volume 168 of {\em LIPIcs}, pages
  59:1--59:18.
\newblock \href {https://doi.org/10.4230/LIPICS.ICALP.2020.59}
  {\path{doi:10.4230/LIPICS.ICALP.2020.59}}.

\bibitem{deterministic-approximation-for-planar-F-deletion}
Anupam Gupta, Euiwoong Lee, Jason Li, Pasin Manurangsi, and Michał
  Włodarczyk.
\newblock {\em Losing Treewidth by Separating Subsets}, pages 1731--1749.
\newblock \href {https://doi.org/10.1137/1.9781611975482.104}
  {\path{doi:10.1137/1.9781611975482.104}}.

\bibitem{linear-parameter-transformation}
Danny Hermelin and Xi~Wu.
\newblock Weak compositions and their applications to polynomial lower bounds
  for kernelization.
\newblock In {\em Proc. 23rd SODA}, pages 104--113, 2012.
\newblock \href {https://doi.org/10.1137/1.9781611973099.9}
  {\path{doi:10.1137/1.9781611973099.9}}.

\bibitem{HolsKP22}
Eva{-}Maria~C. Hols, Stefan Disc.tsch, and Astrid Pieterse.
\newblock Elimination distances, blocking sets, and kernels for vertex cover.
\newblock {\em {SIAM} Journal on Discrete Mathematics}, 36(3):1955--1990, 2022.
\newblock \href {https://doi.org/10.1137/20m1335285}
  {\path{doi:10.1137/20m1335285}}.

\bibitem{HolsK17}
Eva{-}Maria~C. Hols and Stefan Kratsch.
\newblock Smaller parameters for vertex cover kernelization.
\newblock In {\em Proc. of the 12th International Symposium on Parameterized
  and Exact Computation (IPEC)}, volume~89 of {\em LIPIcs}, pages 20:1--20:12,
  2017.
\newblock \href {https://doi.org/10.4230/LIPIcs.IPEC.2017.20}
  {\path{doi:10.4230/LIPIcs.IPEC.2017.20}}.

\bibitem{HolsKP19}
Eva{-}Maria~C. Hols, Stefan Kratsch, and Astrid Pieterse.
\newblock Elimination distances, blocking sets, and kernels for vertex cover.
\newblock In {\em Proc. 37nd STACS}, volume 154 of {\em LIPIcs}, pages
  36:1--36:14, 2020.
\newblock \href {https://doi.org/10.4230/LIPIcs.STACS.2020.36}
  {\path{doi:10.4230/LIPIcs.STACS.2020.36}}.

\bibitem{Bart-personal25}
Bart M.~P. Jansen.
\newblock Personal communication, 2025.

\bibitem{quo-vadis25}
Bart M.~P. Jansen.
\newblock Quo vadis, kernelization?
\newblock Invited survey, to appear in {\em Computer Science Review}, special
  issue dedicated to Michael R. Fellows on the occasion of his 70th birthday,
  2025.
\newblock URL: \url{https://www.sciencedirect.com/special-issue/10540QH6GM2}.

\bibitem{JansenB13}
Bart M.~P. Jansen and Hans~L. Bodlaender.
\newblock Vertex cover kernelization revisited - upper and lower bounds for a
  refined parameter.
\newblock {\em Theory of Computing Systems}, 53(2):263--299, 2013.
\newblock \href {https://doi.org/10.1007/s00224-012-9393-4}
  {\path{doi:10.1007/s00224-012-9393-4}}.

\bibitem{vertex-deletion-parameterized-by-elimination-distance}
Bart M.~P. Jansen, Jari J.~H. de~Kroon, and Michal W{\l}odarczyk.
\newblock Vertex deletion parameterized by elimination distance and even less.
\newblock In {\em Proc. of the 53rd Annual {ACM} Symposium on Theory of
  Computing (STOC)}, pages 1757--1769, 2021.
\newblock \href {https://doi.org/10.1145/3406325.3451068}
  {\path{doi:10.1145/3406325.3451068}}.

\bibitem{vertex-deletion-parameterized-by-elimination-distance-arxiv}
Bart M.~P. Jansen, Jari J.~H. de~Kroon, and Michal W{\l}odarczyk.
\newblock Vertex deletion parameterized by elimination distance and even less.
\newblock {\em CoRR}, abs/2103.09715, 2021.
\newblock \href {https://arxiv.org/abs/2103.09715} {\path{arXiv:2103.09715}}.

\bibitem{minor-hitting}
Bart M.~P. Jansen and Astrid Pieterse.
\newblock Polynomial kernels for hitting forbidden minors under structural
  parameterizations.
\newblock {\em Theoretical Computer Science}, 841:124--166, 2020.
\newblock \href {https://doi.org/10.1016/j.tcs.2020.07.009}
  {\path{doi:10.1016/j.tcs.2020.07.009}}.

\bibitem{vertex-planarization-approximate-kernel}
Bart M.~P. Jansen and Michal W{\l}lodarczyk.
\newblock {Lossy Planarization: {A} Constant-Factor Approximate Kernelization
  for Planar Vertex Deletion}.
\newblock {\em {SIAM} Journal on Computing}, 54(1):1--91, 2025.
\newblock \href {https://doi.org/10.1137/22M152058X}
  {\path{doi:10.1137/22M152058X}}.

\bibitem{JoretPSST14}
Gwena{\"{e}}l Joret, Christophe Paul, Ignasi Sau, Saket Saurabh, and
  St{\'{e}}phan Thomass{\'{e}}.
\newblock Hitting and harvesting pumpkins.
\newblock {\em {SIAM} Journal on Discrete Mathematics}, 28(3):1363--1390, 2014.
\newblock \href {https://doi.org/10.1137/120883736}
  {\path{doi:10.1137/120883736}}.

\bibitem{minor-free-recognition-n2}
Ken{-}ichi Kawarabayashi, Yusuke Kobayashi, and Bruce~A. Reed.
\newblock The disjoint paths problem in quadratic time.
\newblock {\em J. Comb. Theory {B}}, 102(2):424--435, 2012.
\newblock \href {https://doi.org/10.1016/J.JCTB.2011.07.004}
  {\path{doi:10.1016/J.JCTB.2011.07.004}}.

\bibitem{LPRRSS16}
Eun~Jung Kim, Alexander Langer, Christophe Paul, Felix Reidl, Peter Rossmanith,
  Ignasi Sau, and Somnath Sikdar.
\newblock Linear kernels and single-exponential algorithms via protrusion
  decompositions.
\newblock {\em {ACM} Transactions on Algorithms}, 12(2):21:1--21:41, 2016.
\newblock \href {https://doi.org/10.1145/2797140} {\path{doi:10.1145/2797140}}.

\bibitem{minor-containment-almost-linear-time}
Tuukka Korhonen, Michał Pilipczuk, and Giannos Stamoulis.
\newblock Minor containment and disjoint paths in almost-linear time.
\newblock In {\em 2024 IEEE 65th Annual Symposium on Foundations of Computer
  Science (FOCS)}, page 53–61. IEEE, October 2024.
\newblock \href {https://doi.org/10.1109/focs61266.2024.00014}
  {\path{doi:10.1109/focs61266.2024.00014}}.

\bibitem{Kratsch18}
Stefan Kratsch.
\newblock A randomized polynomial kernelization for vertex cover with a smaller
  parameter.
\newblock {\em {SIAM} Journal on Discrete Mathematics}, 32(3):1806--1839, 2018.
\newblock \href {https://doi.org/10.1137/16M1104585}
  {\path{doi:10.1137/16M1104585}}.

\bibitem{KratschW12}
Stefan Kratsch and Magnus Wahlstr{\"o}m.
\newblock Representative sets and irrelevant vertices: New tools for
  kernelization.
\newblock In {\em Proc. 53rd FOCS}, pages 450--459, 2012.
\newblock \href {https://doi.org/10.1109/FOCS.2012.46}
  {\path{doi:10.1109/FOCS.2012.46}}.

\bibitem{planar-minor-characterization}
Casimir Kuratowski.
\newblock Sur le problème des courbes gauches en topologie.
\newblock {\em Fundamenta Mathematicae}, 15:271–283, 1930.
\newblock \href {https://doi.org/10.4064/fm-15-1-271-283}
  {\path{doi:10.4064/fm-15-1-271-283}}.

\bibitem{lossy-kernelization}
Daniel Lokshtanov, Fahad Panolan, M.~S. Ramanujan, and Saket Saurabh.
\newblock Lossy kernelization.
\newblock In {\em Proce. of the 49th Annual {ACM} {SIGACT} Symposium on Theory
  of Computing (STOC)}, pages 224--237. {ACM}, 2017.
\newblock \href {https://doi.org/10.1145/3055399.3055456}
  {\path{doi:10.1145/3055399.3055456}}.

\bibitem{MajumdarRR18}
Diptapriyo Majumdar, Venkatesh Raman, and Saket Saurabh.
\newblock Polynomial kernels for vertex cover parameterized by small degree
  modulators.
\newblock {\em Theory of Computing Systems}, 62(8):1910--1951, 2018.
\newblock \href {https://doi.org/10.1007/s00224-018-9858-1}
  {\path{doi:10.1007/s00224-018-9858-1}}.

\bibitem{book-surfaces}
Bojan Mohar and Carsten Thomassen.
\newblock {\em Graphs on Surfaces}.
\newblock Johns Hopkins University Press, 2001.

\bibitem{parameterized-algorithms-vertex-deletion-minor-closed}
Laure Morelle, Ignasi Sau, Giannos Stamoulis, and Dimitrios~M. Thilikos.
\newblock Faster parameterized algorithms for modification problems to
  minor-closed classes.
\newblock {\em TheoretiCS}, Volume 3, August 2024.
\newblock \href {https://doi.org/10.46298/theoretics.24.19}
  {\path{doi:10.46298/theoretics.24.19}}.

\bibitem{Niedermeier06}
Rolf Niedermeier.
\newblock {\em Invitation to Fixed-Parameter Algorithms}.
\newblock Oxford University Press, 2006.
\newblock \href {https://doi.org/10.1093/acprof:oso/9780198566076.001.0001}
  {\path{doi:10.1093/acprof:oso/9780198566076.001.0001}}.

\bibitem{r-pseudoforest-deletion}
Geevarghese Philip, Ashutosh Rai, and Saket Saurabh.
\newblock Generalized pseudoforest deletion: Algorithms and uniform kernel.
\newblock {\em SIAM Journal on Discrete Mathematics}, 32(2):882--901, 2018.
\newblock \href {https://doi.org/10.1137/16M1100794}
  {\path{doi:10.1137/16M1100794}}.

\bibitem{excluded-minor-theorem}
Neil Robertson and Paul~D. Seymour.
\newblock Graph minors. {V}. {E}xcluding a planar graph.
\newblock {\em J. Comb. Theory, Ser. B}, 41(1):92 -- 114, 1986.
\newblock \href {https://doi.org/10.1016/0095-8956(86)90030-4}
  {\path{doi:10.1016/0095-8956(86)90030-4}}.

\bibitem{minor-free-recognition-n3}
Neil Robertson and Paul~D. Seymour.
\newblock Graph minors. {XIII.} {T}he disjoint paths problem.
\newblock {\em J. Comb. Theory, Ser. B}, 63(1):65--110, 1995.
\newblock \href {https://doi.org/10.1006/jctb.1995.1006}
  {\path{doi:10.1006/jctb.1995.1006}}.

\bibitem{k-apices-LIPIcs}
Ignasi Sau, Giannos Stamoulis, and Dimitrios~M. Thilikos.
\newblock An fpt-algorithm for recognizing k-apices of minor-closed graph
  classes.
\newblock In Artur Czumaj, Anuj Dawar, and Emanuela Merelli, editors, {\em 47th
  International Colloquium on Automata, Languages, and Programming (ICALP
  2020)}, volume 168 of {\em Leibniz International Proceedings in Informatics
  (LIPIcs)}, pages 95:1--95:20, Dagstuhl, Germany, 2020. Schloss Dagstuhl –
  Leibniz-Zentrum für Informatik.
\newblock \href {https://doi.org/10.4230/LIPIcs.ICALP.2020.95}
  {\path{doi:10.4230/LIPIcs.ICALP.2020.95}}.

\bibitem{SauST22}
Ignasi Sau, Giannos Stamoulis, and Dimitrios~M. Thilikos.
\newblock {\emph{k}-apices of Minor-closed Graph Classes. {II.} Parameterized
  Algorithms}.
\newblock {\em {ACM} Transactions on Algorithms}, 18(3):21:1--21:30, 2022.
\newblock \href {https://doi.org/10.1145/3519028} {\path{doi:10.1145/3519028}}.

\bibitem{SauST23}
Ignasi Sau, Giannos Stamoulis, and Dimitrios~M. Thilikos.
\newblock {\emph{k}-apices of minor-closed graph classes. I. Bounding the
  obstructions}.
\newblock {\em Journal of Combinatorial Theory, Series {B}}, 161:180--227,
  2023.
\newblock \href {https://doi.org/10.1016/J.JCTB.2023.02.012}
  {\path{doi:10.1016/J.JCTB.2023.02.012}}.

\bibitem{open-problems-worker}
Saket Saurabh.
\newblock Open problems from the workshop on kernelization (worker 2019, part
  1).
\newblock 2019.
\newblock URL: \url{https://youtu.be/vCjG5zGjQr4?t=124}.

\bibitem{treewidth-2-deletion}
Jeroen~L.G. Schols.
\newblock {Kernelization for Treewidth-2 Vertex Deletion}.
\newblock {\em CoRR}, abs/2203.10070, 2022.
\newblock \href {https://arxiv.org/abs/2203.10070} {\path{arXiv:2203.10070}}.

\bibitem{Tsur23}
Dekel Tsur.
\newblock Faster deterministic algorithm for cactus vertex deletion.
\newblock {\em Information Processing Letters}, 179:106317, 2023.
\newblock \href {https://doi.org/10.1016/J.IPL.2022.106317}
  {\path{doi:10.1016/J.IPL.2022.106317}}.

\end{thebibliography}

\newpage
\appendix

\section{Explicit upper bound on the size of the kernels}
\label{sec:explicit-upper-bounds}

In this section we give an explicit (large) upper bound on the size of the kernels obtained in \autoref{fact:poly-kernel}.

\newrefcommand{\mF}{m_\F}
\renewrefcommand{\nF}{n_\F}
\newcommand{\maxLabels}{\mathsf{maxLabels}}
We begin by upper-bounding the size of the sets $\Q^\star$ and $\RN$ in \namedtheoremref{fact:extra-base-case}.
Define $\mF \coloneqq \max_{H \in \F} \abs{E(H)}$, $\nF \coloneqq \min_{H \in \F} \abs{V(H)}$, and $\norm{\F} \coloneqq \max_{H \in \F} \abs{V(H)}$.
\newrefcommand{\RNBC}{\abs{\RN}_{\mathsf{bc}}}
\begin{lemma}\label{fact:bound-RN-base-case}
    Let $G$ and $S$ be as in \namedtheoremref{fact:extra-base-case}. Under the conditions of \autoref{fact:extra-base-case}, the size of $\RN$ is upper-bounded by
    \begin{align*}
        \RNBC &\coloneqq 2^{\numberof(\abs{\Gboundary} \cdot (\abs{\Gboundary} + 2) \cdot \nF, \abs{\Gboundary}, \mF + \abs{\Gboundary} + 1, \nF)}\\
        &\leq 2^{2^{\polynomial(\abs{\Gboundary} + \nF)}}.
    \end{align*}
\end{lemma}
\begin{proof}
    Following the proof of \autoref{fact:extra-base-case} we can see that
    \begin{equation}
        \label{eq:bound-size-RN}
        \abs{\RN} \leq 2^{\abs*{\folioqt(\GAplusRBS)}}.
    \end{equation}
    Note that the remainders in $\RN$ do not have any $\Q$-minors. Thus, as $\Q$ is $\nF$-saturated, every graph in $\RN$ has less than $\nF$ different labels.
    Combining this observation with the ones made in the proof of \autoref{fact:extra-base-case} we have that
    \begin{equation}\begin{aligned}
        \abs*{\folioqt(\GAplusRBS)} &\leq \numberof(\maxLabels, \abs{\Gboundary}, \mF + \abs{\Gboundary} + 1, \nF)\\
        &= \numberof(\\
        &\qquad    \abs{\Gboundary} \cdot(\nF + \tooManyLabelsReached - 2), \abs{\Gboundary}, \mF + \abs{\Gboundary} + 1, \nF)\\
        &= \numberof(\abs{\Gboundary} \cdot(2\nF +
        \abs{\Gboundary} \cdot (\nF - 1) - 2), \abs{\Gboundary}, \mF + \abs{\Gboundary} + 1, \nF)\\
        &\leq \numberof(\abs{\Gboundary} \cdot
        (\abs{\Gboundary} + 2) \cdot \nF, \abs{\Gboundary}, \mF + \abs{\Gboundary} + 1, \nF)\\
        \label{eq:bound-size-folio-GAplusRBS}
    \end{aligned}\end{equation}

    Restating \autoref{eq:bound-size-RN} in terms of \autoref{eq:bound-size-folio-GAplusRBS} gives the desired bound on the size of $\RN$.
\end{proof}

\newrefcommand{\QBC}{\abs{\Q^\star}_{\mathsf{bc}}}
\begin{lemma}
    \label{fact:bound-Q-base-case}
    Let $G$ and $S$ be as in \namedtheoremref{fact:extra-base-case}. Under the conditions of \autoref{fact:extra-base-case}, the size of $\Q^\star$ is upper-bounded by
    \begin{align*}
        \QBC &\coloneqq \numberof((\abs{\Gboundary} + 1)^2 \cdot \norm{\F}, 0, \mF + 1, (\abs{\Gboundary} + 1)^2 \cdot \norm{\F})\\
        &\leq 2^{\polynomial(\abs{\Gboundary} + \norm{\F})}.
    \end{align*}
\end{lemma}
\begin{proof}
    Following the proof of \autoref{fact:extra-base-case} we can see that the size of $\Q^\star$ is upper-bounded by the number of different graphs with at most $\mF + 1$ vertices and with labels from a set of size at most $\totalMarkedLabels$. In other words,
    \begin{equation}
        \abs{\Q^\star} \leq \numberof(\totalMarkedLabels, 0, \mF + 1, \totalMarkedLabels).
        \label{eq:bound-size-Q-star}
    \end{equation}
    Note that
    \begin{equation}\begin{aligned}
        \totalMarkedLabels &= (\abs{\Gboundary}(\nF - 1) + \nF) \cdot (\abs{\Gboundary} + 1)\\
        &\leq (\abs{\Gboundary} + 1)^2 \cdot \nF.
        \label{eq:bound-size-total-marked-labels}
    \end{aligned}\end{equation}
    Rewriting \autoref{eq:bound-size-Q-star} in terms of \autoref{eq:bound-size-total-marked-labels} gives the desired bound on the size of $\Q^\star$.
\end{proof}

We can now bound the size of the sets $\RN$ and $\Q^\star$ obtained by \autoref{fact:inductive-main-lemma-ed}.

\begin{lemma}[cf. {\cite[Claim 33]{minor-hitting}}]
    \label{fact:explicit-bound-Q-star-inductive}
    Let $G$, $G_A$, and $\Gboundary$ be as in \namedtheoremref{fact:inductive-main-lemma-ed} and define $\eta \coloneqq \EDF(G)$ and $\eta_A \coloneqq \EDF(G_A \setminus S)$. In \autoref{fact:inductive-main-lemma-ed}, the size of the set $\RN$ is upper-bounded by
    \begin{multline*}
        f(\eta_A, \iscon(G_A \setminus S), \mu(G_A,\Pi_A,S), \nu(\Pi_A),\xi(R_B), \norm{\F}, \abs{S}) \coloneqq\\
        \begin{cases}
            1 &\text{if } \sigma > M \text{ or } \sigma \leq 0,\\
            \RNBC &\text{if } \eta_A = 0,\\
            (2C)^{\sigma \cdot [\eta_A \cdot M^{\eta_A} + 1]} \cdot \RNBC^{\sigma \cdot M^{\eta_A - 1}} &\text{if } G_A \setminus S \text{ is disconnected,}\\
            2 \cdot (2C)^{M \cdot [(\eta_A - 1) \cdot M^{\eta_A - 1} + 1]} \cdot \RNBC^{M^{\eta_A - 1}} &\text{otherwise.}
        \end{cases}
    \end{multline*}
    and the size of $\Q^\star$ is upper-bounded by
    \begin{multline*}
        g(\eta_A, \iscon(G_A \setminus S), \mu(G_A,\Pi_A,S), \nu(\Pi_A),\xi(R_B), \norm{\F}, \abs{S}) \coloneqq\\
        \begin{cases}
            1 &\text{if } \sigma > M \text{ or } \sigma \leq 0,\\
            \QBC &\text{if } \eta_A = 0,\\
            \QBC \cdot \bracks*{ 2\cdot (2C)^{(\eta_A - 1)\cdot M^{\eta_A} + M + 1} \cdot \RNBC^{M^{\eta_A - 1}}}^{M \cdot (\eta_A - 1) + \sigma} &\text{if } G_A \setminus S \text{ is disconnected,} \\
            2 \cdot \QBC \cdot \bracks*{ 2\cdot (2C)^{(\eta_A - 2)\cdot M^{\eta_A - 1} + M + 1} \cdot \RNBC^{M^{\eta_A - 2}}}^{M \cdot (\eta_A - 1)} &\text{otherwise.}
        \end{cases}
    \end{multline*}
    where
    \begin{align*}
        C &\coloneqq 2^{\numberof(0, \eta, \eta + \norm{\F}, 0)},\\
        M &\coloneqq \eta + \numberof(0, \eta, \eta + \norm{\F}, 0) + \numberof(\eta \cdot \norm{\F}, \eta, \eta + \norm{\F}^2 + 1, \norm{\F}), \text{ and}\\
        \sigma &\coloneqq \mu(G_A, \Pi_A, S) + \nu(\Pi_A) + \xi(R_B) + 1.
    \end{align*}
\end{lemma}
\begin{proof}
    Notice that we defined $M$ to be smaller than how it was defined by Jansen and Pieterse \cite{minor-hitting} in their Claim 33, replacing $C$ in its definition by $\numberof(0, \eta, \eta + \norm{\F}, 0)$. This allows us to obtain a tighter bound later.

    Let $\Gboundary$ be as in \namedtheoremref{fact:inductive-main-lemma-ed}. Note that $\abs{\Gboundary} \leq \eta$.

    We will follow the proof of Lemma 27 in \cite{minor-hitting} to give explicit bounds to $\abs{\RN}$ and $\abs{\Q^\star}$ step by step. Explicit bounds for these sets in the context of treedepth were given in their Claim 33 \cite{minor-hitting}. A key result enabling their proof is the following.

    \begin{claim}[{\fixedspacingcite[Claim 31]{minor-hitting}}]
        \label{claim:bounds-on-parameters}
        If none of the cases BC2 to BC4 apply, then the following bounds are satisfied:
        \begin{itemize}
        \item $0 \leq \mu(G_A, \Pi_A, S) \leq \abs{S}$,
        \item $0 \leq \nu(\Pi_A) \leq \numberof(0, \abs{S}, \norm{\F} + \abs{S}, 0)$, and
        \item $0 \leq \xi(R_B) \leq \numberof(\abs{S} \cdot \min_{H \in \F}\abs{V(H)}, \abs{S}, \abs{S} + \max_{H \in \Q}\abs{V(H)}, \min_{H\in\F}\abs{V(H)})$.
        \end{itemize}
    \end{claim}

    Using \autoref{claim:bounds-on-parameters} and noting that $\abs{S} \leq \eta$, we conclude the following:
    \begin{claim}
        If $\sigma > M$ or $\sigma \leq 0$, we are in one of the base cases BC2 to BC4, and thus $\abs{\RN} \leq 1$ and $\abs{\Q^\star} \leq 1$.
    \end{claim}

    If $\eta = 0$, then we know by \autoref{fact:bound-RN-base-case} that $\abs{\RN} \leq \RNBC$, and by \autoref{fact:bound-Q-base-case} that $\abs{\Q^\star} \leq \QBC$. We thus assume that we are not in any of the base cases of the induction.

    We will be using the fact that, assuming that $\sigma > 0$ and $\eta_A > 0$, going from a disconnected to a connected graph with the same parameters can only decrease $f$ and $g$. In other words, that
    \begin{align*}
        &f(\eta_A, 1, \mu(G_A,\Pi_A,S), \nu(\Pi_A),\xi(R_B), \norm{\F}, \abs{S}) \leq\\
        &f(\eta_A, 0, \mu(G_A,\Pi_A,S), \nu(\Pi_A),\xi(R_B), \norm{\F}, \abs{S})
    \end{align*}
    and
    \begin{align*}
        &g(\eta_A, 1, \mu(G_A,\Pi_A,S), \nu(\Pi_A),\xi(R_B), \norm{\F}, \abs{S}) \leq\\
        &g(\eta_A, 0, \mu(G_A,\Pi_A,S), \nu(\Pi_A),\xi(R_B), \norm{\F}, \abs{S}).
    \end{align*}

    This can be shown as follows in the case of $f$:
    \begin{align*}
        &f(\eta_A, 1, \mu(G_A,\Pi_A,S), \nu(\Pi_A),\xi(R_B), \norm{\F}, \abs{S})\\
        &\quad= 2 \cdot (2C)^{M \cdot [(\eta_A - 1) \cdot M^{\eta_A - 1} + 1]} \cdot \RNBC^{M^{\eta_A - 1}}\\
        &\quad= 2 \cdot (2C)^{(\eta_A - 1) \cdot M^{\eta_A} + M} \cdot \RNBC^{M^{\eta_A - 1}}\\
        &\quad\leq 2 \cdot (2C)^{\eta_A \cdot M^{\eta_A}} \cdot \RNBC^{M^{\eta_A - 1}}\\
        &\quad\leq (2C)^{\eta_A \cdot M^{\eta_A} + 1} \cdot \RNBC^{M^{\eta_A - 1}}\\
        &\quad \leq (2C)^{\sigma \cdot [\eta_A \cdot M^{\eta_A} + 1]} \cdot \RNBC^{\sigma \cdot M^{\eta_A - 1}}\\
        &\quad= f(\eta_A, 0, \mu(G_A,\Pi_A,S), \nu(\Pi_A),\xi(R_B), \norm{\F}, \abs{S}).
    \end{align*}

    Similarly for $g$:
    \begin{align*}
        &g(\eta_A, 1, \mu(G_A,\Pi_A,S), \nu(\Pi_A),\xi(R_B), \norm{\F}, \abs{S})\\
        &\quad= 2 \cdot \QBC \cdot \bracks*{ 2\cdot (2C)^{(\eta_A - 2)\cdot M^{\eta_A - 1} + M + 1} \cdot \RNBC^{M^{\eta_A - 2}}}^{M \cdot (\eta_A - 1)}\\
        &\quad\leq 2 \cdot \QBC \cdot \bracks*{ 2\cdot (2C)^{(\eta_A - 1)\cdot M^{\eta_A} + M + 1} \cdot \RNBC^{M^{\eta_A - 1}}}^{M \cdot (\eta_A - 1)}\\
        &\quad \leq \QBC \cdot \bracks*{ 2\cdot (2C)^{(\eta_A - 1)\cdot M^{\eta_A} + M + 1} \cdot \RNBC^{M^{\eta_A - 1}}}^{M \cdot (\eta_A - 1) + \sigma}\\
        &\quad= g(\eta_A, 0, \mu(G_A,\Pi_A,S), \nu(\Pi_A),\xi(R_B), \norm{\F}, \abs{S}).
    \end{align*}

    Additionally, observe that decreasing any of the parameters $\eta_A$, $\mu(G_A,\Pi_A,S)$, $\nu(\Pi_A)$, or $\xi(R_B)$ does not increase the values of $f$ and $g$.

    \textbf{($G_A \setminus S$ is connected)} We follow the proof of their Claim 33 and note, as they do, that the size of $\R'_N$ and $\Q'$ are bounded by two calls to $f$ and $g$, respectively, where $\eta_A$ is strictly smaller, and $\sigma$ cannot become larger than $M$ without triggering a base case. Thus,
    \[
    \abs{\R'_N} \leq 2 \cdot (2C)^{M \cdot [(\eta_A - 1) \cdot M^{\eta_A - 1} + 1]} \cdot \RNBC^{M^{\eta_A - 1}},
    \]
    and
    \begin{align*}
    \abs{\Q'} &\leq \max\braces*{\QBC, 2 \cdot \QBC \cdot \bracks*{2\cdot (2C)^{(\eta_A - 2)\cdot M^{\eta_A - 1} + M + 1} \cdot \RNBC^{M^{\eta_A - 2}}}^{M \cdot (\eta_A - 2) + M}}.
    \end{align*}

    The term $(\eta_A - 2)$ might call your attention, as if $\eta_A = 1$, then $\eta_A - 2 = -1$. However, in that case, we have $M \cdot (\eta_A - 2) + M = 0$, which makes the second term in the maximum equal to $2 \cdot \QBC$. Thus, the second term is always greater than $\QBC$, and we can simply write
    \[
    \abs{\Q'} \leq 2 \cdot \QBC \cdot \bracks*{2\cdot (2C)^{(\eta_A - 2)\cdot M^{\eta_A - 1} + M + 1} \cdot \RNBC^{M^{\eta_A - 2}}}^{M \cdot (\eta_A - 1)}.
    \]
    \textbf{($G_A \setminus S$ is diconnected)} Following the proof of their Claim 33, we know that
    \[
    \abs{\R'_N} \leq \sum_\text{chosen\ $\Pi_{A_2}$} \sum_{R \in \tilde{\R}_N^{\Pi_{A_2}, \Pi_{A_1}}} \abs{\hat{\R}^{\Pi_{A_1}, \Pi_{A_2}, R}}.
    \]

    They note that the number of different choices for $\Pi_{A_2}$ is bounded by $C$, and that $\abs{\tilde{\R}_N^{\Pi_{A_2}, \Pi_{A_1}}}$ is bounded by $f$ of a graph where $G_A \setminus S$ is connected and has no greater elimination distance to $\F$-minor-free graphs. Moreover, $\abs{\hat{\R}^{\Pi_{A_1}, \Pi_{A_2}, R}}$ is bounded by $f$ where one of the parameters $\mu$, $\nu$, or $\xi$ has strictly decreased, and none has increased. Also, $\eta_A$ does not increase either. Recall that if $\sigma$ is greater than $M$, then the function evaluates to 1. Therefore,
    \begin{align*}
        \abs{\R'_N} &\leq \underbrace{C}_{\text{\#Options for } \Pi_{A_2}} \cdot \underbrace{2 \cdot (2C)^{M \cdot [(\eta_A - 1)\cdot M^{\eta_A - 1} + 1]} \cdot \RNBC^{M^{\eta_A - 1}}}_{\abs{\tilde{\R}_N^{\Pi_{A_2}, \Pi_{A_1}}}}\\ &\qquad\qquad \cdot \underbrace{(2C)^{(\sigma - 1) \cdot [\eta_A \cdot M^{\eta_A} + 1]} \cdot \RNBC^{(\sigma - 1) \cdot M^{\eta_A - 1}}}_{\abs{\hat{\R}^{\Pi_{A_1}, \Pi_{A_2}, R}}}\\
        &= (2C)^{(\eta_A - 1) \cdot M^{\eta_A} + M} \cdot (2C)^{(\sigma - 1) \cdot \eta_A \cdot M^{\eta_A} + (\sigma - 1) + 1} \cdot \RNBC^{\sigma \cdot M^{\eta_A - 1}}\\
        &\leq (2C)^{\eta_A \cdot M^{\eta_A}} \cdot (2C)^{(\sigma - 1) \cdot \eta_A \cdot M^{\eta_A} + \sigma} \cdot \RNBC^{\sigma \cdot M^{\eta_A - 1}}\\
        &= (2C)^{\sigma \cdot (\eta_A \cdot M^{\eta_A} + 1)} \cdot \RNBC^{\sigma \cdot M^{\eta_A - 1}}.
    \end{align*}

    Similarly, we have that
    \[
    \abs{\Q'} \leq \sum_{\text{chosen }\Pi_{A_2}} \parens*{
        \abs*{\tilde{\Q}^{\Pi_{A_2}, \Pi_{A_1}}} + \sum_{R \in \tilde{\R}_N^{\Pi_{A_2}, \Pi_{A_1}}} \abs*{\hat{\Q}^{\Pi_{A_1}, \Pi_{A_2}, R}}
    }
    \]
    Again, they note that the number of chosen prohibitions $\Pi_{A_2}$ is bounded by $C$. Moreover, $\abs*{\tilde{\Q}^{\Pi_{A_2}, \Pi_{A_1}}}$ is bounded by $g$ of a graph where $G_A \setminus S$ is connected and has no greater elimination distance to $\F$-minor-free graphs. Furthermore, $\abs{\tilde{\R}_N^{\Pi_{A_2}, \Pi_{A_1}}}$ is bounded by $f$ of a graph where $G_A \setminus S$ is connected and has no greater elimination distance to $\F$-minor-free graphs, and $\abs*{\hat{\Q}^{\Pi_{A_1}, \Pi_{A_2}, R}}$ is bounded by $g$ where one of the parameters $\mu$, $\nu$, or $\xi$ has strictly decreased, and none has increased. Also, $\eta_A$ does not increase either.

    Notice that the bound $g$ for $\abs*{\hat{\Q}^{\Pi_{A_1}, \Pi_{A_2}, R}}$ is not smaller than the bound $g$ for $\abs*{\tilde{\Q}^{\Pi_{A_2}, \Pi_{A_1}}}$. We thus simply multiply by two the bound for $\sum_{R \in \tilde{\R}_N^{\Pi_{A_2}, \Pi_{A_1}}} \abs*{\hat{\Q}^{\Pi_{A_1}, \Pi_{A_2}, R}}$ to get an upper bound for
    \[
        \abs*{\tilde{\Q}^{\Pi_{A_2}, \Pi_{A_1}}} + \sum_{R \in \tilde{\R}_N^{\Pi_{A_2}, \Pi_{A_1}}} \abs*{\hat{\Q}^{\Pi_{A_1}, \Pi_{A_2}, R}}.
    \]
    Therefore,
    \begin{align*}
        \abs{\Q'} &\leq \underbrace{C}_{\#\text{Options for } \Pi_{A_2}}
            \cdot 2 \cdot \underbrace{
                    2 \cdot (2C)^{M \cdot [(\eta_A - 1) \cdot M^{\eta_A - 1} + 1]} \cdot \RNBC^{M^{\eta_A - 1}}
                }_{\abs*{\tilde{\R}_N^{\Pi_{A_2}, \Pi_{A_1}}}}\\
            &\qquad\cdot \underbrace{
                    \QBC \cdot \bracks*{ 2\cdot (2C)^{(\eta_A - 1)\cdot M^{\eta_A} + M + 1} \cdot \RNBC^{M^{\eta_A - 1}}}^{M \cdot (\eta_A - 1) + \sigma - 1}
                }_{\abs*{\hat{\Q}^{\Pi_{A_1}, \Pi_{A_2}, R}}}\\
            &= \QBC\cdot 2 \cdot (2C)^{(\eta_A - 1) \cdot M^{\eta_A} + M + 1} \cdot \RNBC^{M^{\eta_A - 1}}\\
                &\qquad \cdot \bracks*{2 \cdot (2C)^{(\eta_A - 1)\cdot M^{\eta_A} + M + 1} \cdot \RNBC^{M^{\eta_A - 1}}}^{M \cdot (\eta_A - 1) + \sigma - 1}\\
            &= \QBC\cdot \bracks*{2 \cdot (2C)^{(\eta_A - 1)\cdot M^{\eta_A} + M + 1} \cdot \RNBC^{M^{\eta_A - 1}}}^{M \cdot (\eta_A - 1) + \sigma}.
    \end{align*}

    In both cases, we have given the desired bounds for $\abs{\RN}$ and $\abs{\Q^\star}$.
\end{proof}

The bound for the sizes of $\RN$ and $\Q^\star$ in \namedtheoremref{fact:inductive-main-lemma-ed} can be used to give a bound on the set $\abs{\Q^\star}$ in \namedtheoremref{fact:main-lemma}.
\begin{lemma}
\label{fact:explicit-bound-Q-star}
Let $G$ be as in \autoref{fact:main-lemma}, and let $\eta \coloneqq \EDF(G)$. In \autoref{fact:main-lemma}, the size of the set $\Q^\star$ is upper-bounded by
\[
    h(\F, \eta) \coloneqq \QBC \cdot \bracks*{ 2\cdot (2C)^{(\eta - 1)\cdot M^{\eta} + M + 1} \cdot \RNBC^{M^{\eta - 1}}}^{M \cdot \eta} \leq 2^{2^{\polynomial(\eta + \norm{\F})}}.
\]
% Calculations:
% &\leq 2^{\polynomial(\eta + \norm{\F})} \cdot \bracks*{ 2\cdot (2C)^{(\eta - 1)\cdot M^{\eta} + M + 1} \cdot 2^{2^{\polynomial(\eta + \norm{\F})}\cdot M^{\eta - 1}}}^{M \cdot \eta}\\
% &\leq 2^{\polynomial(\eta + \norm{\F})} \cdot \bracks*{ (2^{2^{\polynomial(\eta + \norm{\F})}})^{(\eta - 1)\cdot {2^{\polynomial(\eta + \norm{\F})}}^{\eta} + 2^{\polynomial(\eta + \norm{\F})} + 1} \cdot 2^{2^{\polynomial(\eta + \norm{\F})}\cdot {2^{\polynomial(\eta + \norm{\F})}}^{\eta - 1}}}^{2^{\polynomial(\eta + \norm{\F})} \cdot \eta}\\
% &\leq 2^{\polynomial(\eta + \norm{\F})} \cdot \bracks*{ 2^{2^{\polynomial(\eta + \norm{\F})}\cdot(\eta - 1)\cdot 2^{\polynomial(\eta + \norm{\F})\cdot\eta} + 2^{\polynomial(\eta + \norm{\F})} + 1} \cdot 2^{2^{\polynomial(\eta + \norm{\F})}\cdot 2^{\polynomial(\eta + \norm{\F})\cdot\eta - 1}}}^{2^{\polynomial(\eta + \norm{\F})} \cdot \eta}\\
% &\leq \bracks*{ 2^{2^{\polynomial(\eta + \norm{\F})}\cdot 2^{\polynomial(\eta + \norm{\F})} + 2^{\polynomial(\eta + \norm{\F})}} \cdot 2^{2^{\polynomial(\eta + \norm{\F})}\cdot 2^{\polynomial(\eta + \norm{\F})}}}^{2^{\polynomial(\eta + \norm{\F})}}\\

% \begin{align*}
%     C &\coloneqq 2^{\numberof(0, \eta, \eta + \norm{\F}, 0)} \leq 2^{2^{\polynomial(\eta + \norm{\F})}}\\
%     M &\coloneqq \eta + \numberof(0, \eta, \eta + \norm{\F}, 0) + \numberof(\eta \cdot \norm{\F}, \eta, \eta + \norm{\F}^2 + 1, \norm{\F}) \leq 2^{\polynomial(\eta + \norm{\F})}\\
% \end{align*}
\end{lemma}
\begin{proof}
    We simply apply function $g$ from \autoref{fact:explicit-bound-Q-star-inductive} with $G_A \coloneqq G$ and $S \coloneqq \emptyset$.
\end{proof}

In contrast to Jansen and Pieterse \cite[Lemma 3]{minor-hitting}, we obtain a bound of $2^{2^{\polynomial(\eta + \norm{\F})}}$ for the size of $\Q^\star$ instead of $2^{2^{2^{2^{\polynomial(\eta + \norm{\F})}}}}$. This is mainly due to changing the definition of $M$, and a more careful analysis of the size of $\RN$ and $\Q^\star$ in \autoref{fact:explicit-bound-Q-star-inductive}.

We use the bound on the size of $\Q^\star$ to give an explicit bound on the number of connected components in the output graph $G'$ of the algorithm in \namedtheoremref{fact:reduce-components}.

\begin{lemma}
\label{fact:explicit-bound-number-of-components}
Let $\gamma \coloneqq h(\F, \EDF(C))$ be as in \autoref{fact:explicit-bound-Q-star}. The number of connected components in the output graph $G'$ of the algorithm in \namedtheoremref{fact:reduce-components} is upper-bounded by $\delta \cdot \abs{X}^\alpha$, where
\begin{align*}
    \delta &\coloneqq \bracks*{(\mF + 1) \cdot 2^{(m_\F + 2)^2} + 1}^\gamma \cdot \bracks*{\gamma \cdot (\mF + \norm{\F} + 1) + 2} \leq 2^{2^{2^{\polynomial(\EDF(C) + \norm{\F})}}}\\
    \alpha &\coloneqq \nF\cdot(\mF + 1) \cdot \gamma + 1 \leq 2^{2^{\polynomial(\EDF(C) + \norm{\F})}}.
\end{align*}
\end{lemma}
\begin{proof}
We follow the proof of Lemma 6 in \cite{minor-hitting} to reach the bound. There, they iterate over each subset $\Q$ of size at most $\gamma$ of a set $\H$ consisting of connected $\nF$-restricted $X$-labeled graphs that have at most $\mF$ edges. Thus, they
iterate over at most
\begin{equation}
    \label{eq:bound-number-of-sets-Q}
    \bracks*{\numberof(\abs{X}, 0, \mF + 1, \nF) + 1}^\gamma \leq \bracks*{(\mF + 1) \cdot 2^{(m_\F + 1)^2} \cdot (\abs{X}^{\nF} + 1)^{\mF + 1} + 1}^\gamma
\end{equation}
sets $\Q$. For each one of these sets, they mark at most $\tau$ connected components of $G \setminus X$, where
\begin{equation}\begin{aligned}
    \tau &\coloneqq \abs{X} + 1 + \gamma \cdot (\abs{X} + \max_{H \in \F}(\abs{V(H)} + \abs{E(H)}))\\
    &\leq \abs{X} + 1 + \gamma \cdot (\abs{X} + m_\F + \norm{\F}).
    \label{eq:bound-components-marked-per-Q}
\end{aligned}\end{equation}

Multiplying \autoref{eq:bound-number-of-sets-Q} and \autoref{eq:bound-components-marked-per-Q} and assuming $\abs{X} \geq 1$ gives

\begin{align*}
    &\bracks*{(\mF + 1) \cdot 2^{(m_\F + 1)^2} \cdot (\abs{X}^{\nF} + 1)^{\mF + 1} + 1}^\gamma \cdot \bracks*{\abs{X} + 1 + \gamma \cdot (\abs{X} + m_\F + \norm{\F})}\\
    &\quad\leq \bracks*{(\mF + 1) \cdot 2^{(m_\F + 1)^2} + 1}^\gamma \cdot (\abs{X}^{\nF} + 1)^{(\mF + 1) \cdot \gamma} \cdot \bracks*{(\gamma + 1) \cdot \abs{X} + \gamma \cdot (\mF + \norm{\F}) + 1}\\
    &\quad\leq \bracks*{(\mF + 1) \cdot 2^{(m_\F + 1)^2} + 1}^\gamma \cdot (\abs{X}^{\nF} + 1)^{(\mF + 1) \cdot \gamma} \cdot \bracks*{\gamma + 1 + \gamma \cdot (\mF + \norm{\F}) + 1} \cdot \abs{X}\\
    &\quad= \bracks*{(\mF + 1) \cdot 2^{(m_\F + 1)^2} + 1}^\gamma \cdot \bracks*{\gamma \cdot (\mF + \norm{\F} + 1) + 2} \cdot (\abs{X}^{\nF} + 1)^{(\mF + 1) \cdot \gamma} \cdot \abs{X}\\
    &\quad\leq \bracks*{(\mF + 1) \cdot 2^{(m_\F + 1)^2} + 1}^\gamma \cdot \bracks*{\gamma \cdot (\mF + \norm{\F} + 1) + 2} \cdot 2^{(\mF + 1) \cdot \gamma} \cdot \abs{X}^{\nF\cdot(\mF + 1) \cdot \gamma + 1}\\
    &\quad\leq \bracks*{(\mF + 1) \cdot 2^{(m_\F + 2)^2} + 1}^\gamma \cdot \bracks*{\gamma \cdot (\mF + \norm{\F} + 1) + 2} \cdot \abs{X}^{\nF\cdot(\mF + 1) \cdot \gamma + 1}\\
    &\quad= \delta \cdot \abs{X}^\alpha. \qedhere
\end{align*}
\end{proof}

Finally, we use the bound on the number of connected components to give an explicit bound on the size of the kernel obtained by \autoref{fact:poly-kernel}.

\begin{theorem}
    \label{fact:explicit-bound-kernel-size}
    Let $\F$ be a finite family of connected graphs, and let $\C$ be a minor-closed graph class such that $\EDF(\C) = \eta$. Let $\delta$ and $\alpha$ be as in \autoref{fact:explicit-bound-number-of-components}. If there exists a kernel of size $c \cdot \abs{Y}^p$ for \Fdeletion parameterized by the size of a given modulator $Y$ to $\F$-minor-free graphs, then there exists a kernel of size
    \[
        c \cdot \bracks*{(\delta + 1) \cdot \abs{X}}^{p \cdot \alpha ^\eta} \in \bigO\parens*{\abs{X}^{p \cdot2^{2^{\polynomial(\eta + \norm{\F})}}}}
    \]
    for \Fdeletion parameterized by the size of a given $\C$-modulator $X$.
\end{theorem}
\begin{proof}
    Consider an input $(G, X, k)$ to \Fdeletion parameterized by the size of a $\C$-modulator $X$. We will follow the proof in \autoref{fact:poly-kernel} and calculate an explicit bound in each step.

    ($\eta = 0$) Here, we simply apply the kernel for \Fdeletion parameterized by the size of a modulator to $\F$-minor-free graphs, which has size at most
    \[
        c \cdot \abs{X}^p \leq c \cdot ((\delta + 1)\cdot \abs{X})^{p \cdot \alpha^0}
    \] for some constant $c$ by hypothesis.

    ($\eta \geq 1$) In this case, we apply \namedtheoremref{fact:reduce-components} on the input to obtain a graph $G'$ and an integer $\Delta$. By \autoref{fact:explicit-bound-number-of-components}, the number of connected components of $G' \setminus X$ can be upper-bounded by $\delta \cdot \abs{X}^\alpha$.
    We then add to $X$ at most one vertex from each of the connected components of $G' \setminus X$, resulting in a modulator $X'$ to graphs with elimination distance at most $\eta - 1$ to $\F$-minor-free graphs. We thus have that
    \[
        \abs{X'} \leq \abs{X} + \delta \cdot \abs{X}^\alpha.
    \]

    Next, we apply the inductive hypothesis on $G'$, $X'$ and $k - \Delta$ to obtain a kernel $(G^\star, X^\star, k^\star)$ for the original input. By our inductive hypothesis, the size of this kernel can be upper-bounded by
    \begin{align*}
        c \cdot \bracks*{(\delta + 1) \cdot \abs{X'}}^{p \cdot \alpha ^{\eta - 1}}
        &\leq c \cdot \bracks*{
            (\delta + 1) \cdot (\abs{X} + \delta \cdot \abs{X}^\alpha)
        }^{p \cdot \alpha ^{\eta - 1}}\\
        &\leq c \cdot \bracks*{
            (\delta + 1)^2 \cdot \abs{X}^\alpha
        }^{p \cdot \alpha ^{\eta - 1}}\\
        &\leq c \cdot \bracks*{
            (\delta + 1) \cdot \abs{X}
        }^{p \cdot \alpha ^{\eta}},
    \end{align*}
    which is what we wanted to show.
\end{proof}

\section{Lower bounds for forest-like graph families}
\label{sec:lower-bounds-treelike}
We present in \autoref{fact:lower-bound-treelike} another example of families of graphs $\F$ that are not necessarily biconnected, and are different from the ones considered in \autoref{sec:lower-bounds-not-biconnected}. These families also do not admit polynomial kernels for \Fdeletion parameterized by a given modulator to $\C$, when $\EDF(\C)$ is unbounded, unless $\NP \subseteq \coNP/\text{poly}$.

The difference with the families presented in \autoref{sec:lower-bounds-not-biconnected} is that all biconnected components must have less than three vertices, except one. In other words, each graph must be a forest with an ``extra'' biconnected component of at least three vertices.

\newcommand{\treelikeConnectedCondition}{\hyperref[item:treelike-connected]{\conditionTitle{(Connected)}}}
\newcommand{\treelikeNonMinorBlocksCondition}{\hyperref[item:treelike-non-minor-blocks]{\conditionTitle{(Non-minor components)}}}
\newcommand{\treelikePlanarCondition}{\hyperref[item:treelike-planar]{\conditionTitle{(Planar)}}}
\begin{theorem}
    \label{fact:lower-bound-treelike}
    Let ${\mathcal C}$ be a minor-closed class of graphs and let $\F$ be a finite family of graphs meeting the following conditions:

    \begin{enumerate}[label=(\roman*), align=left, leftmargin=!]
        \item[\conditionTitle{(Connected)}]\label{item:treelike-connected} Every graph $H \in \F$ is connected.
        \item[\conditionTitle{(Planar)}]\label{item:treelike-planar} $\F$ contains at least one planar graph.
        \item[\conditionTitle{(One big component)}] For every graph $H \in \F$, there is exactly one biconnected component of $H$ with at least three vertices.
        \item[\conditionTitle{(Non-minor components)}]\label{item:treelike-non-minor-blocks} Let $H$ and $H'$ be two (not necessarily distinct) graphs in $\F$. Denote by $B$ and $B'$ the unique biconnected components of $H$ and $H'$, respectively, that have at least three vertices. For every edge $e \in E(B)$, the biconnected component $B'$ is not a minor of $H - e$.
    \end{enumerate}

    If ${\mathcal C}$ has unbounded elimination distance to an $\F$-minor-free graph, then \Fdeletion does not admit a polynomial kernel in the size of a ${\mathcal C}$-modulator, unless $\NP \subseteq \coNP/\poly$.
\end{theorem}

\begin{figure}[ht]
    \centering
    \includegraphics[width=0.4\textwidth]{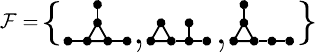}
    \caption{An example of a family $\F$ satisfying the conditions of \autoref{fact:lower-bound-treelike}. In this example, the unique biconnected component with at least three vertices is the same for each graph in $\F$, but this is not a requirement.}
    \label{fig:treelike-family}
\end{figure}

An example of a family $\F$ satisfying the conditions of \autoref{fact:lower-bound-treelike} can be seen in \autoref{fig:treelike-family}. Notice that the families presented in \autoref{fact:lower-bound-treelike} are incomparable with the ones from \autoref{fact:lower-bound-not-biconnected}: the former can contain biconnected components of size two while the latter cannot, and the latter can contain many biconnected components of size at least three while the former can contain only one.

Another (simpler) example of applicability of \autoref{fact:lower-bound-treelike} is given by $\F = \{$ \begin{tikzpicture}[baseline=0.5ex, scale=0.4]
    % Style for the vertices
    \tikzstyle{v} = [circle, draw, fill=black, inner sep=0pt, minimum size=3pt]

    % Vertices
    \node[v] (triangle-base-left) at (0,0) {};
    \node[v] (triangle-base-right) at (0.866, 0.5) {};
    \node[v] (triangle-top) at (0, 1) {};
    \node[v] (extra-vertex) at (2,0.5) {};

    % Edges
    \draw (triangle-base-left) -- (triangle-top);
    \draw (triangle-base-right) -- (triangle-top);
    \draw (triangle-base-left) -- (triangle-base-right);
    \draw (triangle-base-right) -- (extra-vertex);
\end{tikzpicture} $\}$, which correponds to the \textsc{(Cycles or Trees)-Vertex Deletion} problem.

The following observation will be useful for the proof of \autoref{fact:lower-bound-treelike}.

\begin{observation}
    \label{fact:H-minus-B-is-a-forest}
    Let $\F$ be a family of graphs satisfying conditions in \autoref{fact:lower-bound-treelike}. For every graph $H \in \F$, the graph $H \setminus V(B)$ is a forest, where $B$ is the unique biconnected component of $H$ with at least three vertices.
\end{observation}

\begin{proof}[Proof of \autoref{fact:lower-bound-treelike}]
    The proof will rely heavily on the one of \autoref{fact:lower-bound-not-biconnected}. We will again show a polynomial parameter transformation from \textsc{CNF-SAT} parameterized by the number of variables.

    We assume that $\F$ is minor-minimal, as otherwise we can replace $\F$ by the family of minor-minimal graphs in $\F$ without changing the result of \Fdeletion.

    By \autoref{prop:long-uniform-necklaces} and condition \planarCondition, the graph class $\C$ contains arbitrarily long uniform $\F$-necklaces; let $(H,\leftEndpoint,\rightEndpoint)$ be the structure of such necklaces. Denote by $B$ the unique biconnected component of $H$ with at least three vertices.

    We construct the graph $G$ and the budget $k$ in the same way as in the proof of \autoref{fact:lower-bound-not-biconnected}. To represent each variable $x_i$ we will use copies $\VariableGadget_i$ of $\NonBiconnectedVariableGadget_{H, B}$ from \autoref{def:NonBiconnectedVariableGadget}, and to represent each clause $C_j$ we will use adequately long $\F$-necklaces with structure $(H,\leftEndpoint,\rightEndpoint)$, which we will denote $\ClauseGadget_j$. The literal vertex $v_{j, \ell}$ of $\ClauseGadget_j$ corresponding to the literal $\ell$ in $C_j$ will be any vertex different from $\leftEndpoint_{j, \ell}$ and $\rightEndpoint_{j, \ell}$ in the copy of $B$ in the bead $\bead{j}{\ell}$ corresponding to $\ell$, which exists because $B$ has at least three vertices. Additionally, we add a copy of $H - e$ for some edge $e \in E(B)$ as the extra gadget. The connections between the variable gadgets, the clause gadgets, and the extra gadget will also be the same as in the proof of \autoref{fact:lower-bound-not-biconnected}. It remains to show that $G$ admits an \Fdeletion set of size $k$ if and only if the original formula $F$ is satisfiable.

    The backwards implication is the same as in the proof of \autoref{fact:lower-bound-not-biconnected}, so we only show the forward implication.

    $\implies)$ Suppose $F$ is satisfiable. We build the \Fdeletion set $Y$ similarly to the proof of \autoref{fact:lower-bound-not-biconnected}.

    For each variable $x_i$, if $x_i$ is set to true, we add all vertices in $V_{x_i}$ to $Y$, and if $x_i$ is set to false, we add all vertices in $V_{\neg x_i}$ to $Y$.
    For each clause $C_j$, we pick a literal $\ell$ in $C_j$ that is satisfied by the assignment. We add the vertex $v_{j, \ell'}$ to $Y$ for every literal $\ell' \neq \ell$ in $C_j$. For the bead corresponding to $\ell$, in contrast to the proof of \autoref{fact:lower-bound-not-biconnected}, we do not necessarily add the vertex $\leftEndpoint_{j, \ell}$ to $Y$. Indeed, \autoref{fig:treelike-cases} shows an example where adding $\leftEndpoint_{j, \ell}$ to $Y$ would leave an $H$-minor model in $G \setminus Y$. Instead we do the following:
    \begin{enumerate}
        \item\label{item:add-vertex-in-B} If every path from $\leftEndpoint_{j, \ell}$ to $\rightEndpoint_{j, \ell}$ in $\bead{j}{\ell}$ contains a vertex $w \in B$, then we add $w$ to $Y$.
        \item\label{item:add-vertex-separating-B} Otherwise, neither $\leftEndpoint_{j, \ell}$ nor $\rightEndpoint_{j, \ell}$ belong to $B$. By \autoref{fact:H-minus-B-is-a-forest}, there exists a unique path $P$ in $\bead{j}{\ell}$ connecting $\leftEndpoint_{j, \ell}$ and $\rightEndpoint_{j, \ell}$ that does not contain any vertex from $B$. Additionally, there exists a unique vertex $z \in P$ that disconnects $B$ from both $\leftEndpoint_{j, \ell}$ and $\rightEndpoint_{j, \ell}$. We add this vertex $z$ to $Y$.
    \end{enumerate}

    An illustration of the two cases can be seen in \autoref{fig:treelike-cases}.
    \begin{figure}[ht]
        \centering
        \includegraphics[width=0.58\textwidth]{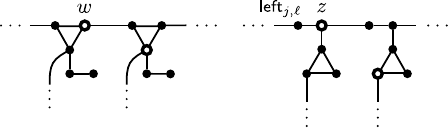}
        \caption{The two cases for adding a vertex to $Y$ for $H = $ \begin{tikzpicture}[baseline=0.5ex, scale=0.4]
    % Style for the vertices
    \tikzstyle{v} = [circle, draw, fill=black, inner sep=0pt, minimum size=3pt]
    
    % Vertices
    \node[v] (triangle-base-left) at (0,0) {};
    \node[v] (triangle-base-right) at (0.866, 0.5) {};
    \node[v] (triangle-top) at (0, 1) {};
    \node[v] (extra-vertex) at (1.666,0.5) {};
    \node[v] (extra-vertex-2) at (2.466,0.5) {};

    % Edges
    \draw (triangle-base-left) -- (triangle-top);
    \draw (triangle-base-right) -- (triangle-top);
    \draw (triangle-base-left) -- (triangle-base-right);
    \draw (triangle-base-right) -- (extra-vertex);
    \draw (extra-vertex) -- (extra-vertex-2);
\end{tikzpicture}. The hollowed-out vertices are the ones added to $Y$. Note that adding $\leftEndpoint_{j, \ell}$ to $Y$ instead of $z$ in the right case would leave an $H$-minor model in $G \setminus Y$.}
        \label{fig:treelike-cases}
    \end{figure}

    If case \ref{item:add-vertex-in-B} happens, then $Y$ hits all copies of $B$ in the variable and clause gadgets. Moreover, $Y$ separates $\leftEndpoint$ from $\rightEndpoint$ in the first and last beads of each clause gadget, respectively, thus hitting the last copies of $B$ formed between the extra gadget and the clause gadgets. Therefore, by \autoref{fact:biconnected-components-minor-model} and condition \treelikeNonMinorBlocksCondition, $Y$ is an \Fdeletion set.

    If case \ref{item:add-vertex-separating-B} happens, then $Y$ also hits all copies of $B$ in $G$, except for the ones in the beads from which we took the vertex $z$. These copies of $B$, however, are separated from most of the graph by $z$ and the corresponding $V_\ell$, and belong to a connected component of $G \setminus Y$ that is a minor of $H \setminus \set{z}$. This connected component has no $\F$-minor, as we assumed $\F$ to be minor-minimal. Thus, again by \autoref{fact:biconnected-components-minor-model}, and conditions \treelikeConnectedCondition\ and \treelikeNonMinorBlocksCondition, $Y$ is an \Fdeletion set of $G$.

    Lastly, as in the proof of \autoref{fact:lower-bound-not-biconnected}, this solution $Y$ has size exactly $k$. This concludes the forward direction and the proof of the theorem.
\end{proof}

\end{document}